\documentclass[12pt]{report}%
\usepackage{ amsmath, graphics, rotating}%

\begin{document}
\large
\begin{center}{\large\bf DE SITTER INVARIANCE AND 
A POSSIBLE MECHANISM OF GRAVITY}\end{center}

\begin{center} {\large Felix M. Lev} \end{center}
\begin{center} {\it Artwork Conversion Software Inc.,
1201 Morningside Drive, Manhattan Beach, CA 90266, USA (Email: felixlev314@gmail.com}) \end{center}

\begin{flushleft} {\it Abstract:}\end{flushleft}
It is believed that gravity will be explained in the framework of the existing quantum theory 
when one succeeds in eliminating divergencies at large momenta or small distances 
(although the phenomenon of gravity has been observed only at nonrelativistic momenta and large 
distances). We consider a quantum-mechanical description of systems of two free particles in de Sitter 
invariant quantum theory (i.e. the paper contains nothing but the two-body de Sitter kinematics). 
In our pure algebraic approach the cosmological constant problem 
does not arise. It is shown that a system can be simultaneously quasiclassical in relative 
momentum and energy only if
the cosmological constant is not anomalously small. We explicitly construct the relative
distance operator. The corresponding eigenvectors differ from standard ones at both, large and
small momenta. At large momenta they ensure fast convergence of quasiclassical wave functions. 
There also exists an anomalously large (but finite) contribution 
from small momenta, which is a consequence of the fact that the cosmological constant is finite. 
We argue that gravity might be a manifestation of this contribution.

\begin{flushleft} PACS: 11.30Cp, 11.30.Ly\end{flushleft}

\begin{flushleft} Keywords: quantum theory, de Sitter invariance, gravity\end{flushleft}

\vfill\eject

\tableofcontents

\vfill\eject

\chapter{Introduction}
\label{Ch1}

\begin{flushright} {\it To Skiff Nikolaevich Sokolov} \end{flushright}

\section{Motivation}
\label{S0}

Skiff Nikolaevich Sokolov has contributed to several areas of physics. I was very impressed by his
method of packing operators in relativistic theory of direct interactions \cite{pack} and his
proof with Shatny \cite{equiv} that all the three Dirac's forms of relativistic dynamics are
unitarily equivalent. And what is much more important, Skiff Nikolaevich is a man with
extremely high moral standards.

Many years ago Skiff Nikolaevich told me about his idea that gravity might be a direct interaction
and we discussed whether or not this might be in agreement with the results on the binary pulsar 
PSR B1913+16 (the present status of the problem is described e.g. in Refs. \cite{Taylor,WT}). 
If I remember Skiff Nikolaevich's arguments correctly, he 
said that until a direct and unique explanation of gravity is given, different possibilities should 
be investigated.

The results on binary pulsars are
treated as a strong indirect indication of the existence of gravitons. In reality only some 
radiation is seen. Then one describes this radiation assuming that it belongs to a binary pulsar. 
One constructs models
where the masses of the pulsars, their distances from each other etc. are adjustable 
parameters and it is assumed that the interaction of the pulsars with the interstellar matter
is weak. Then by fitting the parameters, the Einstein quadrupole formula is reproduced with a 
high accuracy. This is indeed a strong argument in favor of gravitational waves. At the same
time, this cannot be treated as a direct proof. In addition, the analysis performed
in the above references is based only on classical General Relativity (GR) and quantum effects are
not considered. Therefore, if the Einstein quadrupole formula is indeed valid, the fact that
on quantum level the main contribution to the standard nonrelativistic gravity is 
the graviton exchange, is an additional assumption based on 
belief that gravity can be described in quantum field theory (QFT) or its generalizations
(string theory, loop quantum gravity etc.). 

One might wonder why this belief is so strong in spite of the fact that numerous efforts to construct 
quantum theory of gravity without infinities has not been successful yet. Probably this belief is
based on the fact that in QED, electroweak theory and QCD the problem of infinities can be somehow 
circumvented and there are well known examples (especially in QED) of striking agreement between
theory and experiment. This is considered as much more important than the fact that from the
mathematical point of view, operators in QFT are not well defined. Therefore the usual 
statement is that the standard model describes almost everything and to include gravity one
needs only to generalize the existing theory to Planck distances. The present paper will be of no 
interest for physicists having such a philosophy.

At the same time, some famous physicists are not so optimistic.
For example, Weinberg believes \cite{Wein2} that the new theory
may be 'centuries away'. Although he has contributed much to QFT and believes that it can be treated 
{\it the way it is}, he also believes that it is a {\it low energy approximation to a deeper theory 
that may not even be a field theory, but something different like a string theory} \cite{Wein}.

Dirac was probably the least optimistic famous physicist. In his opinion \cite{DirMath}: {\it The agreement
with observation is presumably by coincidence, just like the original calculation of the hydrogen 
spectrum with Bohr orbits.
Such coincidences are no reason for turning a blind eye to the
faults of the theory. Quantum electrodynamics is rather like
Klein-Gordon equation. It was built up from physical ideas
that were not correctly incorporated into the theory and it
has no sound mathematical foundation}. 

The main problem is the choice of strategy for constructing a 
new quantum theory. Since nobody knows for sure which strategy 
is the best one, different approaches should be investigated. Our philosophy is based on
Dirac's advice given in Ref. \cite{DirMath}: {\it I learned
to distrust all physical concepts as a basis for a theory. 
Instead one should put one's trust in a mathematical scheme, 
even if the scheme does not appear at first sight to be
connected with physics. One should concentrate on
getting an interesting mathematics}.

The main motivation for the present work was to investigate whether or not gravity might be
simply a manifestation of the fact that nature is discrete and finite and therefore quantum
theory should be based not on complex numbers but on a finite field. We did not succeed
in giving a definite conclusion yet. However, we have realized that even in STANDARD
de Sitter invariant quantum theory KINEMATICS of a free two-body system has several very interesting
features, which seem to fully contradict our experience based on standard quantum theory.

One usually believes that there is no need to describe macroscopic systems (e.g. the Sun - Earth
or the Earth - Moon systems) in the framework of quantum mechanics since quasiclassical
approximation works for such systems with a very high accuracy. This belief is based mainly
on our experience in standard nonrelativistic quantum mechanics. In this theory there exist
well defined coordinate and momentum operators and it is easy to construct quasiclassical
wave functions having a nice behavior when the relative distance or relative momentum go
to infinity (there is a well known phrase that the Fourier transform of a Gaussian wave functions
also is a Gaussian wave function). That's why nobody poses a question of what a real wave
function of the Sun - Earth or Earth - Moon system is. 

However, since 30th of the last century it is well known that, when quantum mechanics 
is combined with relativity, there is no operator satisfying all the properties of the spatial 
position operator (see e.g. Ref. \cite{NW}). In other words, the coordinate cannot be exactly 
measured by itself even in 
situations when exact measurement is allowed by the nonrelativistic uncertainty principle.

In the introductory section of the well-known textbook \cite{BLP} simple arguments are given 
that for a particle with the mass $m$ the coordinate cannot be measured with the accuracy 
better than the Compton wave length $\hbar/mc$.
Therefore exact measurement is possible only either in the nonrelativistic limit 
(when $c\rightarrow \infty$) or
classical limit (when $\hbar\rightarrow 0$). 

It has also been known for many years (see e.g. Ref. \cite{time})
that there is no good operator, which can be related to time. In particular, we cannot construct 
a state which is the eigenvector of the time operator with the eigenvalue -5000 years BC or 
2010 years AD.

In particular, the quantity $x$ in the Lagrangian density 
$L(x)$ is only a parameter which becomes the spacetime coordinate in 
the nonrelativistic or classical limit. Note that even in the standard formulation of QFT, 
the Lagrangian is only an auxiliary tool for constructing Hilbert spaces
and operators. After this construction has been done, one can safely forget about Lagrangian and concentrate 
his or her efforts on calculating different observables. As Rosenfeld writes in his memoirs about Bohr 
\cite{Rosenfeld}: {\it His (Bohr) first remark ... was that field components taken at definite space-time
points are used in the formalism as idealization without immediate physical meaning; the only meaningful
statements of the theory concern averages of such fields 
components over finite space-time regions....Bohr 
certainly never showed any respect for the noble elegance of a Lagrangian principle.}

The above ideas became very popular in 60th (recall the
famous Heisenberg S-matrix program). The authors of Ref. \cite{BLP} claim that spacetime, Lagrangian and local 
quantum fields are rudimentary notions which will disappear in the ultimate quantum theory.
Since that time, no arguments questioning those ideas have been given, but in view of the great success of gauge
theories in 70th and 80th, these ideas are now almost forgotten.

The recent experimental data indicate (see e.g. Ref. \cite{Perlmutter}) that the cosmological
constant is not exactly zero and is positive. It is quite natural to treat this 
situation as an indication that de Sitter invariance
describes nature better than Poincare invariance. However, at present this symmetry
is not popular for several reasons. Firstly, the works where de Sitter
symmetry has been intensively investigated by group theoretical methods can be found only
in journals published mainly in 60th and 70th of the last century. 
In particular, it is rather odd that the excellent book by Mensky \cite{Men} on 
a theory of induced representations for physicists has been published only in Russian.
Secondly, from the point of view of QFT and its generalizations, de Sitter symmetry encounters serious
difficulties and the case of anti de Sitter symmetry looks much more preferable (see
e.g. Refs. \cite{Narlikar,Witten}). As noted by Witten \cite{Witten} {\it 'I don't know any clear-cut way to
get de Sitter space from string theory or M-theory'}. In Sect. \ref{S3} this question is discussed in
greater details. In any case, to the best of our knowledge, the problem of quasiclassical states in de Sitter 
invariant quantum theories has not been investigated at all.

One might say that since the Sun - Earth or Earth - Moon systems are classical and nonrelativistic, 
nonrelativistic quantum theory is a special case of relativistic one when $c\rightarrow \infty$ and
relativistic theory is a special case of de Sitter invariant theory when $R\rightarrow\infty$,
where $R$ is the radius of the Universe, our experience can be applied to such systems and
nothing unusual should be expected. However, in the spirit of the above Dirac's advice, it is
always desirable not only to rely on physical intuition but to verify mathematically everything
what can be verified. The results of the present paper show that the standard
intuition does not work and anomalously large (but finite) contributions to quasiclassical
wave functions arise not from the region of large momenta (as one might expect) but from the region
of very small momenta. This might be an indication that the true quantum theory is
based not on the field of complex numbers but on a finite field (or, in the spirit of 
Ref. \cite{Planat}, even on a finite ring). We suppose to consider this possibility in future 
publications, while in the present paper
we consider in detail the construction of quasiclassical states only in standard de
Sitter invariant quantum theory. 

The paper is organized as follows. In the next section and in Chap.
\ref{Ch2} we argue that gravity might be a special phenomenon, which is not described by analogy with 
the known local quantum field theories. We consider in detail a well known example of de Sitter
antigravity and note that this is an example of true direct interaction (i.e. an interaction,
which is not a consequence of exchange by intermediate particles). In Chap. \ref{Ch3} we develop
a pure algebraic approach for describing free two-body systems in de Sitter invariant theory.
In Chap. \ref{Ch4} the uncertainty relation between the de Sitter momentum and energy is discussed. 
It is shown that
a system can be simultanenously quasiclassical in momentum and energy only if the cosmological
constant is not anomalously small. The main results of the paper are obtained in Chap. \ref{Ch5}.
We explicitly construct a relative distance operator in dS theory and calculate its spectrum.
The properties of the corresponding eigenstates considerably differ from ones in the standard theory 
at both, large and small momenta. At large momenta the falloff of the eigenstates is very fast and
therefore quasiclassical wave functions rapidly decrease at such momenta. 
At the same time, there arises an anomalously large (but finite) contribution from the region of
small momenta. This is a consequence of the fact that the cosmological constant is finite. There
is no analog of such a behavior in standard theory. Finally, Chap. \ref{Ch6} is discussion.  

\section{Is gravity a special phenomenon or it should be combined with other interactions?}
\label{S1}

The main stream of the modern fundamental physics is that gravity should be treated by
analogy with electromagnetic, weak and strong interactions and eventually all these interactions
will be unified. In such an approach, gravitational interaction is treated as a consequence of 
the graviton exchange and in the nonrelativistic approximation the Newton gravitational law is obtained 
from the diagram of the one-graviton exchange in the same way as the Coulomb law is obtained from the
diagram of the one-photon exchange. On the other hand, since in units where $\hbar=c=1$ the gravitational 
constant has the dimension of length squared, a gravitational theory constructed by analogy with QED
contains strong divergencies. Physicists adhering to QFT do not treat this fact as a major drawback
of the theory but rather as a technical difficulty, which will be resolved by modifying the theory
at small distances.  

A rather unusual property of gravity is that it is the most known and, at 
the same time, the least understandable. In several aspects gravity is fully different from the electroweak 
and strong interactions. The phenomenon of gravity is known only at macroscopic level and only in nonrelativistic 
and post-Newtonian approximations. At this level gravity is universal, i.e. it applies to any bodies
regardless of whether they have electric charges, magnetic moments etc. Therefore it is interesting 
to investigate a possibility that the mechanism of gravity is more fundamental (and more simple)
than the mechanism of other interactions. 

If one accepts that de Sitter invariance is a good (approximate or exact) symmetry then there exists
another universal phenomenon, which is well known: the universal de Sitter antigravity. 
Consider a very simple example of two free particles. Then each particle is described by an 
irreducible representation (IR) of de Sitter (dS) algebra, and the fact that the particles
are free means (by definition) that the two-body system is described by the tensor product of the respective IRs. 
However, when we describe such a system in Poincare invariant terms, it looks like the particle
interact although no interaction have been introduced 'by hands'. On classical level the nature of this 
"interaction" is obvious since
it is well known that in the dS space the force of repulsion between two
particles is proportional (not inversely proportional!) to the distance between them and the relative acceleration of
"free" particles in the dS space under no circumstances can be zero. 

It is worth noting that the dS antigravity might be treated as a true direct interaction, i.e. it is not 
a manifestation of
exchange by any virtual particles. It is rather strange that in the vast literature on de Sitter
antigravity nobody pointed out to this fact. A typical objection against direct interactions 
is that any interaction
should propagate with a final velocity. In particular, if a position of a particle is changed then the other
particle should feel this change not instantly but only after some period of time. This seems to be not the
case if the interaction depends only on the relative distance between the particles. However, the 
classical dS antigravity
is not a typical interaction on the Poincare background but simply an inherent property of the dS space.
There is no way to exclude this "interaction" by letting some interaction constant to be zero. The relative
acceleration of two particles in the dS space is proportional to 1/$R^2$ where $R$ is the radius of the dS
space (a detailed discussion is given in the next chapter). Therefore one might have a temptation to treat 
1/$R^2$ as an interaction constant. However, this 
"interaction constant" disapears only in the formal limit $R\rightarrow\infty$ when the dS space does not exist 
anymore (it becomes the Poincare one). In other words, it is reasonable to say that the universal de Sitter
antigravity is not an interaction at all but simply an inherent property of de Sitter invariance.

The above example poses a problem of whether all the existing interactions in nature are in fact not true 
interactions but effective interactions arising as a result of transition from a higher symmetry to a lower
one. This idea has something in common with ones in the string theory that the existing interactions
represent a way how the extra dimensions are compactified. However, in our formulation we do not consider
geometries on extremely small distances, since as noted above, they have no physical meaning.

Since in many aspects gravity is fully different from the electroweak and strong 
interactions (see the above discussion), it is interesting to investigate a possibility that gravity is not 
an interaction at all but simply an inherent property of nature, 
which cannot be excluded by letting some interaction constant to be zero. The
results of the paper might be an indication that this is the case.  
      
\section{Elementary particles in quantum theory}
\label{S2}

Let us consider how one should define the notion of elementary particles. 
Although particles are observable and fields are not, 
in the spirit of QFT, fields are more fundamental 
than particles, and a possible definition is as 
follows \cite{Wein1}: {\it It is simply a particle whose
field appears in the Lagrangian. It does not matter if
it's stable, unstable, heavy, light --- if its field 
appears in the Lagrangian then it's elementary, 
otherwise it's composite.}
    
Another approach has been developed by Wigner in his 
investigations of unitary irreducible representations (UIRs)
of the Poincare group \cite{Wigner}. In view 
of this approach, one might postulate that a particle 
is elementary if the set 
of its wave functions is the space of a UIR 
of the symmetry group in the given theory. 

Although in standard well-known theories (QED, electroweak 
theory and QCD) the above approaches are equivalent, 
the following problem arises. The symmetry 
group is usually chosen as a group of motions of 
some classical 
manifold. How does this agree with the above discussion 
that quantum theory in the operator formulation should 
not contain spacetime? A possible answer is as follows. 
One can notice that for calculating observables (e.g. the
spectrum of the Hamiltonian) we need in fact not a 
representation of the group but a representation of its 
Lie algebra by Hermitian operators. After such a 
representation has been constructed, we have only 
operators acting in the Hilbert space and this is all 
we need in the operator approach. The representation 
operators of the group are needed only if it is 
necessary to calculate some macroscopic 
transformation, e.g. time evolution. In the approximation 
when classical time is a good approximate parameter, 
one can calculate evolution, but nothing guarantees 
that this is always the case (e.g. at the very early
stage of the Universe).  
Let us also note that in
the stationary formulation of scattering theory, the
S-matrix can be defined without any mentioning of
time (see e.g. Ref. \cite{Kato}). For these reasons 
we can assume that on quantum level the symmetry 
algebra is more fundamental than the symmetry group.
We will consider only representations of Lie algebras by Hermitian
operators and IRs will imply irreducible representations.

Consider for illustration a well-known example of 
nonrelativistic quantum mechanics. Usually
the existence of the Galilei spacetime is assumed from
the beginning. Let $({\bf r},t)$ be the space-time coordinates
of a particle in that spacetime. Then
the particle momentum operator is $-i\partial/\partial {\bf r}$
and the Hamiltonian describes evolution by the Schroedinger equation.
We believe that the notion of empty space is not physical since the
existence of space without particles contradicts basic principles of
quantum theory. Indeed, if we accept existence of a space without particles
than nothing can be measured in such a space. In addition, the assumption
that the notion of classical space $R^3$ can be applied on quantum level
contradicts the existence of particles with half-integer spins.
Indeed, in $R^3$ there can be no continuous functions acquiring the
factor -1 after rotation by $2\pi$. The only way to describe such
particles in the standard approach is to replace the group SO(3) by its
covering group SU(2) i.e. in fact to acknowledge that the notion of $R^3$
on quantum level is not applicable.

In our approach one starts from IR of the Galilei algebra.
The momentum operator and the Hamiltonian represent four of ten
generators of such a representation. If it is implemented in 
a space of functions 
$\psi({\bf p})$ then the momentum operator is simply the 
operator of multiplication by ${\bf p}$. Then the
position operator can be {\it defined} as 
$i\partial/\partial {\bf p}$
and time can be {\it defined} as an evolution parameter such that 
evolution is described by the Schroedinger equation with
the given Hamiltonian. In this approach the problem with half-integer
spins does not arise since the Lie algebras so(3) and su(2) are equivalent.
Mathematically the both approaches are related to each other by the Fourier
transform. However, the philosophies behind them are
essentially different. In the second approach there is 
no empty spacetime and the spacetime coordinates have a physical 
meaning only if
there are particles for which the coordinates can be measured.

Summarizing our discussion, we assume that, 
{\it by definition}, on quantum level a Lie algebra is 
the symmetry algebra if there exist physical
observables such that their operators  
satisfy the commutation relations characterizing the
algebra. Then, a particle is called elementary if the 
set of its wave functions is a space of IR of this algebra. 
Such an approach is in the spirit of that considered 
by Dirac in Ref. \cite{Dir}. In the present paper we consider
de Sitter invariant quantum theory (dS theory). Therefore elementary
particles in this theory should be described by IRs of the
de Sitter algebra so(1,4). In the next chapter we give a
detailed description of such IRs.

\chapter{Basic properties of de Sitter invariant 
quantum theories}
\label{Ch2}

\section{de Sitter invariance}
\label{S3}
 
As already mentioned, the original motivation for this work was
to investigate whether
the standard gravitational effects can be obtained in the
framework of a free theory. In the standard nonrelativistic
approximation gravity can be described by adding the
term $-Gm_1m_2/r$ to the nonrelativistic Hamiltonian,
where $G$ is the gravitational constant, $m_1$ and $m_2$
are the particle masses and $r$ is the distance between
the particles. Since the kinetic energy is always positive,
the free nonrelativistic Hamiltonian is positive definite
and therefore there is no way to obtain gravity in the
framework of the free theory. Analogously, in Poincare
invariant theory the spectrum of the free two-body mass
operator belongs to the interval $[m_1+m_2,\infty )$ 
while the existence of gravity necessarily requires
that the spectrum should contain values less than
$m_1+m_2$.

In theories where the invariance group is the anti
de Sitter (AdS) group SO(2,3), the structure of IRs
of the so(2,3) algebra is well known (see e.g. Ref.
\cite{Evans}). In particular, for positive
energy IRs the AdS Hamiltonian has the spectrum in
the interval $[m,\infty )$ and $m$ has the meaning 
of the mass. Therefore the situation is pretty much
analogous to that in Poincare invariant theories.
In particular, the free two-body mass operator again
has the spectrum in the interval $[m_1+m_2,\infty )$
and therefore there is no way to reproduce gravitational
effects in the free AdS invariant theory.

Consider now the case when the dS algebra 
so(1,4) is chosen as the symmetry algebra. It is 
well known that in IRs of the dS 
algebra, the dS Hamiltonian is not positive definite and
has the spectrum in the interval $(-\infty,+\infty)$ 
see e.g. Refs. \cite{Men,Dobrev,Moy,lev1,lev1a,lev3}). 
Note also that in contrast to the AdS
algebra so(2,3), the dS one does not have a supersymmetric
generalization. For this and other reasons it was believed 
that the dS group or algebra were not suitable for constructing 
elementary particle theory. 

In the framework of QFT in curved spacetime (see e.g.
Refs. \cite{Narlikar,Witten} and references therein) the choice
of SO(1,4) as the symmetry group encounters 
serious difficulties. Our approach fully differs from that in Refs. \cite{Narlikar,Witten}.
In particular, we do not require the existence of empty spacetime (see
the discussion in Sect. \ref{S2}). It has been also shown in Ref.
\cite{jpa1}, that a correct interpretation of IRs of the
so(1,4) algebra requires that one IR should describe a particle and
its antiparticle simultaneously. Then the theory 
with the dS symmetry become consistent (see Ref. \cite{jpa1}
for details). As already noted, the existing experimental data
indicate that the cosmological constant is positive. The nomenclature is such that the 
positive cosmological
constant corresponds to the SO(1,4) symmetry and the negative one -
to the SO(2,3) symmetry.

It is well known that the group SO(1,4) is the symmetry 
group of the four-dimensional manifold in the five-dimensional 
space, defined by the equation 
\begin{equation}
 x_1^2+x_2^2+x_3^2+x_4^2-x_0^2=R^2
\label{1}
\end{equation}
where a constant $R$ has the dimension of length.
The quantity $R^2$ is often written as $R^2=3/\Lambda$ where
$\Lambda$ is the cosmological constant. 
Elements of a map of the point $(0,0,0,0,R)$ (or $(0,0,0,0,-R)$) 
can be parametrized by 
coordinates $(x_0,x_1,x_2,x_3)$. If $R$ is very large then such a
map  proceeds to Minkowski space and the action of the dS group 
on this map --- to  the action of the Poincare group.

\begin{sloppypar}
In units $\hbar/2=c=1$, the spin of any particle is
always an integer and for the normal relation between spin and 
statistics, the spin of fermions is odd and the spin of bosons
is even. In this system of units the representation generators 
of the SO(1,4) group
$M^{ab}$ ($a,b=0,1,2,3,4$, $M^{ab}=-M^{ba}$) should satisfy the
commutation relations
\begin{equation}
[M^{ab},M^{cd}]=-2i (\eta^{ac}M^{bd}+\eta^{bd}M^{as}-
\eta^{ad}M^{bc}-\eta^{bc}M^{ad})
\label{2}
\end{equation}
where $\eta^{ab}$ is the diagonal metric tensor such that
$\eta^{00}=-\eta^{11}=-\eta^{22}=-\eta^{33}=-\eta^{44}=1$.
\end{sloppypar}

An important observation is as follows. If we accept that the
symmetry on quantum level means that proper commutation 
relations are satisfied (see Sect. \ref{S2}) then Eq. (\ref{2})
can be treated as the {\it definition} of the dS symmetry on 
that level. In this formulation the dS symmetry looks much more natural
than the Poincare symmetry where six operators are angular momenta and
the remaining four - linear momenta. Mirmovich has proposed a 
hypothesis \cite{Mirmovich} that only
quantities having the dimension of the angular momentum
can be fundamental. Our definition of the dS symmetry on
quantum level does not involve the cosmological constant at 
all. It appears only if one is interested in interpreting results in terms of
the dS spacetime or in the Poincare limit. 

\section{Comments on popular statements about fundamental physics}
\label{Comments}

In view of our definition of the dS symmetry on quantum level, we
would like to comment several popular statements widely discussed
in the literature.

If one assumes that spacetime is fundamental then in the spirit of GR 
it is natural to think that the empty spacetime is flat, i.e. that the cosmological 
constant is equal to zero. This was the subject of the well-known dispute between 
Einstein and de Sitter. In QFT with the Poincare background, the cosmological constant 
is given by a contribution of vacuum diagrams, and the problem is to explain why it 
is so small. The theory is based on the existence of empty spacetime and on the
assumption that the gravitational constant $G$ is fundamental. Meanwhile, as noted in Ref.
\cite{Uzan}, {\it "Contrary to most of the other fundamental constants, as the precision 
of the measurements increased, the disparity between the measured values of $G$ also
increased. This led the CODATA in 1998 to raise the relative uncertainty for $G$ from
0.013\% to 0.15\%".} Several well known physicists, including Dirac, discussed a possibility
that $G$ is time dependent. In view of the fact that only three digits in $G$ are known for sure,
its value has been measured only within a period of 210 years and the Universe exists at least
thirteen billion years, such a possibility
by no way can be excluded (this observation was pointed out to me by Volodia Netchitailo).
However, physicists believing in QFT assume that $G$ is fundamental, and as argued by the third
author in Ref. \cite{Okun}, in the string theory the fundamental role is played by the string
constant $\lambda_s$. In particular, there is a belief that quantum gravity will manifest itself at
Planck distances. Since the pure gravitational theory based on QFT contains only the constant
$G$, and $\Lambda$ has the dimension $(length)^{-2}$, the value of $\Lambda$ in this theory
is of order $G^{-2}$ and the discrepancy with observations is approximately 122 orders
of magnitude. In the spirit of the modern philosophy that divergencies and mathematical
inconsistencies of QFT do not represent fundamental drawbacks of the theory but only 
technical difficulties,
a conclusion is made that the cosmological constant problem is a fundamental problem
of modern physics. We believe, however, that this problem (sometimes it is called the old 
cosmological constant problem \cite{Ishak}) is fully artificial since it is not reasonable
to draw conclusions from inconsistent theories.

If one accepts that symmetry on quantum level in our formulation is more fundamental, 
then the cosmological constant problem does not arise since Eq. (\ref{2}) does not
contain $\Lambda$ at all. As noted above, it appears only if one is interested in 
interpreting results in terms of
the dS spacetime or in the Poincare limit. Then one might wonder why nowadays Poincare symmetry 
is so good approximate symmetry. This is rather a problem of cosmology but not quantum physics.
In particular, the cosmological constant might be not a constant at all.

If one accepts one of the main principles of quantum theory that any physical quantity
is described by an operator, then, as noted in the preceding chapter, the notions of space
and time are not fundamental on quantum level. Meanwhile, the success of gauge theories 
and new results in the string theory have revived the hope that Einstein's 
dream about geometrization of physics could be implemented. Einstein said
that the left-hand-side of his equation of GR,
containing the Ricci tensor, is made from gold while the right-hand-side containing the 
energy-momentum tensor of the matter is made from wood. Since that time a lot of efforts have 
been made to derive physics from geometry of spacetime.
As already noted, the modern ideas in the superstring theory are such that quantum gravity 
comes into play at Plank distances and all the existing interactions can be described if one 
finds how the extra dimensions are compactified. These investigations
involve very sophisticated methods of topology, algebraic geometry etc.

We believe that such investigations might be of mathematical interests and might give 
interesting results but cannot
lead to ultimate quantum theory. It is rather obvious that geometrical and topological ideas 
originate from our macroscopic experience. For example, the water in the ocean seems to be 
continuous and is described
with a good accuracy by equations of hydrodynamics. At the same time, we understand that this is 
only an approximation
and in fact the water is discrete. The notion of spacetime at Planck distances does not have 
any physical significance. In particular, strings and manifolds at such distances 
have no physical meaning. Therefore methods involving geometry and topology at Planck distances 
cannot give a reasonable physics. 

While the notion of spacetime coordinates for real bodies can be only a good approximation at 
some conditions, the notion 
of empty spacetime fully contradicts the basic principles of quantum theory that only measurable 
quantities can have a physical meaning. Indeed, coordinates of points which exist
only in our imagination cannot be related to any measurement. Note that even in GR, which is a pure
classical (i.e. non-quantum) theory, the meaning of reference frame is not quite clear. In standard 
textbooks (see e.g.
Ref. \cite{LLII}) the reference frame in GR is defined as a collection of weightless bodies, each
of which is characterized by three numbers (coordinates) and is supplied by a clock. It is
obvious that such a notion (which resembles ether) is not physical.
Meanwhile the modern theories typically begin with the 
background empty spacetime. Many years ago, when quantum
theory was not known, Mach proposed his famous principle, according to which the properties of 
space at a given point depend on the distribution of masses in the whole Universe. This principle is 
fully in the spirit of quantum theory.

The problem of choosing background spacetime is a subject of debates
between adherents of loop quantum gravity and superstring theory.
The former state that background independence is fundamental and the choice of the flat backgound 
in the string theory is not in spirit of GR. Many of those physicists believe that 
fundamental physical theory, where gravity is unified with other interactions,
should resemble main features of GR but on quantum level.  

There is no doubt that GR is a great achievement of theoretical physics and has achieved 
great successes in describing 
experimental data. At the same time, it is a pure classical theory 
fully based on classical spacetime. Therefore it is unrealistic to expect that successful 
quantum theory of gravity will be based on
quantization of GR. The results of GR should follow from quantum theory of gravity only 
in situations when spacetime coordinates of real bodies is a good approximation while in general the 
formulation of quantum theory might not involve spacetime at all.

In the literature the following question is also widely discussed.
How many independent dimensionful constants are needed for a complete description of nature? 
A recent paper \cite{Okun} represents a trialogue between three well known scientists:
M.J. Duff, L.B. Okun and G. Veneziano. The results of their discussions are summarized as
follows: {\it LBO develops the traditional approach with three constants, GV argues in favor of 
at most two (within superstring theory), while MJD advocates zero.} According to Weinberg 
\cite{W-units}, a possible definition of a fundamental constant might be such that it cannot
be calculated in the existing theory. 

Below we would like to give additional arguments that the fundamental physical theory should 
not contain any dimensionful constants at all.  The present status of
fundamental constants is described in a wide literature
(see e.g. Ref. \cite{Uzan}). The authors often note that even the term
"constant" for them is not adequate since variations of those "constants"
is not prohibited by our knowledge. Indeed, any dimensionful constant
is related to a macroscopic measurement carried out at certain
macroscopic conditions (at the present state of the Universe, in the
given place of the Universe etc.). So there is no guarantee that
any dimensionful constant is really constant.

Consider, for example the measurement of angular momentum. The 
result depends on the system of units. As noted above, in units
$\hbar/2=1$ the result is given by an integer. But we can
reverse the order of units and say that in units where the momentum
is a dimensional integer $l$, its value in $g\cdot cm^2/sec$ is
$(1.05457162\cdot 10^{-27}\cdot l/2)g\cdot  cm^2/sec$. Which of those two values
has more physical significance? The meaning of $l$ 
is clear: it shows how big the angular momentum is in comparison with the minimum 
nonzero value 1. At the same time, the measurement of the angular momentum in units 
$g\cdot cm^2/sec$ reflects only a historic fact that in macroscopic
conditions on the Earth in the period between the 18th and 21st
centuries people measured the angular momentum in such units.
If tomorrow morning we decide to conduct a new measurement of $\hbar$
and realize that its value is now $2.0\cdot 10^{-27}\cdot g\cdot  cm^2/sec$
then the same experiment will give for the angular momentum the same
value $l$ in the dimensional units since the commutation relations
between the angular momentum operators in the form 
$$[M_x,M_y]=2iM_z\quad [M_z,M_x]=2iM_y\quad [M_y,M_z]=2iM_x$$
are the same. On the other hand, in units $g\cdot cm^2/sec$ we will
have a new value $1.0\cdot 10^{-27}\cdot l\cdot g\cdot  cm^2/sec$,
which has any meaning only if we compare this value
with other angular momenta measured in units $g\cdot cm^2/sec$.  

As follows from Eq. (\ref{2}), in our formulation the free de Sitter theory  does not involve 
any arbitrary parameters (neither dimensionful nor dimesionless). One might say that 
commutation relations for the representation operators of the Poincare algebra also
can be written without any arbitrary parameters (for example, in Planck units) and
therefore Poincare invariance might be more fundamental than de Sitter one. As already noted,
an argument in favor of de Sitter invariance is that the dS algebra is more symmetric than the
Poincare algebra and, as already noted, the experimental data indicate that de Sitter
invariance is more relevant than Poincare one. We believe that there also exists the following
strong argument. 
If one accepts the idea that an ultimate quantum theory will be discrete and finite
then the only possibility is that the theory will be based on Galois fields (or
even Galois rings \cite{jmp0,lev3,Planat}). As argued in Ref. \cite{jmp0,lev3}, de Sitter invariant theory
can be generalized to a theory based on a Galois field while Poincare invariant theory
cannot. This is a consequence of the fact that de Sitter invariant theory can be formulated
in such a basis that all physical quantities have only discrete spectrum while in Poincare
invariant theories it is not possible to avoid quantities with continuous spectrum. 

The "real" elements of Galois fields can be represented as 0,1,...$p-1$, where $p$ is the 
characteristic of the field.
In that case the theory depends on the number $p$ and, as argued in Ref. \cite{FF},
such a situation is natural. At the same time, such a theory cannot contain dimensionful 
quantities in principle and the meaning of each measurable quantity is clear by analogy 
with the above example with the angular momentum (this is also in the spirit of Mirmovich's
idea \cite{Mirmovich} that only angular momenta are fundamental). 

In summary, we agree with the first author of Ref. \cite{Okun} that fundamental physics
should not contain dimensionful quantities at all and its predictions
should give only relations between dimensionless physical quantities. Dimensionful
units arise only when we want to express those quantities in terms of macroscopic
measurements on the Earth in the given historical period of time and in units we are
comfortable with. We also believe that de Sitter invariance is more fundamental than
Poincare invariance. There are no physical arguments in favor of treating Poincare invariance 
as superior over de Sitter invariance and de Sitter invariance - as a result of breaking down
Poincare invariance as a consequence of nonzero vacuum energy in the
theory with the Poincare background. 

\section{IRs of the so(1,4) algebra}
\label{S4}

As noted above, in our approach the description of  
elementary particles involves construction of IRs of the symmetry Lie algebra,
not the Lie group. However, in some cases it is technically convenient to derive 
properties of IRs of Lie algebras from well known results on UIRs of the corresponding
Lie groups. There exists a wide literature devoted to 
UIRs of the dS group and IRs of its Lie algebra
(see e.g. Refs. \cite{Dix1,Tak,Hann,Str,Schwarz,
Men,Moy,Dobrev,Mielke,Klimyk,lev1,lev1a}).
In particular the first complete mathematical classification of the
UIRs has been given in Ref. \cite{Dix1}, three well known
implementations of the UIRs have been first considered in Ref. \cite{Tak}
and their physical context has been first discussed in Ref. \cite{Hann}.

It is well known that for classification of UIRs of the dS group,
one should, strictly speaking, consider not the group SO(1,4) 
itself but its
universal covering group. The investigation carried out in
Refs. \cite{Dix1,Tak,Hann,Str,Moy} has shown that this 
involves only replacement of the SO(3) group by its universal 
covering group SU(2). Since this procedure is well known then
for illustrations we will work with the SO(1,4) group itself and 
follow a very elegant presentation for physicists in terms
of induced representations, given in the book \cite{Men}.
The elements of the SO(1,4) group can be described in the
block form
\begin{equation}
g=\left\|\begin{array}{ccc}
g_0^0&{\bf a}^T&g_4^0\\
{\bf b}&r&{\bf c}\\
g_0^4&{\bf d}^T&g_4^4
\end{array}\right\|\ 
\label{3}
\end{equation}
where 
\begin{equation}
\label{4}
{\bf a}=\left\|\begin{array}{c}a^1\\a^2\\a^3\end{array}\right\| \quad
{\bf b}^T=\left\|\begin{array}{ccc}b_1&b_2&b_3\end{array}\right\|
\quad r\in SO(3)
\end{equation}
(the subscript $^T$ means a transposed vector).

UIRs of the SO(1,4) 
group are induced from UIRs of the subgroup $H$ defined
as follows \cite{Str,Men,Dobrev}. Each element of $H$ can be uniquely
represented as a product of elements of the subgroups
SO(3), $A$ and ${\bf T}$: $h=r\tau_A{\bf a}_{\bf T}$ where 
\begin{equation}
\tau_A=\left\|\begin{array}{ccc}
cosh(\tau)&0&sinh(\tau)\\
0&1&0\\
sinh(\tau)&0&cosh(\tau)
\end{array}\right\|\ \quad
{\bf a}_{\bf T}=\left\|\begin{array}{ccc}
1+{\bf a}^2/2&-{\bf a}^T&{\bf a}^2/2\\
-{\bf a}&1&-{\bf a}\\
-{\bf a}^2/2&{\bf a}^T&1-{\bf a}^2/2
\end{array}\right\|\ 
\label{5}
\end{equation}

The subgroup $A$ is one-dimensional and the three-dimensional
group ${\bf T}$ is the dS analog of the conventional
translation group (see e.g. Ref. \cite{Men}). We hope it 
should not cause misunderstandings when 1 is used in its
usual meaning and when to denote the unit element of the
SO(3) group. It should also be clear when $r$ is a true
element of the SO(3) group or belongs to the SO(3) subgroup
of the SO(1,4) group. 

Let $r\rightarrow \Delta(r;{\bf s})$ be a UIR of the group
SO(3) with the spin ${\bf s}$ and 
$\tau_A\rightarrow exp(i\mu\tau)$ be a
one-dimensional UIR of the group $A$, where $\mu$ is a real
parameter. Then UIRs of the group $H$ used for inducing to
the SO(1,4) group, have the form
\begin{equation}
\Delta(r\tau_A{\bf a}_{\bf T};\mu,{\bf s})=
exp(i\mu\tau)\Delta(r;{\bf s})
\label{6}
\end{equation} 
We will see below that $\mu$ has the meaning of the dS
mass and therefore UIRs of the SO(1,4) group are
defined by the mass and spin, by analogy with UIRs
in Poincare invariant theory.

Let $G$=SO(1,4) and $X=G/H$ be a factor space (or
coset space) of $G$ over $H$. The notion of the factor 
space is well known (see e.g. Refs. 
\cite{Dobrev,Str,Men}).
Each element $x\in X$ is a class containing the
elements $x_Gh$ where $h\in H$, and $x_G\in G$ is a
representative of the class $x$. The choice of
representatives is not unique since if $x_G$ is
a representative of the class $x\in G/H$ then
$x_Gh_0$, where $h_0$ is an arbitrary element
from $H$, also is a representative of the same 
class. It is well known that $X$ can be treated 
as a left $G$ space. This means that if $x\in X$
then the action of the group $G$ on $X$ can be
defined as follows: if $g\in G$ then $gx$ is a
class containing $gx_G$ (it is easy to verify
that such an action is correctly defined). 

As noted above, although we
can use well known facts about group representations,
our final goal is the construction of the
generators. The explicit form of the generators $M^{ab}$
depends on the choice of representatives in
the space $G/H$. As explained in several
papers devoted to UIRs of the SO(1,4) group
(see e.g. Ref. \cite{Men}), to obtain
the  possible closest analogy between UIRs of
the SO(1,4) and Poincare groups, one should proceed
as follows. Let ${\bf v}_L$ be a representative 
of the Lorentz group in the factor space SO(1,3)/SO(3)
(strictly speaking, we should consider $SL(2,c)/SU(2)$).
This space can be represented as the well known velocity
hyperboloid with the Lorentz invariant measure
\begin{equation}
d\rho({\bf v})=d^3{\bf v}/v_0
\label{7}
\end{equation}
where $v_0=(1+{\bf v}^2)^{1/2}$. Let $I\in SO(1,4)$ be a
matrix which formally has the same form as
the metric tensor $\eta$. One can show 
(see e.g. Ref. \cite{Men} for details) that 
$X=G/H$ can be represented as a union of three
spaces, $X_+$, $X_-$ and $X_0$ such that 
$X_+$ contains classes ${\bf v}_Lh$, $X_-$
contains classes ${\bf v}_LIh$ and $X_0$ is of
no interest for UIRs describing elementary particles 
since it has measure zero relative to the spaces
$X_+$ and $X_-$.

As a consequence, the space of IR
of the so(1,4) algebra can be implemented as follows.  
If $s$ is the spin of the particle under 
consideration, then we
use $||...||$ to denote the norm in the space of 
IR of the su(2) algebra with the spin $s$. 
Then the space of IR in question is the space of 
functions $\{f_1({\bf v}),f_2({\bf v})\}$ on
two Lorentz hyperboloids with the range in the space of
IR of the su(2) algebra with the spin $s$ and such that
\begin{equation}
\int\nolimits [||f_1({\bf v})||^2+
||f_2({\bf v})||^2]d\rho({\bf v}) <\infty
\label{8}
\end{equation}

We see that, in contrast with IRs of the Poincare 
algebra (and AdS one), where IRs are implemented on
one Lorentz hyperboloid, IRs of the dS algebra can be
implemented only on two Lorentz hyperboloids, $X_+$
and $X_-$. As shown in Ref. \cite{jpa1}, this
fact (which is well known) has a natural explanation
if it is required that one IR should describe a
particle and its antiparticle simultaneously.

In the case of the Poincare and AdS algebras, the positive
energy IRs are implemented on an analog of $X_+$ and 
negative energy IRs - on an analog of $X_-$. Since the 
Poincare and AdS groups
do not contain elements transforming these spaces
to one another, the positive and negative energy IRs 
are fully independent. At the same time, the dS 
group contains 
such elements (e.g. $I$ \cite{Men,Dobrev,Mielke}) and for 
this reason its IRs cannot be implemented only on 
one hyperboloid. 

\begin{sloppypar}
In Ref. \cite{jpa1} we have described all the technical details
needed for computing the explicit form of the generators $M^{ab}$.
The action  of the generators on functions with the supporter in
$X_+$ has the form
\begin{eqnarray}
&&{\bf M}^{(+)}=2l({\bf v})+{\bf s},\quad {\bf N}^{(+)}==-2i v_0
\frac{\partial}{\partial {\bf v}}+\frac{{\bf s}\times {\bf v}}
{v_0+1}, \nonumber\\
&& {\bf B}^{(+)}=\mu {\bf v}+2i [\frac{\partial}{\partial {\bf v}}+
{\bf v}({\bf v}\frac{\partial}{\partial {\bf v}})+\frac{3}{2}{\bf v}]+
\frac{{\bf s}\times {\bf v}}{v_0+1},\nonumber\\
&& M_{04}^{(+)}=\mu v_0+2i v_0({\bf v}
\frac{\partial}{\partial {\bf v}}+\frac{3}{2})
\label{9}
\end{eqnarray}
where ${\bf M}=\{M^{23},M^{31},M^{12}\}$,
${\bf N}=\{M^{01},M^{02},M^{03}\}$,
${\bf B}=-\{M^{14},M^{24},M^{34}\}$, ${\bf s}$ is the spin operator,
and ${\bf l}({\bf v})=-i{\bf v}
\times \partial/\partial {\bf v}$.
At the same time, the action of the generators on 
functions with the supporter 
in $X_-$ is given by
\begin{eqnarray}
&&{\bf M}^{(-)}=2l({\bf v})+{\bf s},\quad {\bf N}^{(-)}==-2i v_0
\frac{\partial}{\partial {\bf v}}+\frac{{\bf s}\times {\bf v}}
{v_0+1}, \nonumber\\
&& {\bf B}^{(-)}=-\mu {\bf v}-2i [\frac{\partial}{\partial {\bf v}}+
{\bf v}({\bf v}\frac{\partial}{\partial {\bf v}})+\frac{3}{2}{\bf v}]-
\frac{{\bf s}\times {\bf v}}{v_0+1},\nonumber\\
&& M_{04}^{(-)}=-\mu v_0-2i v_0({\bf v}
\frac{\partial}{\partial {\bf v}}+\frac{3}{2})
\label{10}
\end{eqnarray}
\end{sloppypar}

In view of the fact that SO(1,4)=SO(4)$AT$ and $H$=SO(3)$AT$, there
also exists a choice of representatives which is probably even
more natural than that described above \cite{Men,Dobrev,Moy}. 
Namely, we can choose as
representatives the elements from the coset space SO(4)/SO(3).
Since the universal covering group for SO(4) is SU(2)$\times$SU(2)
and for SO(3) --- SU(2), we can choose as representatives the
elements of the first multiplier in the product SU(2)$\times$SU(2). 
Elements of SU(2) can be represented by the points $u=({\bf u},u_4)$
of the three-dimensional sphere $S^3$ in the four-dimensional
space as $u_4+i{\bf \sigma u}$ where ${\bf \sigma}$ are the Pauli
matrices and $u_4=\pm (1-{\bf u}^2)^{1/2}$ for the upper and
lower hemispheres, respectively. Then the calculation of the
generators is similar to that described above and the results 
are as follows.

The Hilbert space is now 
the space of functions $\varphi (u)$ on $S^3$ 
with the range in the space of the IR of the su(2) algebra 
with the spin $s$ and such that
\begin{equation}
\int\nolimits ||\varphi(u)||^2du <\infty
\label{11}
\end{equation}
where $du$ is the SO(4) invariant volume element on $S^3$.
The explicit calculation  shows  that  the  generators for  this
realization have the form
\begin{eqnarray}
&&{\bf M}=2l({\bf u})+{\bf s},\quad {\bf B}=2\imath u_4
\frac{\partial}{\partial {\bf u}}-{\bf s}, \nonumber\\
&& {\bf N}=-2\imath [\frac{\partial}{\partial {\bf u}}-
{\bf u}({\bf u}\frac{\partial}{\partial {\bf u}})]
+(\mu +3\imath){\bf u}-{\bf u}\times {\bf s}+u_4{\bf s},\nonumber\\
&& M_{04}=(\mu +3\imath)u_4+2\imath u_4{\bf u}
\frac{\partial}{\partial {\bf u}}
\label{12}
\end{eqnarray}

Since Eqs. (\ref{8}-\ref{10}) on the one hand and
Eqs. (\ref{11}) and (\ref{12}) on  the
other  are  the  different implementations of one  
and   the   same
representation, there exists a unitary operator transforming
functions $f(v)$ into $\varphi (u)$ and operators 
(\ref{9},\ref{10}) into
operators (\ref{12}). For example in the spinless case the
operators (\ref{9}) and (\ref{12}) are related to each other
by a unitary transformation 
\begin{equation}
\varphi (u)=exp(-\frac{\imath}{2}\,\mu \,lnv_0)v_0^{3/2}f(v)
\label{13}
\end{equation}
where ${\bf u}={\bf v}/v_0$. 

\section{Poincare limit}
\label{S5}

A general notion of contraction has been developed in 
Ref. \cite{IW}. In our case it can be performed
as follows. Let us assume that $\mu > 0$ and denote
$m=\mu /2R$, ${\bf P}={\bf B}/2R$ and $E=M_{04}/2R$.
Then, as follows from  Eq. (\ref{9}), in the limit
when $R\rightarrow \infty$, $\mu\rightarrow \infty$
but $\mu /R$ is finite,   
one obtains a standard representation of the
Poincare algebra for a particle with the mass $m$ such 
that ${\bf P}=m{\bf v}$ is the particle momentum
and $E=mv_0$ is the particle energy. In that case
the generators of the Lorentz algebra have the same form
for the Poincare and dS algebras. Analogously the
operators given by Eq. (\ref{10}) are contracted to
ones describing negative energy IRs of the Poincare
algebra.

In Sect. \ref{Comments} we argued that fundamental physical
theory should not contain dimensional parameters at all.
In this connection it is interesting to note that the
de Sitter mass $\mu$ has a clear meaning: it is a ratio
of the radius of the Universe $R$ (or $2R$) to the 
Compton wave length of the particle under consideration.
Therefore even for elementary particles the de Sitter
masses are very large. For example, if $R$ is of order
$10^{28}cm$ then the de Sitter masses of the electron,
the Earth and the Sun are of order $10^{39}$, $10^{93}$ 
and $10^{99}$, respectively.

In the standard interpretation of IRs
it is assumed that each element of the full 
representation space represents a possible physical
state for the elementary particle in question. 
It is also well known (see e.g. Ref. 
\cite{Dobrev,Men,Moy,Mielke})
that the dS group contains elements (e.g. $I$)
such that the corresponding representation operator
transforms positive energy states to negative energy
ones and {\it vice versa}. Are these properties 
compatible with the fact that in the Poincare
limit there exist states with negative energies?
This problem is discussed in detail in Ref. \cite{jpa1}.
It has been shown that the interpretation of IRs of the
so(1,4) algebra is consistent only we accept that one IR
describes a particle and its antiparticle simultaneously.
In this case the states with the negative energies can be
converted to states with the positive energy by using the second
quantization (by analogy with the interpretation of negative
energy solutions of the Dirac equation as positrons).

At present the phenomenon of gravity has been observed
only on macroscopic level, i.e. for particles which
cannot be treated as elementary. Then the question
arises whether they can be described by using the
results on IRs. The usual assumption is as follows.
In the approximation when it is possible to neglect
the internal structure of the particles (e.g. when
the distance between them is much greater than their
sizes), the structure of the internal wave function
is not important and one can consider only the part
of the wave function describing the motion of the
particle as the whole. This part is described by the
same parameters as the wave function of the elementary
particle. For this reason it is usually sufficient to describe
the motion of the macroscopic system as a whole by
using wave functions of IRs with zero spin.  

\section{de Sitter antigravity}
\label{S6}

Consider now the Poincare limit in the approximation
when $R$ is large but the first order corrections 
in $1/R$ to the conventional energy and momentum are
taken into account. By using the definitions of the
Poincare mass, energy and momentum from the 
preceding section and taking the dS generators in
the form (\ref{9}), one can obtain the expressions
for the conventional energy and momentum in first
order in $1/R$. For simplicity we assume that the
particles are spinless and nonrelativistic. Consider
a system of two free particles with the masses 
$m_1$ and $m_2$. Then the momentum and energy
operators for each particle are given by 
\begin{eqnarray}
&& {\bf P}_j={\bf p}_j+\frac{im_j}{R} \frac{\partial}
{\partial {\bf p}_j}\nonumber\\
&& E_j=m_j + \frac{{\bf p}_j^2}{2m_j} +\frac{i}{R}({\bf p}_j
\frac{\partial}{\partial {\bf p}}_j+\frac{3}{2})
\end{eqnarray}
\label{AG1}
where ${\bf p}_j=m{\bf v}_j$ and $j=1,2$.

The fact that the particle do not interact with each other 
implies (by definition) that the generators for the two-body system 
are equal to sums of the corresponding single-particle
generators. Adding the corresponding operators and
introducing the standard total and relative momenta
\begin{equation}
{\bf P}={\bf p}_1+{\bf p}_2\quad 
{\bf q} = (m_2{\bf p}_1-m_1{\bf p}_2)/(m_1+m_2)
\label{AG2}
\end{equation}
one can obtain the expressions for the momentum ${\bf P}$ 
and energy $E$ of the two-body system as a whole. Let $M$ be
the mass operator of the two-body system defined as
$M^2=E^2-{\bf P}^2$. Then a simple calculation shows
that in our approximation
\begin{equation}
M = m_1+m_2 +\frac{{\bf q}^2}{2m_{12}}+
\frac{i}{R}({\bf q}\frac{\partial}{\partial 
{\bf q}}+\frac{3}{2})
\label{AG3}
\end{equation}
where $m_{12}=m_1m_2/(m_1+m_2)$ is the reduced mass.

In spherical coordinates the nonrelativistic mass
operator can be written as
\begin{equation}
M_{nr}=\frac{q^2}{2m_{12}} + V,\quad 
V=\frac{i}{R}(q\frac{\partial}{\partial q}+\frac{3}{2})
\label{AG4}
\end{equation}
where $q=|{\bf q}|$.
Although this expression has been obtained in first
order in $1/R$, let us consider for illustrative
purposes the spectrum of this operator. 
It acts in the space of functions $\psi(q)$ such that
\begin{equation}
\int_{0}^{\infty}|\psi(q)|^2q^2dq <\infty
\label{AG5}
\end{equation}
and the eigenfunction $\psi_K$ of $M_{nr}$ with the
eigenvalue $K$ satisfies the equation
\begin{equation}
q\frac{d\psi_K}{dq}=\frac{iRq^2}{2m_{12}}\psi_K-
(\frac{3}{2}+iRK)\psi_K
\label{AG6}
\end{equation}
The solution of this equation is
\begin{equation}
\psi_K=\sqrt{\frac{R}{2\pi}}q^{-3/2}
exp(\frac{iRq^2}{4m_{12}}-iRKlnq)
\label{AG7}
\end{equation}
and the normalization condition is
\begin{equation}
(\psi_K,\psi_{K'})=\delta(K-K')
\label{AG8}
\end{equation}

The spectrum of the operator $M_{nr}$ obviously
belongs to the interval $(-\infty,\infty)$ and one
might think that this is unacceptable. Suppose
however that $f(q)$ is a wave function of some
state. As follows from Eq. (\ref{AG7}), the probability 
to have the value of the kinetic energy $K$ in this 
state is given by
\begin{equation}
c_K=\sqrt{\frac{R}{2\pi}}\int_{0}^{\infty}
exp(-\frac{iRq^2}{4m_{12}}+iRKlnq)f(q)\sqrt{q}dq
\label{AG9}
\end{equation}
If $f(q)$ does not on $R$ and $R$ is very large 
then $c_K$ will practically be
different from zero only if the integrand in Eq. 
(\ref{AG9}) has a stationary point $q_0$. It is
obvious that the stationary point is defined by
the condition $K=q_0^2/2m_{12}$. Therefore, for
negative $K$, when the stationary point is absent,
the value of $c_K$ will be very small.

We see that if one works only with a subset of 
wave functions not depending on $R$ (which is 
typically the case), then the existence of the
points of the spectrum of the two-body mass
operator with the values less than $m_1+m_2$
does not play an important role. 

\section{de Sitter antigravity in quasiclassical
approximation}
\label{classical}

In conventional quantum mechanics the motion of a
particle is quasiclassical if at each moment of time
$t=t_0$ the particle wave function satisfies the
following conditions (see e.g. Ref. \cite{LL}). In the
coordinate representation the function has a sharp
maximum at some ${\bf r}={\bf r}_0$, and the uncertainty
of the position $\Delta {\bf r}$ is much less than 
${\bf r}_0$. At the same time, in the velocity 
representation it should have a sharp maximum at some 
${\bf v}={\bf v}_0$, and the uncertainty
of the velocity $\Delta {\bf v}$ should be much less than 
${\bf v}_0$. In particular, the particle cannot be 
quasiclassical if it is at rest, i.e. ${\bf v}_0=0$.

As follows from this definition, the notion of
quasiclassical approximation necessarily implies that
the position and velocity operators are well-defined
and have a clear physical meaning. This is indeed the
case in conventional nonrelativistic quantum mechanics.
As already noted, in relativistic quantum theory there is 
no operator satisfying all the requirements for the position 
operator. In dS theories there
exists an analogous problem. In particular, as seen
from Eq. (\ref{9}), the operator ${\bf v}$ by itself
does not define the dS momentum (which is a physical
operator) uniquely, and the operator 
$i\partial /(m\partial {\bf v})$, which in nonrelativistic
quantum mechanics is the position operator in velocity
representation, does not define the physical Lorentz boost
operators uniquely. However, as noted in Sect. \ref{S5},
when $R$ is very large, the generators (\ref{9}) can
be contracted to standard generators of the IR of the
Poincare algebra. In this case the momentum ${\bf P}$ is 
exactly proportional to ${\bf v}$ and the proportionality
coefficient is the mass. Moreover, when the particle is
nonrelativistic, then, as follows from Eq. (\ref{9}),
the Lorentz boost operators are proportional to the
corresponding coordinate operators in velocity 
representation. 

We see that, at least when $R$ is large and
$|{\bf v}|\ll 1$, there exists a well-defined
quasiclassical approximation in the representation
when the generators are given by Eq. (\ref{9}). 
For example, the wave function can be chosen in the
form 
\begin{equation}
f({\bf v})=a({\bf v})exp(-im{\bf v}{\bf r}_0/2) 
\label{49}
\end{equation}
where $a({\bf v})$ has a sharp maximum at ${\bf v}={\bf v}_0$
with a width $|\Delta {\bf v}|\ll |{\bf v}_0|$ and such
that $|\partial a({\bf v})/\partial {\bf v}|\ll m|{\bf r}_0|$.
Note that the factor 1/2 in the exponent is a consequence of the
fact that we are working with units where $\hbar/2=1$.
A possible choice of $a({\bf v})$ is 
\begin{equation}
a({\bf v})=cv_0^{1/2}exp[-\frac{b^2}{2}({\bf v}-{\bf v}_0)^2]
\label{50}
\end{equation}
where the function is normalized to one 
(see Eq. (\ref{8})) if 
$$c=\sqrt{2}b^{3/2}/\pi^{3/4}.$$ 
Then the condition 
$|\Delta {\bf v}|\ll |{\bf v}_0|$ is satisfied if
$b|{\bf v}_0|\gg 1$ and the condition 
$|\Delta {\bf r}|\ll |{\bf r}_0|$ is satisfied if 
$b\ll m|{\bf r}_0|$. For macroscopic particles there 
exists a wide range of values $b$ such that these
conditions can be satisfied. Also, since the function (\ref{49})
has the Gaussian form, the integral defining the mean values
of the velocity and mass operators is rapidly convergent when
${\bf v}$ is far away from ${\bf v}_0$.

On classical level the effect of the additional
term in Eq. (\ref{AG3}) in comparison with the
standard free nonrelativistic expression can be
investigated as follows. We define the position
operator ${\bf r}$ as 
${\bf r}=2i(\partial /\partial {\bf q} )$.
Then the classical Hamiltonian of the internal
motion corresponding to the operator (\ref{AG3}) is
\begin{equation}
H({\bf r}, {\bf q}) =\frac{{\bf q}^2}{2m_{12}} +
\frac{{\bf r}{\bf q}}{R}
\label{AG10}
\end{equation}
From classical equations of motion it follows that
$d^2{\bf r}/dt^2={\bf r}/R^2$. It is well known that
in classical dS space there exists a universal 
repulsion (antigravity) the force is which is
proportional to the distance between particles.
Therefore the operator $V$ indeed corresponds to
the dS antigravity. 

\begin{sloppypar}
In classical mechanics there exist transformations 
of the Hamiltonian, which do not change classical equations 
of motions. One can easily verify
that classical equations of motion for the Hamiltonian
\begin{equation}
H({\bf r}, {\bf q}) =\frac{{\bf q}^2}{2m_{12}} -
\frac{m_{12}{\bf r}^2}{R^2}
\label{AG11}
\end{equation}
are the same as for the Hamiltonian (\ref{AG10}). 
\end{sloppypar}

Although the example of the dS antigravity is extremely simple, 
we can draw the following very important conclusions.

The first conclusion is that the standard classical
dS antigravity has been obtained from a quantum
operator without introducing any classical background.
When the position operator is defined as 
${\bf r}=2i(\partial /\partial {\bf q} )$ and time is
defined by the condition that the Hamiltonian is the
evolution operator then one recovers the classical
result obtained by considering a motion of particles
in the classical dS spacetime. This is an illustration
of the discussion in Chap. \ref{Ch1} and Sect. \ref{Comments} 
about the difference
between the standard approach, where the classical dS
spacetime is introduced from the beginning, and our one.

The second conclusion is as follows. We have considered
the particles as free, i.e. no interaction into the
two-body system has been introduced. However, we have
realized that when the two-body system in the dS 
theory is considered from the point of view of 
the Galilei invariant theory, the particles interact 
with each other. Although the reason of the effective 
interaction in our example is obvious, the existence of
the dS antigravity poses the problem whether other 
interactions, e.g. gravity, can be treated as a result of 
transition from a higher symmetry to Poincare or Galilei 
one.

The third conclusion is that the dS antigravity is a true direct
interaction since it is not a consequence of the exchange of 
virtual particles.

Finally, the fourth conclusion is that if in a free
theory the spectrum of the mass operator has values less
than $m_1+m_2$, this does not necessarily mean that the 
theory is unphysical.

The above discussion shows that in dS theories our
intuition based on nonrelativistic quantum mechanics still works at least when 
$R$ is large and ${\bf v}\ll 1$ if decomposition in powers of $1/R$ and ${\bf v}$ is
legitimate. In particular, the Lorentz boost
operator can be treated as an operator proportional to the position operator.
The problem arises how one should describe quasiclassical approximation without
decomposition in powers of $1/R$ and ${\bf v}$.
The only operators in our disposal are those defined by Eqs. (\ref{9}) or (\ref{12}).
In quasiclassical approximation the action of the Lorentz boost operator (\ref{12})
on the wave function (\ref{49}) is given by
\begin{equation}
{\bf N}f({\bf v})=-m\sqrt{1+{\bf v}^2}{\bf r}_0f({\bf v})
\label{AG12}
\end{equation} 
Therefore the position operator cannot be proportional anymore to the Lorentz boost operator. 
We will see below that it is possible to describe the two-body mass operator on pure algebraic
level without using decomposition in powers of ${\bf v}$ and $1/R$.

\chapter{de Sitter invariant quantum theory in 
su(2)$\times$su(2) basis}
\label{Ch3}

\section{IRs in the su(2)$\times$su(2) basis}
\label{S8}

Proceeding from the method of su(2)$\times$su(2) shift
operators, developed by Hughes \cite{Hug} for constructing
UIRs of the group SO(5), and following Ref. \cite{lev3},
we now give a pure algebraic description of IRs of
the so(1,4) algebra. It will be convenient for us to  deal
with the set of operators $({\bf J}',{\bf J}",R_{ij})$ ($i,j=1,2$)
instead  of $M^{ab}$.
Here ${\bf J}'$ and ${\bf J}"$ are two independent su(2) algebras
(i.e. $[{\bf J}',{\bf J}"]=0$).
In each of them one chooses as the basis the operators  
$(J_+,J_-,J_3)$ such that  
$J_1=J_++J_-$, $J_2=-\imath (J_+-J_-)$ and the commutation
relations have the form
\begin{equation}
[J_3,J_+]=2J_+,\quad [J_3,J_-]=-2J_-,\quad [J_+,J_-]=J_3
\label{14}
\end{equation}
The commutation relations of the operators ${\bf J}'$ and
${\bf J}"$  with $R_{ij}$ have the form
\begin{eqnarray}
&&[J_3',R_{1j}]=R_{1j},\quad [J_3',R_{2j}]=-R_{2j},\quad
[J_3",R_{i1}]=R_{i1},\nonumber\\
&& [J_3",R_{i2}]=-R_{i2},\quad
[J_+',R_{2j}]=R_{1j},\quad [J_+",R_{i2}]=R_{i1},\nonumber\\
&&[J_-',R_{1j}]=R_{2j},\quad [J_-",R_{i1}]=R_{i2},\quad
[J_+',R_{1j}]=\nonumber\\
&&[J_+",R_{i1}]=[J_-',R_{2j}]=[J_-",R_{i2}]=0,\nonumber\\
\label{15}
\end{eqnarray}
and the commutation relations of the operators $R_{ij}$ 
with each other have the form
\begin{eqnarray}
&&[R_{11},R_{12}]=2J_+',\quad 
[R_{11},R_{21}]=2J_+",\nonumber\\
&& [R_{11},R_{22}]=-(J_3'+J_3"),\quad 
[R_{12},R_{21}]=J_3'-J_3"\nonumber\\
&& [R_{11},R_{22}]=-2J_-",\quad [R_{21},R_{22}]=-2J_-'
\label{16}
\end{eqnarray}
The relation between the sets $({\bf J}',{\bf J}",R_{ij})$ and
$M^{ab}$  is given by
\begin{eqnarray}
&&{\bf M}={\bf J}'+{\bf J}", \quad {\bf B}={\bf J}'-{\bf J}",
\quad M_{01}=\imath (R_{11}-R_{22}), \nonumber\\
&& M_{02}=R_{11}+R_{22}, \quad  
M_{03}=-i(R_{12}+R_{21}),\nonumber\\
&&  M_{04}=R_{12}-R_{21}
\label{17}
\end{eqnarray}
Then it is easy to see that Eq. (\ref{2}) 
follows from Eqs. (\ref{15}-\ref{17}) and {\it vice versa}.

Consider the space of maximal  $su(2)\times su(2)$  vectors,
i.e.  such vectors $x$ that $J_+'x=J_+"x=0$. Then from
Eqs. (\ref{15}) and (\ref{16}) it follows that the operators
\begin{eqnarray}
&&A^{++}=R_{11}\quad  A^{+-}=R_{12}(J_3"+1)-
J_-"R_{11},  \nonumber\\
&&A^{-+}=R_{21}(J_3'+1)-J_-'R_{11}\nonumber\\
&&A^{--}=-R_{22}(J_3'+1)(J_3"+1)+J_-"R_{21}(J_3'+1)+\nonumber\\
&&J_-'R_{12}(J_3"+1)-J_-'J_-"R_{11}
\label{18}
\end{eqnarray}
act invariantly on this space.
The notations are related to the property  that
if $x^{kl}$  ($k,l>0$) is the maximal su(2)$\times$su(2)
vector and simultaneously
the eigenvector of operators $J_3'$ and $J_3"$ with the 
eigenvalues $k$ and $l$, respectively, then  $A^{++}x^{kl}$
is  the  eigenvector  of  the  same
operators with the values $k+1$ and $l+1$, $A^{+-}x^{kl}$ - the
eigenvector  with
the values $k+1$ and $l-1$, $A^{-+}x^{kl}$ - 
the eigenvector with the values  $k-1$ and $l+1$ and 
$A^{--}x^{kl}$ - the eigenvector with the
values $k-1$ and $l-1$.

As follows from Eq. (\ref{14}), the vector $x_{ij}^{kl}
=(J_-')^i(J_-")^jx^{kl}$ is
the eigenvector of the operators $J_3'$ and $J_3"$ with 
the eigenvalues $k-2i$ and $l-2j$, respectively. 
Since 
$${\bf J}^2=J_3^2-2J_3+4J_+J_-=J_3^2+2J_3+4J_-J_+$$
is the Casimir operator for the ${\bf J}$  algebra, and the
Hermiticity condition can be written as $J_-^*=J_+$, 
it  follows  in  addition that 
\begin{equation}
{\bf J}^{'2}x_{ij}^{kl}=k(k+2)x_{ij}^{kl},\quad
{\bf J}^{"2}x_{ij}^{kl}=l(l+2)x_{ij}^{kl}
\label{19}
\end{equation}
\begin{equation}
J_+'x_{ij}^{kl}=i (k+1-i)x_{i -1,j}^{kl},
\quad  J_+"x_{ij}^{kl}=j (l+1-j)x_{i,j -1}^{kl}
\label{20}
\end{equation}
\begin{equation}
(x_{ij}^{kl},x_{ij}^{kl}))=
\frac{i !j !k!l!}{(k-i !)(l-j !)}(x^{kl},x^{kl})
\label{21}
\end{equation}
where $(...,...)$ is the scalar product in the representation space.
From these formulas it follows that the action of 
the operators ${\bf J}'$
and ${\bf J}"$ on $x^{kl}$  generates a space with the  
dimension $(k+1)(l+1)$  and  the
basis $x_{ij}^{kl}$ ($i=0,1,...k$, $j=0,1,...l$).
Note that the vectors $x_{ij}^{kl}$   are
orthogonal but in this section we do not normalize them to one.

The Casimir operator of the second order for the 
algebra  (\ref{2}) can be written as
\begin{eqnarray}
&&I_2 =-\frac{1}{2}\sum_{ab} M_{ab}M^{ab}=\nonumber\\
&&4(R_{22}R_{11}-R_{21}R_{12}-J_3')-
2({\bf J}^{'2}+{\bf J}^{"2})
\label{22}
\end{eqnarray}
A direct calculation shows that for the
generators given by Eqs. (\ref{9}), (\ref{10}) and 
(\ref{12}), $I_2$  has the numerical  value
\begin{equation}
I_2 =w-s(s+2)+9
\label{23}
\end{equation}
where $w=\mu^2$. As noted in Sect. \ref{S5},
$\mu = 2mR$ where $m$ is the conventional mass. If $m \neq 0$
then $\mu$ is very large since $R$ is very large. We conclude
that for massive IRs the quantity $I_2$ is a large positive
number.

 The basis in the representation space
can be explicitly constructed assuming that there exists a
vector $e^0$ which is the maximal su(2)$\times$su(2)
vector such that
\begin{equation}
J_3'e_0=n_1e_0\quad J_3"e_0=n_2e_0
\label{24}
\end{equation}
and $n_1$ is the minimum possible eigenvalue of $J_3'$ in
the space of the maximal vectors. Then $e_0$ should also
satisfy the conditions
\begin{equation}
A^{--}e_0=A^{-+}e_0=0
\label{25}
\end{equation}
We use ${\tilde I}$ to denote the operator 
$R_{22}R_{11}-R_{21}R_{12}$.
Then as follows from Eqs. (\ref{15}), (\ref{16}), (\ref{18}),
(\ref{22}), (\ref{24}) and (\ref{25}),
$${\tilde I}n_1e_0=2n_1(n_1+1)e_0.$$
Therefore, if $n_1\neq 0$ the vector $e_0$ is the eigenvector
of the operator ${\tilde I}$ with the eigenvalue 
$2(n_1+1)$ and the
eigenvector of the operator $I_2$ with the eigenvalue
$$-2[(n_1+2)(n_2-2)+n_2(n_2+2)].$$ 
The latter is obviously incompatible with Eq. (\ref{23})
for massive IRs. Therefore the compatibility can be
achieved only if $n_1=0$. In that case we use $s$ to denote
$n_2$ since it will be clear soon that the value of $n_2$
indeed has the meaning of spin. Then, as follows from 
Eqs. (\ref{23}) and (\ref{24}), the vector $e_0$ should 
satisfy the conditions  
\begin{eqnarray}
&&{\bf J}'e^0=J_+"e^0=0,\quad J_3"e^0 =se^0, \nonumber\\
&&I_2e^0 =[w-s(s+2)+9] e^0
\label{26}
\end{eqnarray}
where $w,s>0$ and $s$ is an integer.  

Define the vectors
\begin{equation}
e^{nr}=(A^{++})^n(A^{+-})^re^0
\label{27}
\end{equation}
Then a direct calculation taking into account Eqs.
(\ref{14})-(\ref{16}), (\ref{18}), (\ref{19}), (\ref{22}), 
(\ref{25}) and (\ref{26}) gives
\begin{equation}
A^{--}A^{++}e^{nr}=-\frac{1}{4}(n+1)(n+s+2)[w+(2n+s+3)^2]e^{nr}
\label{28}
\end{equation}
\begin{equation}
A^{-+}A^{+-}e^{nr}=-\frac{1}{4}(r+1)(s-r)[w+1+(2r-s)
(2r+2-s)]e^{nr}
\label{29}
\end{equation}
\begin{equation}
(e^{n+1,r},e^{n+1,r})=\frac{(n+1)(n+s+2)[w+(2n+s+3)^2]}
{4(n+r+2)(s-r+n+2)}(e^{nr},e^{nr})
\label{28a}
\end{equation}
\begin{eqnarray}
&&(e^{n,r+1},e^{n,r+1})=\frac{1}{4}(r+1)(s-r)[w+1+(2r-s)
(2r+2-s)]\nonumber\\
&&\frac{s-r+n+1}{s-r+n+2}(e^{nr},e^{nr})
\label{29a}
\end{eqnarray}
As follows from Eqs. (\ref{28}) and (\ref{28a}), 
the possible values of $n$ are $n=0,1,2,...$ while,
as follows from Eqs. (\ref{29}) and (\ref{29a}), 
$r$ can take only the values of $0,1,....s$
(and therefore $s$ indeed has the meaning of the 
particle spin). 
Since $e^{nr}$ is the maximal $su(2)\times su(2)$ vector with the
eigenvalues of the operators ${\bf J}'$ and ${\bf J}"$ equal to
$n+r$ and $n+s-r$, respectively, then
as a basis of the representation space one can take the vectors
$e_{ij}^{nr}=(J_-')^i (J_-")^je^{nr}$
where, for the given $n$ and $s$, the quantity $i$ can
take the values of $0,1,...n+r$ and $j$ - the values 
of $0,1,...n+s-r$. 

One can show \cite{lev3} that the construction discussed in this section is
is an implementation of the generators (\ref{12}) but not (\ref{9}) and (\ref{10}).

Below we will discuss in detail a system of two spinless particles. 
If $s=0$ then there exist only the maximal $su(2)\times su(2)$ vectors $x^{kl}$
with $k=l$ and therefore the basis of the  representation  space  is
formed by the vectors $e_{\alpha\beta}^n\equiv e_{\alpha\beta}^{n0}$
where $n=0,1,2,...$; $\alpha,\beta=0,1,...n$. The explicit 
expressions for the  action  of  operators $R_{ij}$ in this
basis can be calculated by using Eq. (\ref{15}), and the
result is
\begin{eqnarray}
&& R_{11}e_{\alpha\beta}^n=\frac{(n+1-\alpha)(n+1-\beta)}
{(n+1)^2}e_{\alpha\beta}^{n+1}+ \nonumber\\
&&\frac{\alpha\beta n}{4(n+1)}
[w+(2n+1)^2]e_{\alpha -1,\beta-1}^{n-1},\nonumber\\
&&R_{12}e_{\alpha\beta}^n=\frac{n+1-\alpha}{(n+1)^2}
e_{\alpha,\beta+1}^{n+1}-\frac{\alpha n}{4(n+1)}[w+(2n+1)^2]
e_{\alpha -1,\beta}^{n-1},\nonumber\\
&& R_{21}e_{\alpha\beta}^n=\frac{n+1-\beta}{(n+1)^2}
e_{\alpha+1,\beta}^{n+1}-\frac{\beta n}{4(n+1)}[w+(2n+1)^2]
e_{\alpha,\beta-1}^{n-1},\nonumber\\
&&R_{22}e_{\alpha\beta}^n=\frac{1}{(n+1)^2}
e_{\alpha+1,\beta+1}^{n+1}+\nonumber\\
&&\frac{n}{4(n+1)}[w+(2n+1)^2]e_{\alpha \beta}^{n-1}
\label{34}
\end{eqnarray}
As follows from Eqs. (\ref{21}) and (\ref{28a})
\begin{equation}
(e_{\alpha\beta}^n,e_{\alpha\beta}^n)=
\frac{(n!)^2\alpha ! \beta !}{4^n(n+1)(n-\alpha)!(n-\beta)!}
\prod_{j=1}^{n}  [w+(2j+1)^2]
\label{35}
\end{equation}
if $(e_0,e_0)=1$ and, as follows from Eqs. (\ref{17}) and (\ref{19})
\begin{equation}
({\bf B}^2+{\bf M}^2)e_{\alpha\beta}^n=4n(n+2)e_{\alpha\beta}^n
\label{36}
\end{equation}
As noted in Sect. \ref{S5}, in Poincare limit ${\bf B}/2R$ becomes the momentum operator. 
Therefore in Poincare limit the eigenvalues of ${\bf B}^2$ are much greater than those
for ${\bf M}^2$. For quasiclassical states the values of $n$ are very large and therefore
it follows from Eq. (\ref{36}) that the quantum number $n$ has a meaning that 
$n/R$ becomes the magnitude of the momentum in Poincare limit.

\section{Free two-body mass operator}
\label{S11}

We now consider a two-body system and assume that it is free, i.e. there is no interaction between
the particles. How can one distinguish quantum-mechanical descriptions of free and interacting systems? 
In the literature (see e.g. Refs. \cite{pack,rqm}) a system is called free if its 
generators are sums of the single-particle generators for the particles comprising
the system. In our case this implies that $M_{ab}=M_{ab}^{(1)}+M_{ab}^{(2)}$
where $M_{ab}^{(1)}$ are the generators for the first particle and
$M_{ab}^{(2)}$ - for the second one. Each generator acts over the variables of its 
"own" particle, as described  
in Sect. \ref{S4} or Sect. \ref{S8}, and over the variables of another particle it acts 
as the identity 
operator. In other words, the representation describing the two-body system is the 
tensor product of single-particle IRs. Several authors (see e.g. Refs. \cite{pack,equiv})
define a system as free if its S-matrix is an identical operator and interacting otherwise.
In Poincare invariant theories the S-matrix is well defined but in de Sitter invariant
theories it is probably not possible to define an analog of Poincare invariant S-matrix
(but this does not mean that dS theories are unphysical). So we accept the
first definition of free and interacting systems. 

However, in this case the following problem arises. As noted in Sect. \ref{S4}, the
generators of IR are defined up to unitary equivalence. For example, the operators
defined by Eqs. (\ref{9}) and (\ref{10}) on one hand and by Eq. (\ref{12}) on the other, are
unitarily equivalent and the unitary transformation implementing the equivalence is
defined by Eq. (\ref{13}). Suppose that the operators of IR are implemented in two
forms, such that $G_a$ and $G_b$ denote the sets of operators in the form $a$ and $b$,
respectively. They are related by a unitary transformation $U_{ba}$ such that 
$G_b=U_{ba}G_aU_{ba}^{-1}$. Let now $G_a^{(j)},G_b^{(j)},U_{ba}^{(j)}$ ($j=1,2$) be the
corresponding operators for particles 1 and 2, respectively and $G_a,G_b$ be the
two-body operators in the forms $a$ and $b$, respectively. If we assume that the
two-body operators are sums of the single-particle operators in the form $a$ then
$G_a=G_a^{(1)}+G_a^{(2)}$ while if we assume that the
two-body operators are sums of the single-particle operators in the form $b$ then
$$G_b=G_b^{(1)}+G_b^{(2)}=U_{ba}^{(1)}G_a^{(1)}(U_{ba}^{(1)})^{-1}+
U_{ba}^{(2)}G_a^{(2)}(U_{ba}^{(2)})^{-1}$$ 
In this case the two-body operators $G_a$ and $G_b$ are not unitarily equivalent since
the operators $U_{ba}^{(1)}$ and $U_{ba}^{(2)}$ are different.

We conclude that the notion of free or interacting system depends on the choice of the
form of IR. With one choice the system can be treated as free but then with the other
choice it will be treated as interacting. Therefore the problem arises, which form of
IRs of the dS algebra is more physical. As noted in Sect. \ref{S5}, the form of the
operators defined by Eqs. (\ref{9}) and (\ref{10}) is convenient since in this case the Poincare
limit is straightforward. On the other hand, the form of the operators defined by 
Eq. (\ref{12}) has its own advantages (for example, the Hilbert space is the space of
functions on $S^3$ rather than the space of functions on two Lorentz hyperboloids).
We believe that the latter choice is more fundamental from the following considerations.
As shown in the preceding section, this choice allows a pure algebraic description.
If we believe that the ultimate quantum theory will be discrete and finite then we
should be looking for such implementations of the standard theory, which can be
generalized to a discrete and finite theory. As shown in Ref. \cite{lev3}, if the
operators of IR are described as in the preceding section, then the theory can be
directly generalized to a case when quantum theory is defined over a Galois field
rather then the field of complex numbers. At the same time, such a generalization is
not possible for operators defined by Eqs. (\ref{9}) and (\ref{10}) since the velocity operator
has a pure continuous spectrum. In summary, we believe that the above arguments in
favor of the representation described in the preceding section are much more important than
the convenience of transition to the Poincare limit. We will see below that this 
representation does not contradict experiment but in some cases it leads to 
consequences, which are far from what one might expect from naive considerations.

Denote by $\mu_1$  and $\mu_2$  $(\mu_1,\mu_2 >0)$ the dS
masses of the corresponding particles and assume that they are
spinless. Then, as follows from Eq. (\ref{23}), 
$I_2^{(1)}=2(w_1+9)$,
$I_2^{(2)}=2(w_2+9)$, where $w_1=\mu_1^2$,  $w_2=\mu_2^2$.
The tensor product of IRs can be decomposed into the direct 
integral of IRs and there exists a well elaborated general
theory \cite{Naimark}. In terminology of the theory of
induced UIRs, UIRs discussed in Sect. \ref{S4} belong to
the principal series of UIRs. In general, the decomposition
of the tensor product of UIRs belonging to the principal
series may contain not only UIRs of the principal series
(i.e. it may also contain UIRs not having "rest states"
defined by Eq. (\ref{26})). We will consider only a part
of the tensor product containing the "rest states"
and show that even for this part the spectrum of the
mass operator is not bounded below by the value of
$(\mu_1+\mu_2)^2$.

\begin{sloppypar}
It is clear that only IRs with $s=0,2,4...$ can enter the  tensor
product of two spinless representations. Therefore in 
order to find which values of $w$ are possible for the 
given $s$ one can act as
follows. First define the two-body operators ${\bf J}'$ and ${\bf J}"$ 
as sums of the corresponding single-particle operators 
defined in the preceding section, i.e.
$${\bf J}'= {\bf J}'^{(1)}+{\bf J}'^{(2)},\quad
{\bf J}"= {\bf J}"^{(1)}+{\bf J}"^{(2)}$$
Construct $H_s$ ---  a space  of elements $x$,
satisfying the condition (compare with Eq. (\ref{26}))
\begin{equation}
 {\bf J}'x=J_+"x=0,\quad   {\bf J}^{"2} x=s(s+2)x
\label{59}
\end{equation}
We now use $I_2$ to denote the two-body operator constructed
by analogy with Eq. (\ref{22}). Since it commutes with all the 
two-body representation operators
then $H_s$ is invariant under the action of $I_2$. 
Since $I_{2s} =2[W -s(s+2)+9]$, the operator $W$ also can be
reduced onto $H_s$ and the spectrum of $W$ in $H_s$
defines possible values of $w$ for the given $s$. 
\end{sloppypar}

Note that although $W$ is the dS analog of the mass
operator squared in Poincare invariant theory, it is
not a square of any operator. Therefore one cannot exclude
a possibility that $W$ has even a negative part of the
spectrum. However, the part of $W$ corresponding to the
principle series IRs has only the positive spectrum.

To construct a basis in the space $H_s$, we have first
to ascertain which linear  combinations  of  the  elements
$e_{\alpha_1\beta_1}^{(1)n_1}e_{\alpha_2\beta_2}^{(2)n_2}$
belong to $H_s$. Since $e_{\alpha_1\beta_1}^{(1)n_1}$ is 
the spinor with the spin $n_1$ with respect to the 
algebra ${\bf J}^{'(1)}$ as well as to ${\bf J}^{"(1)}$,
and  analogously for $e_{\alpha_2\beta_2}^{(2)n_2}$,  
then zero  eigenvalues of  the  operator ${\bf J}'$  can  be
obtained only if $n_1=n_2$, and the value s for the spin 
relative to the ${\bf J}"$ algebra can be obtained only if 
$n_1,n_2\geq s/2$. One can verify directly that the vectors
\begin{eqnarray}
\Phi_n&=& \frac{4^{n+j}(1+2j)!}{[(n+j)!]^2j!}
\sum_{\alpha=0}^{n+j}\sum_{\beta=0}^n
(-1)^{\alpha+\beta}\times\nonumber\\
&&\frac{(j+\beta)!(n+j-\beta)!}{\beta!(n-\beta)!}
e_{\alpha\beta}^{(1)n+j}
e_{n+j-\alpha,n-\beta}^{(2)n+j}
\label{60}
\end{eqnarray}
where $j=s/2$ and $n=0,1,2...$, belong to $H_s$.
The value of $j$ is obviously equal to the spin
of the two-body system in conventional units. The
fact that the vectors $e^{(1)}$ and $e^{(2)}$
enter Eq. (\ref{60}) with the same value of the
quantum number $n$ confirms the interpretation
of $n/R$ as the magnitude of the momentum 
(see the preceding section) since, by analogy
with Poincare invariant theory, one would
expect that the magnitudes of particle momenta in 
their common c.m. frame are equal to each other.

It is obvious that the  vectors
$\Phi_n$   with different $n'$s are orthogonal to each other.
The result of the calculation of the norm of the vector $\Phi_n$
(see Ref. \cite{lev3} for details) is
\begin{eqnarray}
(\Phi_n,\Phi_n)&=&\{\prod_{l=1}^{n+j} [w_1+(2l+1)^2]
[w_2+(2l+1)^2]\}\times\nonumber\\
&&\frac{(n+2j+1)!(1+2j)!}{(n+j+1)n!}
\label{61}
\end{eqnarray}

Our next goal is to find how the operator $W$  acts 
in the basis $\{\Phi_n\}$. As follows from Eq. (\ref{22}),
\begin{eqnarray}
&&I_2^s \Phi_n=8[(R_{22}^{(1)}+R_{22}^{(2)})(R_{11}^{(1)}+
R_{11}^{(2)})-(R_{21}^{(1)}+R_{21}^{(2)}) \times\nonumber\\
&&(R_{12}^{(1)}+R_{12}^{(2)})]\Phi^n -4s(s+2)\Phi_n,
\label{62}
\end{eqnarray}
\begin{eqnarray}
&&I_2^{(1)}e_{\alpha\beta}^{(1)n}=
8(R_{22}^{(1)}R_{11}^{(1)}-R_{21}^{(1)}R_{12}^{(1)})
e_{\alpha\beta}^{(1)n}-8[n(n+2)+ \nonumber\\
&&(n-2\alpha)]e_{\alpha\beta}^{(1)n}=2(w_1 +9)e_{\alpha\beta}^{(1)n}
\label{63}
\end{eqnarray}
and an analogous formula takes place for
$I_2^{(2)}e_{\alpha\beta}^{(2)n}$.

Taking into account Eqs. (\ref{34}), (\ref{60}), (\ref{62}),
(\ref{63}) and the definition of the operator $W$, a
direct calculation shows that
\begin{equation}
  W\Phi_n=\sum_{l=0}^{\infty}\Phi_l W_{ln}
\label{64}
\end{equation}
where the matrix $||W_{ln}||$ has only the following components
different from zero:
\begin{eqnarray}
&& W_{n+1,n}^s=\frac{n+1}{n+1+j},\quad
W_{nn}^s=w_1+w_2 +8(n+1)^2+ \nonumber\\
&&2s(4n+3)+s^2 +1,\quad W_{n,n+1}^s=\frac{n+2j+2}
{n+j+2}\times\nonumber\\
&&[w_1+(2n+2j+3)^2][w_2+(2n+2j+3)^2]
\label{65}
\end{eqnarray}
Such a matrix is called three-diagonal and in fact, 
only the terms with $l=n-1,n,n+1$
contribute to the sum (\ref{64}). Note that
the operator $W$   is  certainly  Hermitian,  but  
since  the  basis elements are not normalized to one, 
the Hermiticity  condition
has not the usual form $W_{nl}=W_{ln}^*$, but
$||\Phi_n||^2W_{nl}=||\Phi_l||^2W_{ln}^*$.

The matrix of the operator $W-\lambda$ has the matrix elements
$||W_{nl} -\lambda\delta_{nl}||$. We use $\Delta^n(\lambda)$
to denote the determinant of the matrix obtained from this
one by taking into account only the rows and columns with the
numbers $0,1,...n$. It  is well known (and can be verified 
directly)  that  for  the  three-diagonal matrix
the following relation is valid:
\begin{equation}
\Delta^{n+1}(\lambda)=(W_{n+1,n+1}-\lambda)\Delta^n(\lambda)-
W_{n+1,n}W_{n,n+1} \Delta^{n-1}(\lambda)
\label{66}
\end{equation}
where it is formally assumed  that $\Delta^{-1}(\lambda)=1$. Since
$\Delta^0(\lambda)=W_{00}-\lambda$, Eqs. (\ref{65}) and (\ref{66}) make
it possible to calculate $\Delta^n(\lambda)$ for any $n=1,2,3...$. The result 
can be represented as follows. Denote
\begin{equation}
\lambda_l=[\mu_1+\mu_2+i(s+4l+3)]^2\quad (l=0,1,2...)
\label{66a}
\end{equation}
Then
\begin{eqnarray}
&&\Delta^n(\lambda) = (-1)^{n+1}\sum_{k=0}^{n+1}C_{n+1}^k[\prod_{l=0}^{k-1}
(\lambda-\lambda_l)]
\prod_{l=k}^n\frac{s+k+l+2}{j+l+1}\nonumber\\
&&[\mu_1+i(s+2l+3)][\mu_2+i(s+2l+3)]
\label{66b}
\end{eqnarray}

Since we consider only representations of the  principle series,
we are interested only in the region of positive
$\lambda$'s. Therefore we can represent $\lambda$ as
$\lambda=(\mu_1+\mu_2+\sigma)^2$  where
$\sigma$ has the meaning of the dS kinetic energy. However as 
we will see below, not  only  $\sigma\geq 0$,
but also $\sigma <0$ is possible. Let $(a)_n=a(a+1)\cdots (a+n-1)$ be the
Pochhammer symbol. Then the expression (\ref{66b}) can be written in terms of the
hypergeometric series as
\begin{eqnarray}
&&\Delta^n(\lambda)=4^{n+1}\frac{(s+2+n)!j!}{(1+s)!(j+n+1)!}(\frac{s+3-i\mu_1}{2})_{n+1}
(\frac{s+3-i\mu_2}{2})_{n+1}\nonumber\\
&&_4F_3(\frac{s+3+i\sigma}{4},\frac{s+3-i(2\mu_1+2\mu_2+\sigma)}{4},
s+n+3,-(n+1);\nonumber\\
&&\frac{s+3-i\mu_1}{2},\frac{s+3-i\mu_1}{2},\frac{s+3}{2};1)
\label{66c}
\end{eqnarray}
Consider a vector
\begin{equation}
\chi(\lambda,N)=\sum_{n=0}^N (-1)^n
\Delta^{n-1}(\lambda)[\prod_{l=0}^{n-1}W_{l,l+1}]^{-1}\Phi_n
\label{67}
\end{equation}
where $N$ is a natural number. Let $F(\lambda',\lambda, N)=(\chi(\lambda',N),\chi(\lambda,N))$
and $F(\lambda',\lambda)$ be the
limit of $F(\lambda',\lambda, N)$ when $N\to\infty$. If the limit in Eq.
(\ref{67}) exists then, as follows from Eqs. (\ref{65}) and (\ref{66}), it is an 
eigenvector of
the operator $W$  with the eigenvalue $\lambda$. 
In this case $\lambda$ is the true eigenvalue i.e. it belongs to the discrete spectrum of 
the operator $W$. There also exists a possibility that the limit does not represent a vector 
belonging to the Hilbert space but is a generalized eigenvector, i.e.
$F(\lambda',\lambda)$ is proportional to $\delta(\lambda-\lambda')$. Then $\lambda$ 
belongs to the continuous  spectrum  of the operator $W$.  
As follows from Eqs. (\ref{61}), (\ref{66c}) and (\ref{67}),
\begin{eqnarray}
&&F(\lambda',\lambda, N)=\frac{[(1+s)!]^2}{(1+j)^2}
\{\prod_{l=1}^j[w_1+(2l+1)^2][w_1+(2l+1)^2]\}\nonumber\\
&&\sum_{n=0}^N(n+j+1)C_{n+s+1}^{1+s}\, _4F_3(\frac{s+3-i\sigma'}{4},
\frac{s+3+i(2\mu_1+2\mu_2+\sigma')}{4},\nonumber\\
&&s+n+2,-n;\frac{s+3+i\mu_1}{2},\frac{s+3+i\mu_1}{2},\frac{s+3}{2};1)\nonumber\\
&&_4F_3(\frac{s+3+i\sigma}{4},\frac{s+3-i(2\mu_1+2\mu_2+\sigma)}{4},
s+n+2,-n;\nonumber\\
&&\frac{s+3+i\mu_1}{2},\frac{s+3+i\mu_1}{2},\frac{s+3}{2};1)
\label{67a}
\end{eqnarray}
where $\lambda'=\mu_1+\mu_2+\sigma'$, and $C_n^k$ is the binomial coefficient.
By using the relations (see e.g. Ref. \cite{BE}) 
\begin{eqnarray}
&&_{q+1}F_q(\beta,\alpha_1,...\alpha_q;\gamma,\rho_1,...\rho_{q-1};z)=
\frac{\Gamma(\gamma)}{\Gamma(\beta)\Gamma(\gamma-\beta)}\nonumber\\
&&\int_0^1t^{\beta-1}(1-t)^{\gamma-\beta-1}\, _qF_{q-1}(\alpha_1,...\alpha_q;\rho_1,...\rho_{q-1};tz)dt
\label{67b}
\end{eqnarray}
\begin{equation}
n!C_n^{\lambda}(x)=(2\lambda)_nF(-n,n+2\lambda;\lambda+1/2;(1-x)/2)
\label{67c}
\end{equation}
where $C_n^{\lambda}(x)$ is the Gegenbauer polynomial and in the case of the hypergeometric 
function $_2F_1$ we write simply $F$, the expression (\ref{67a}) can be
represented as
\begin{eqnarray}
&&F(\lambda',\lambda, N)=\frac{[(1+s)!]^2}{(1+j)^2}
\{\prod_{l=1}^j[w_1+(2l+1)^2][w_1+(2l+1)^2]\}\nonumber\\
&&|\Gamma(\frac{s+3-i\mu_1}{2})\Gamma(\frac{s+3-i\mu_2}{2})|^2
\{\Gamma(\frac{s+3}{2})^2\Gamma(\frac{s+3-i\sigma'}{4})\nonumber\\
&&\Gamma(\frac{s+3+i\sigma}{4})\Gamma(\frac{s+3+i(2\mu_1+2\mu_2+\sigma')}{4})\nonumber\\
&&\Gamma(\frac{s+3-i(2\mu_1+2\mu_2+\sigma)}{4})\}^{-1}
\sum_{n=0}^N\frac{(n+j+1)}{C_{n+s+1}^{1+s}}\int_0^1u^{(s-1-i\sigma')/4}\nonumber\\
&&(1-u)^{(s+1)/2}C_n^{1+j}(1-2u)F(\frac{s+3-i(2\mu_1+\sigma')}{4},\nonumber\\
&&\frac{s+3-i(2\mu_2+\sigma')}{4};\frac{s+3}{2};1-u)du\int_0^1
u^{(s-1+i\sigma)/4}(1-u)^{(s+1)/2}\nonumber\\
&&C_n^{1+j}(1-2u)F(\frac{s+3+i(2\mu_1+\sigma')}{4},\frac{s+3+i(2\mu_2+\sigma)}{4};\nonumber\\
&&\frac{s+3}{2};1-u)du
\label{67d}
\end{eqnarray}

To calculate the limit if this expression when $N\to\infty$ we use the fact that the system
of Gegenbauer polynomials form a complete orthogonal set in the space of functions 
quadratically integrable on $[-1,1]$ with some weight (see e.g. Ref. \cite{BE}). As a result,
the completeness relation for the Gegenbauer polynomials can be written as
\begin{equation}
\sum_{n=0}^{\infty} C_n^{1+j}(x)C_n^{1+j}(y)
\frac{(n+1+j)\Gamma(1+j)}{\sqrt{\pi}C_{n+s+1}^{1+s}\Gamma((s+3)/2)}=
\frac{\delta(x-y)}{(1-y^2)^{(s+1)/2}}
\label{67e}
\end{equation}
and the limit of the expression (\ref{67d}) when $N\to\infty$ can be written as
\begin{eqnarray}
&&F(\lambda',\lambda)=\frac{\sqrt{\pi}[(1+s)!]^2}{2^{s+2}(1+j)^2j!}
\{\prod_{l=1}^j[w_1+(2l+1)^2][w_1+(2l+1)^2]\}\nonumber\\
&&|\Gamma(\frac{s+3-i\mu_1}{2})\Gamma(\frac{s+3-i\mu_2}{2})|^2
\{\Gamma(\frac{s+3}{2})^2\Gamma(\frac{s+3-i\sigma'}{4})\nonumber\\
&&\Gamma(\frac{s+3+i\sigma}{4})\Gamma(\frac{s+3+i(2\mu_1+2\mu_2+\sigma')}{4})\nonumber\\
&&\Gamma(\frac{s+3-i(2\mu_1+2\mu_2+\sigma)}{4})\}^{-1}
\int_0^1u^{i(\sigma-\sigma')/4-1}(1-u)^{(s+1)/2}\nonumber\\
&&F(\frac{s+3-i(2\mu_1+\sigma')}{4},\frac{s+3-i(2\mu_2+\sigma')}{4};
\frac{s+3}{2};1-u)\nonumber\\
&&F(\frac{s+3+i(2\mu_1+\sigma')}{4},\frac{s+3+i(2\mu_2+\sigma)}{4};\frac{s+3}{2};1-u)du
\label{67g}
\end{eqnarray}
Since the integral diverges when $u\to 0$, we can set $u=0$ in the nonsingular expressions
in this intergal and the final result is
\begin{eqnarray}
&&F(\lambda',\lambda)=\frac{\pi^{3/2}[(1+s)!]^2}{2^{s+2}(1+j)^2j!}
\{\prod_{l=1}^j[w_1+(2l+1)^2][w_1+(2l+1)^2]\}\nonumber\\
&&|\frac{\Gamma(\frac{s+3+i\mu_1}{2})\Gamma(\frac{s+3+i\mu_2}{2})
\Gamma(\frac{i}{2}(\mu_1+\mu+\sigma))}
{\Gamma(\frac{s+3+i\sigma}{4})\Gamma(\frac{s+3+i(2\mu_1+\sigma)}{4})
\Gamma(\frac{s+3+i(2\mu_2+\sigma)}{4})}|^2\nonumber\\
&&\Gamma(\frac{s+3}{2})\delta(\sigma -\sigma')
\label{68}
\end{eqnarray}
Therefore the discrete spectrum is absent and
continuous one fills all the interval $\lambda\in (0,\infty)$.

In Ref. \cite{lev1}, where the explicit expression 
for the two-body mass operator in the form  of
a differential operator in some space  of  functions 
has been has been found, a similar result has been obtained. 
The result (\ref{68}), however is in fact algebraic since the operator 
$W$ has been 
considered in the form of an infinite matrix. In Sect.
\ref{S6} we have shown that the dS mass operator has
the infinite spectrum in the range $(-\infty,\infty)$.
However, this result has been obtained in the nonrelativistic
approximation and in first order in $1/R$. On the contrary,
no approximation has been assumed in deriving Eq. (\ref{68}).

\section{Mean value of the two-body mass operator}
\label{S12}

It is well known that in nonrelativistic quantum mechanics, dynamics
of a two-body system is fully defined by the two-body mass operator,
which acts only over the internal variables of the system. In a wide
literature on Poincare invariant dynamics (see e.g. Ref. \cite{pack,rqm})
the following problem has been widely discussed: how to define 
interactions in a many-body system such that the commutation relations
of the Poincare algebra are preserved. It has been shown that it is possible
to define internal and external variables in such a way that interactions 
are present only in the mass operator, which is unitarily
equivalent to one acting only in the internal Hilbert space. Then the
commutation relations are preserved if the mass operator commutes with spin
operator of the system.

The problem arises what is the internal Hilbert space in dS theories. 
In the preceding section we have investigated the structure of the Hilbert 
space $H_s$, which might be treated as an analog of the internal Hilbert
space in Poincare invariant theory if the spin of the systems equals $s$. Indeed, 
by analogy with the construction 
of IRs in Sect. \ref{S8}, one can construct the full two-body 
Hilbert space by acting by the two-body operators $R_{ij}$ ($i,j=1,2$) on spaces
$H_s$ with all possible values of $s$. In this case, a subspace corresponding to a
given $s$ is a direct sum of spaces identical to $H_s$. Since the mass operator commutes with 
all the representation operators, it also commutes with all the projectors onto 
such spaces and its spectra in each space identical to $H_s$ are the same. Therefore 
the properties of the dS mass operator is fully defined by its action in $H_s$ for
all possible values of $s$. Our first goal is to understand when the mean value of the 
dS mass operators corresponds to the well known mean value of the Poincare invariant mass
operator. 

Instead of the states $\Phi_n$ in $H_s$, we introduce the states
\begin{equation}
e_n=\Phi_n(\Phi_n,\Phi_n)^{-1/2}
\label{80}
\end{equation}
Then the elements $\{e_n\}$ form the orthonormal
basis in $H_s$. As follows from Eqs. (\ref{61}) and 
(\ref{65}), the nonzero matrix elements of 
the operator $W$ in the basis $\{e_n\}$ are 
given by
\begin{eqnarray}
&&W(n,j)=\{[w_1+(2n+2j+3)^2][w_2+(2n+2j+3)^2] \nonumber\\
&&\frac{(n+2j+2)(n+1)}{(n+1+j)(n+2+j)}\}^{1/2}\quad
W_{n+1,n}=W_{n,n+1}=W(n,j)\nonumber\\
&&W_{nn}=w_1+w_2+8(n+1)^2+4j(4n+3)+4j^2+1
\label{81}
\end{eqnarray} 
where $j=s/2$. Let
\begin{equation}
\Psi = \sum_{n=0}^{\infty} c_ne_n\quad 
(\sum_{n=0}^{\infty} |c_n|^2=1)
\label{82}
\end{equation}
be a state in $H_s$. Then the mean value of the
operator $W$ in the state $\Psi$ is given by
\begin{eqnarray}
&&(\Psi,W\Psi)=\sum_{n=0}^{\infty} 
[W_{nn}|c_n|^2 +2W_{n,n+1}Re(c_{n+1}c_n^*)]
\label{83}
\end{eqnarray}
where $W_{nn}$ and $W_{n,n+1}$ are given by Eq.
(\ref{81}). 

This result has been obtained without any approximation, so it is valid
for any wave function ${c_n}$ in the $n$-representation. If we
are interested in comparing the results with Poincare invariant theory, 
the mass operator squared in Poincare terms should be defined as
$M^2=W/4R^2$ (see Sects. \ref{S5} and \ref{S11}). 
We also take into account that typical values of $n$ are very large
since, as noted in the preceding section, $n$ is of order $qR$ where ${\bf q}$ is 
the relative momentum and $q=|{\bf q}|$. In particular, $n\gg j$ in the
approximation when $R$ is very large. Then taking into account the
relation between the dS and Poincare masses (see Sect. \ref{S5}), we get from 
Eq. (\ref{83})
\begin{eqnarray}
&&(\Psi,M^2\Psi)=\sum_{n=0}^{\infty} 
\{[m_1^2+m_2^2+2(n/R)^2]|c_n|^2 +\nonumber\\
&&2[m_1^2+(n/R)^2]^{1/2}[m_2^2+(n/R)^2]^{1/2}
Re(c_{n+1}c_n^*)\}
\label{84}
\end{eqnarray}
i.e. the dependence on $s$ formally dissapears.

One usually expects that the quasiclassical wave function in the relative momentum representation 
has a sharp maximum and therefore one might expect that in the $n$-representation  
the quasiclassical wave function has a sharp maximum at some value $n_0$ such that $q_0=n_0/R$ has 
the meaning of relativistic relative
momentum. Then, if we assume additinally that $c_n$ does not change significantly when $n$
is replaced by $n+1$, i.e. $c_n\approx c_{n+1}$, then Eq. (\ref{84}) gives the standard
relativistic result:
\begin{equation}
(\Psi,M^2\Psi)=[\epsilon_1(q)+\epsilon_2(q)]^2
\label{85}
\end{equation}
where $\epsilon_j(q)=\sqrt{m_j^2+q^2}$ (j=1,2).
This result is an additional argument in favor of calling $n$ the de Sitter relative momentum.
Therefore the set $\{c_n\}$ has the meaning of the internal wave function in de Sitter momentum
representation. In contrast with the internal relativistic wave function in momentum 
representation $c(q)$, where $q$ is a continuos variable in the range $[0,\infty)$, the
quantity $n$ takes the values 0,1,2,.., i.e. it is a nonnegative integer. In other words,
the de Sitter relative momentum is quantized with the quantum equal to 1. If Poincare invariant
theory is treated as a limit of de Sitter invariant theory when $R$ is large, then the
relativistic relative momentum becomes a quantity quantized with the quantum $1/R$. This
is an extremely small value and therefore the discrete and continuous cases are practically
indistinguishable. The result (\ref{85}) is also an advancement in comparison with the
resuls of Sect. \ref{classical} since a relation between Poincare and de Sitter invariant
theories is obtained without assuming that the velocity is small. Nevertheless, we are still
very far from our main goal - understanding quasiclassical approximation in de Sitter
invariant theory without decomposition in powers of $1/R$. Our next task is to understand
how the result (\ref{84}) can recover de Sitter antigravity if we do not assume that the
velocity is nonrelativistic. We first recall well known facts about quasiclassical approximation
in the standard theory.  

\section{Discussion of standard quasiclassical approximation}
\label{S13}

In Sect. \ref{classical} we have discussed quasiclassical approximation in the nonrelativistic
approximation assuming that the operators describing a particle are given by Eq. (\ref{9}).
Our goal in subsequent sections is to describe quasiclassical approximation in the framework 
of the algebraic
approach developed in this chapter and not assuming that $|{\bf v}|\ll 1$. In this section
we recall well known facts about quasiclassical approximation in the standard theory. 
 
Let $\psi(x)$ be a one-dimensional wave function in the coordinate representation in
nonrelativistic quantum mechanics. This function can be normalized as
\begin{equation}
\int_{-\infty}^{\infty} |\psi(x)|^2dx=1
\label{86}
\end{equation}
If $\varphi(p)$ is the wave function in momentum representation then
\begin{eqnarray}
\varphi(p)=\frac{1}{\sqrt{2\pi}}\int_{-\infty}^{\infty}\psi(x)exp(-ipx)dx\quad
\int_{-\infty}^{\infty} |\varphi(p)|^2dp=1
\label{87}
\end{eqnarray}
and in this section we are working with the usual system of units where $\hbar=1$.

The usual assumptions about the wave function in coordinate representation are as follows.
The function $\psi(x)$ has the form $\psi(x)=a(x)exp(ip_0x)$ where $p_0$ is the classical
momentum of the particle; the amplitude $a(x)$ is a real function, which has a sharp 
maximum at $x=x_0$, where $x_0$ is the classical coordinate of the particle; the width of
the maximum, $\delta x$, is such that $\delta x \ll |x_0|$ and $|p_0|\delta x \gg 1$,
such that the wave functions makes many oscillations on the interval $\delta x$. A well
known example is the Gaussian wave function
\begin{equation}
\psi(x)=\frac{1}{\pi^{1/4}(\delta x)^{1/2}}exp[-\frac{(x-x_0)^2}{2(\delta x)^2}+ip_0x]
\label{88}
\end{equation}
Then, as follows from Eq. (\ref{87})
\begin{equation}
\varphi(p) = \frac{(\delta x)^{1/2}}{\pi^{1/4}}exp[-\frac{1}{2}(p-p_0)^2(\delta x)^2
+i(p_0-p)x_0]
\label{89}
\end{equation}

Eqs. (\ref{88}) and (\ref{89}) can be easily generalized to the three-dimensional case
and to the internal wave function of the two-body system. If we assume for simplicity that
the uncertainties in all the three relative coordinates are the same and equal $\delta$
then a possible generalization of Eq. (\ref{89}) is
\begin{equation}
\varphi({\bf q}) = \frac{\delta^{3/2}}{\pi^{3/4}}
exp[-\frac{1}{2}({\bf q}-{\bf q}_0)^2\delta^2+i({\bf q}_0-{\bf q}){\bf r}_0]
\label{90}
\end{equation}
where ${\bf q}_0$ and ${\bf r}_0$ are the classical relative momentum and the classical
distance, respectively. Therefore one can conclude the following. The wave function in 
the relative momentum representation has a very sharp maximum at ${\bf q}={\bf q}_0$ and 
therefore the mean value of any function $f({\bf q})$, which does not rise exponentially at
infinity, approximately equals $f({\bf q}_0)$. In particular, the
mean value of the nonrelativistic energy operator $H={\bf q}^2/2m_{12}$ approximately equals 
$H={\bf q}_0^2/2m_{12}$, the mean value of the free relativistic mass operator
$\epsilon_1(q)+\epsilon_2(q)$ approximately equals 
$\epsilon_1(q_0)+\epsilon_2(q_0)$ etc. This is fully in agreement with
our intuition in nonrelativistic quantum mechanics.

The above conclusion is essentially based on the fact that in standard theory all the spatial
coordinates are in the range $(-\infty,\infty)$. As a consequence, any quasiclassical state
contains an infinite number of angular momenta. However, as follows from Eq. (\ref{1}), in
the dS space all the spatial coordinates are finite. We also noted in the preceding section
that the quantity $n=qR$ has the meaning of the dS relative momentum and, since the 
internal angular momentum is of order $l=qr$ or less, then $l\ll n$. Therefore one might expect
that for the description of quasiclassical states in the dS theory, only a finite
number of internal angular momentum states is important. For this reason we first recall
how quasiclassical states with a fixed value of the internal angular momentum are treated
in standard quantum mechanics.

Consider the internal wave function $\varphi({\bf q})=f(q)Y_{lm}({\bf q}/q)$ where 
$q=|{\bf q}|$ and $Y_{lm}$ is the spherical function. The normalization integral for
this function is
\begin{equation}
||\varphi||^2 = \int_0^{\infty}q^2|f(q)|^2dq=1
\label{91}
\end{equation}
The operator $A$ of the relative distance squared acts on $\varphi({\bf q})$ as
\begin{equation}
A\varphi({\bf q})=-\Delta\varphi({\bf q})=[-\frac{1}{q^2}\frac{\partial}{\partial q}
(q^2\frac{\partial}{\partial q})+\frac{l(l+1)}{q^2}]f(q)Y_{lm}({\bf q}/q)
\label{92}
\end{equation} 
where $\Delta=\partial^2/\partial {\bf q}^2$ is the Laplacian. It is well known (and can be
easily verified) that with the substitution $f(q)=\chi(q)/q$, the operator $A$ acts as
\begin{equation}
A\chi(q) = [-\frac{d^2}{dq^2}+\frac{l(l+1)}{q^2}]\chi(q)
\label{93}
\end{equation}
This operator is now defined in the Hilbert space of functions with the scalar product
\begin{equation}
(\chi_1,\chi_2)=\int_0^{\infty}\chi_1(q)^*\chi_2(q)dq
\label{94}
\end{equation}
The term with $l(l+1)$ is called centrifugal one. We first consider the case $l=0$ and the 
general case $l\neq 0$ will be discussed at the end of the section.

The physical wave function $\chi(q)$ should be twice differentiable inside the interval
$(0,\infty)$ and, in view of Eq. (\ref{91}) and the definition of $\chi(q)$, $\chi(0)=0$. 
Let $D$ be a set of such functions.
The operator $A$ with the domain $D$ is positive definite, as it should be for the operator 
representing distance squared. It is well known that any positive definite self-adjoint
operator in the Hilbert space has a positive definite self-adjoint square root. One might
think that in the case $l=0$, $A$ is a square of a self-adjoint operator $i\partial/\partial q$,
but this is not so. Indeed, let $D_0$ be a set of differentiable functions $\chi(q)$ such that $\chi(0)=0$
and $D_1$ be a set of differentiable functions $\chi(q)$ such that $\chi(0)=0$ is not required.
Let $B_0$ be the operator $i\partial/\partial q$ with the domain $D_0$ and $B_1$ be the
operator $i\partial/\partial q$ but with the domain $D_1$. If $\chi_1\in D_0$ and $\chi_2\in D_0$ then
$(\chi_1,B_0\chi_2)=(B_0\chi_1,\chi_2)$, i.e. the operator $B_0$ is Hermitian. However, in
infinite-dimensional Hilbert spaces the property of an operator to be Hermitian is necessary
but not sufficient to be self-adjoint. If $\chi_1\in D_1$ and $\chi_2\in D_0$ then
$(\chi_1,B_0\chi_2)=(B_1\chi_1,\chi_2)$. Therefore $B_1$ is adjoint to $B_0$ ($B_1=B_0^*$),
$B_1$ is an extention of $B_0$ and $B_1\neq B_0$. Therefore $B_0$ is not self-adjoint (the
operator $T$ is self-adjoint if and only if $T^*=T$). In particular, there is no guarantee that
$B_0$ or $B_1$ have a spectral decomposition. The function $\chi_a(q)=exp(-iqa)$ is formally
a generalized eigenfunction of $B_1$ with the eigenvalue $a$ but, since $\chi_a(0)\neq 0$, this
function does not satisfy $(\chi_a,B_1\chi_a)=(B_1\chi_1,\chi_2)$. At the same time, the
operator $A$ is self-adjoint since $(B\chi_1,\chi_2)=(\chi_1,A\chi_2)$ is possible only if $B=A$.
The function $sin(qa)$ is the generalized eigenfunction of $A$ with the eigenvalue $a^2$.
From the physical point of view the fact that $A$ is self-adjoint while $B_0$ and $B_1$ are
not, is reasonable since $B_0$ and $B_1$ are not positive definite while an
operator representing the distance or distance squared should be positive definite. 
In the above discussion of the operators $B_0$, $B_1$ and $A$ we followed the book by Kato
\cite{Kato}.

In standard quantum mechanics the relation between $\chi(q)$ and the radial wave function $\psi(r)$
is given by
\begin{equation}
\psi(r)=\sqrt{\frac{2}{\pi}}\int_0^{\infty}\chi(q)sin(qr)dq, \quad
\chi(q)=\sqrt{\frac{2}{\pi}}\int_0^{\infty}\psi(r)sin(qr)dr
\label{95}
\end{equation}
The compatibility of these relations follows from the fact that 
$$\int_0^{\infty}sin(qr)sin(qr')dq=(\pi/2)\delta(r-r')$$

By analogy with Eq. (\ref{88}), consider a radial wave function
\begin{equation}
\psi(r)=const\, exp[-\frac{(r-r_0)^2}{2\delta^2}]sin(q_0r)
\label{96}
\end{equation}
where $const$ can be found from the normalization condition. Here and below in this section the
explicit values of constants will not be important for us. As follows from Eq. (\ref{95}),
the corresponding wave function in momentum space is given by
\begin{equation}
\chi(q)=const\,\int_0^{\infty}exp[-\frac{(r-r_0)^2}{2\delta^2}]sin(q_0r)sin(qr)dr
\label{97}
\end{equation}
We can represent this expression as 
\begin{equation}
\chi(q)=const\,(I_1-I_2)
\label{98}
\end{equation}
where
\begin{eqnarray}
&&I_1=\int_{-\infty}^{\infty}exp[-\frac{(r-r_0)^2}{2\delta^2}]sin(q_0r)sin(qr)dr\nonumber\\
&&I_2=\int_{-\infty}^0exp[-\frac{(r-r_0)^2}{2\delta^2}]sin(q_0r)sin(qr)dr
\label{99}
\end{eqnarray}
A simple calculation gives
\begin{eqnarray}
&&I_1=const\,\{cos[(q-q_0)r_0]exp[-\frac{1}{2}(q-q_0)^2\delta^2]-\nonumber\\
&&cos[(q+q_0)r_0]exp[-\frac{1}{2}(q+q_0)^2\delta^2]\}
\label{100}
\end{eqnarray}
Therefore, in contrast with the standard one-dimensional case discussed above (see Eq. (\ref{89})),
the wave function contains not only a contribution depending on $(q-q_0)$ but also one 
depending on $(q+q_0)$. However, since we require that $q_0\delta \gg 1$, this contribution is
exponentially small. 

One might think that the term with $I_2$ in Eq. (\ref{99}) is 
exponentially small since $r_0\gg \delta$. Indeed, $I_2$ will always contain a factor, which is
less than $exp(-r_0^2/2\delta^2)$. Nevertheless, we should investigate the dependence of $I_2$
on $q$ when $q$ is large. Since
\begin{eqnarray}
&&sin(q_0r)sin(qr)=-\frac{1}{4}\{exp[i(q+q_0)r]+exp[-i(q+q_0)r]-\nonumber\\
&&exp[i(q-q_0)r]-exp[i(q_0-q)r]\}
\label{101}
\end{eqnarray}
we can represent $I_2$ as a sum of four integrals containing imaginary exponents. In each of them
we can integrate those exponents by parts. Integrating twice we get 
\begin{eqnarray}
&& I_2=const\{\int_{-\infty}^0\frac{d^2}{dr^2}(exp[-\frac{(r-r_0)^2}{2\delta^2}])
[\frac{cos((q+q_0)r)}{(q+q_0)^2}-\nonumber\\
&&\frac{cos((q-q_0)r)}{(q-q_0)^2}]dr-exp(-\frac{r_0^2}{2\delta^2})
[\frac{1}{(q+q_0)^2}-\frac{1}{(q-q_0)^2}]\}
\label{102}
\end{eqnarray} 
Therefore, when $q$ is large, $I_2<const\,exp(-r_0^2/2\delta^2)/q^3$. Although the falloff of
$I_2^2$ at large $q$ is not exponential, it is clear that the contribution of $I_2$ to the
mean value of the mass operator is well defined (since $|I_2|^2<const/q^6$) and small.

We conclude that the quasiclassical description of the state (\ref{96}) is essentially the
same as the description of the state (\ref{88}). However, the following  
question arises: is the wave function (\ref{96}) realistic for description of macroscopic
bodies? For example, in the Sun - Earth system we know that the relative wave function in
the coordinate space has a very sharp maximum at $r\approx 150\cdot 10^6km$. There is no
doubt that the probability to find the Earth on the Venus or Mars orbits is
extremely small. But is this probability exactly zero? One might think that this question 
is of academic interest only. However, this question is 
extremely important and the problem is as follows. If we accept the reduction of the
wave function in accordance with the Copenhagen formulation of quantum theory, then after
each measurement of the Sun - Earth distance, the function $\psi(r)$ for the Sun - Earth
system is not equal to zero only in a very small vicinity of $r\approx 150\cdot 10^6km$. 
In the Copenhagen formulation, the measurement is treated as an interaction with a
classical object. The theory of quantum measurements is not well understood yet. So 
it is not quite clear how often the quantities $r$ and $q$ in the Sun - Earth 
system are measured. If the reduction of the wave function applies to the coordinate wave
function then the corresponding wave function in momentum space contains all possible momenta 
while if the reduction applies to the momentum wave function then the coordinate
wave function contains all possible coordinates. Probably our intuition tells us that the
first possibility takes place but at present this hypothesis cannot be proved. 

On the other hand, at present the Everett multiworld interpretation of
quantum theory is widely discussed. Adherents of this interpretation claim that the reduction
postulate contradicts unitarity (see e.g. Ref. \cite{Mensky2}). In this approach the reduction 
of the wave function is not accepted but it is assumed that different outcomes of measurement are 
observable in different worlds. Does it give any hint on the above problems? In any case, it is 
interesting to investigate a possibility when the coordinate wave function
$\psi(r)$ is not equal to zero only if $r\in [r_1,r_2]$. 

For example, consider a model wave function 
\begin{equation}
\psi(r)=\frac{1}{\sqrt{2d}}e(r)exp(iq_0r)
\label{103}
\end{equation}
where $d=(r_2-r_1)/2$ and $e(r)$ is the characteristic function of the set $[r_1,r_2]$. If $r_1>0$ then the 
function satisfies the condition $\psi(0)=0$. Such a function is not differentiable at $r=r_1$
and $r=r_2$ but it can be approximated (in the sense of the norm of the Hilbert space) with
any desired accuracy by infinitely differentiable functions with the supporter in $[r_1,r_2]$.
For example, as shown in standard textbooks on distributions (e.g. \cite{Vladimirov}), for any 
$\epsilon > 0$, $\epsilon < d$, 
it is possible to construct an infinitely differentiable function $e_{\epsilon}(r)$ such that
$e_{\epsilon}(r)=1$ at $r\in [r_1+\epsilon, r_2-\epsilon]$, $0\leq e_{\epsilon}(r)\leq 1$ if 
$r\in [r_1,r_1+\epsilon]$ and $r\in [r_2-\epsilon,r_2]$ and $e_{\epsilon}(r)=0$ if $r\leq r_1$
and $r \geq r_2$. Therefore one might expect that the wave function (\ref{103}) describes a 
quasicassical state if $r_0=(r_1+r_2)\gg d$ and $q_0d\gg 1$. 

As follows from Eq. (\ref{95}), the
corresponding wave function in momentum space is
\begin{eqnarray}
&&\chi(q)=\frac{i}{\sqrt{\pi d}}\{\frac{exp[i(q+q_0)r_0]}{q+q_0}
sin[(q+q_0)d]-\nonumber\\
&&\frac{exp[i(q_0-q)r_0]}{q-q_0}sin[(q-q_0)d]\}
\label{104}
\end{eqnarray}
This function is obviously normalized to one since it is a unitary transformation of the function
given by Eq. (\ref{103}). At the same time, since
\begin{equation} 
\frac{1}{\pi d}\int_{-\infty}^{\infty}\{\frac{sin[(q-q_0)d]}{q-q_0}\}^2dq=1
\label{105}
\end{equation}    
it is clear that for macroscopic bodies the first term in Eq. (\ref{104}) is negligible in
comparison with the second one, and the wave function in momentum space is not small only if
$|(q-q_0)|\leq 1/d$. Therefore one might think that Eqs. (\ref{103}) and (\ref{104})
give a good quasiclassical description. On the other hand, it is obvious that, since at large
$q$, $|\chi(q)|^2=O(1/q^2)$, the mean value of the relativistic mass operator 
$\sqrt{m_1^2+q^2}+\sqrt{m_2^2+q^2}$ is divergent, to say nothing about the mean value of
the nonrelativistic mass operator $q^2/2m_{12}$. 

If $e(r)$ is replaced by $e_{\epsilon}(r)$ 
then the mean value of the mass operator becomes finite.
This is clear from the fact that the mean value of $(q-q_0)^2$ over the Fourier transform is proportional
to $\int_{r_1}^{r_2} [de_{\epsilon}(r)/dr]^2dr$. However, as follows from the Lebesgue theorem, 
if $\epsilon\to 0$, the mean value goes to infinity. Therefore, $\epsilon$ cannot be very small and we 
conclude that in
standard quantum theory the wave function (\ref{103}) is not realistic. One might investigate
orther wave functions. For example, by analogy with a function
discussed in textbooks on distributions, instead of $e_{\epsilon}(r)$ one might consider
a function, which has a supporter $[r_1,r_2]$ and is proportional to  $exp[-d^2/(d^2-(r-r_0)^2)]$ at
$r\in [r_1,r_2]$. In any case, although quasiclassical
approximation has been discussed in numerous works, we are not aware of examples where it has been 
explicitly shown that a wave function with the supporter $[r_1,r_2]$ such that
$b \ll r_0$ is such that the uncertainty of momentum is much less than $q_0$
and the uncertainty of the kinetic energy is much less than $q_0^2/2m_{12}$. As it is clear from the 
example with the state (\ref{103}), our intuition is not sufficient to make a conclusion without 
explicit calculations. An interesting pedagogical example has been proposed by Alik Makarov.
Let $E_n$ be the $n$th state of the harmonic oscilator ($n=0,1,...$), $\omega(n)$ is the energy
of the $n$th state and $\alpha$ is a number such that $1/2 < \alpha < 1$. Then the state 
$\sum_n E_n/\omega(n)^{\alpha}$ is normalizable but the mean value of the energy is infinite.  
We will see in Chap. \ref{Ch5} that in the dS theory an analog of (\ref{103}) is a state with a finite
energy. 

Consider now the case $l\neq 0$. Let $a^2$ be an eigenvalue of the operator defined by
Eq. (\ref{93}). The corresponding eigenfunction is a solution of the equation
\begin{equation}
[-\frac{d^2}{dq^2}+\frac{l(l+1)}{q^2}]\chi(q)=a^2\chi(q)
\label{lneq0}
\end{equation}
It is well known (see e.g. Ref. \cite{LL}) that the solution regular at $q=0$ can be written as 
\begin{equation}
\chi_a(q)=\sqrt{aq}J_{l+1/2}(aq)
\label{cfug1}
\end{equation}
where $a>0$ and $J_{l+1/2}$ is the Bessel function. Since we are interested in the behavior 
of quasiclassical wave functions at large momenta, it suffices to consider the solution
at $aq\gg l$. The asymptotics of the Bessel function $J_{l+1/2}(z)$ at $z\gg l$ 
is given by \cite{BE}
\begin{equation}
J_{l+1/2}(z)=\sqrt{\frac{2}{\pi z}}sin(z-\frac{l\pi}{2})
\label{cfug2}
\end{equation}
and therefore $\chi_a(q)\approx \sqrt{2/\pi}sin(aq-l\pi /2)$ at $aq\gg l$. This function
depends on $l$ only in the phase of the sine function while it is the eigenfunction of the operator
$-d^2/dq^2$ with the eigenvalue $a^2$. 

The fact that the centrifugal term is not important at $aq\gg l$ can also be understood
as follows \cite{LL}.The solution of Eq. (\ref{93}) can be written as
\begin{equation}
\chi_a(q)=(-1)^l\frac{q^{l+1}}{a^l}(\frac{1}{q}\frac{d}{dq})^l(\frac{sin(aq)}{q})
\label{cfug3}
\end{equation}
When $aq\gg l$, the main contribution arises from the differentiation of the sine and we
again obtain a function proportional to $sin(aq-l\pi /2)$. This example shows that each
differentiation gives a factor $a$. Therefore $-d^2/dq^2$ gives a factor $a^2$. This
is much greater than the centrifugal term $l(l+1)/q^2$ if $qa\gg l$. It is also clear that
if $aq\gg l$ then the centrifugal term in Eq. (\ref{lneq0}) is much smaller than the
term with $a^2$, therefore at this condition the centrifugal term can be neglected.

In summary, if $l\neq 0$,
we can draw the same conclusions about the behavior of quasiclassical wave functions at
large momenta as in the case $l=0$. In particular, the state (\ref{103}) is not legitimate.
It will be shown in Chap. \ref{Ch5} that in the dS theory the behavior of eigenstates of the 
relative distance operator at large momenta strongly depends on $l$: the greater is $l$,
the faster is the falloff of the wave function. The behavior of the eigenstates at small
momenta is also considerably different since there exist not only solutions with 
$J_{l+1/2}$ but also ones with $J_{-(l+1/2)}$.

The above discussion sheds some light on a possible form of a quasiclasical wave function in
the dS theory. Since the relative distance operator in this theory has not been
discussed yet, we cannot guarantee that $r_0$ is the relative distance. Suppose
however that this is the case. As noted in the preceding section, the analog of relative momentum
in the dS theory is a quantity $n$ such that $q=n/R$. Therefore, by analogy with
Eqs. (\ref{100})) and (\ref{104}), one might expect that a quasiclassical wave function should
contain a rapidly oscillating term $exp[i(n-n_0)r_0/R]$ or $cos[(n-n_0)r_0/R]$. Suppose, for
example, that the quantities $c_n$ in Eq. (\ref{84}) have the form $c_n=a(n)cos[(n-n_0)r_0/R]$,
where the amplitude $a(n)$ has a sharp maximum at $n=n_0$. In the spirit of quasiclassical
approximation, we assume that when $n$ changes by one, it suffices to consider only the 
change in the rapidly changing argument of $cos[(n-n_0)r_0/R]$. Then
\begin{eqnarray}
&&c_{n+1}c_n*\approx |a(n)|^2\{(1-r_0^2/2R^2)cos^2[(n-n_0)r_0/R]-\nonumber\\
&&(r_0/2R)sin[2(n-n_0)r_0/R]\}  
\label{106}
\end{eqnarray} 
Since $sin[2(n-n_0)r_0/R]$ is a rapidly oscillating function, we assume that the contribution
of the last term to the mean value in Eq. (\ref{84}) is negligible. Therefore, if we take into 
account the correction of order $r_0^2/R^2$, then instead of Eq. (\ref{85}), we get 
\begin{equation}
(\Psi,M^2\Psi)=[\epsilon_1(q_0)+\epsilon_2(q_0)]^2-
\frac{r_0^2}{R^2}\epsilon_1(q_0)\epsilon_2(q_0)
\label{107}
\end{equation} 
The same result can be obtained if $c_n=a(n)exp[i(n-n_0)r_0/R]$. Therefore, the correction of
order $r_0^2/R^2$ to the mean value of the nonrelativistic mass operator is $-m_{12}r_0^2/2R^2$ 
in agreement with Eq. (\ref{AG11}). 

The results (\ref{85}) and (\ref{107}) give an additional argument that the quantum number $n$
has the meaning of the dS relative momentum. At the same time, quasiclassical approximation
implies that not only the mean value of the mass operator is in agreement with our expectations,
but the wave function in the mass representation has a sharp maximum around the mean value.
This problem is studied in the next chapter.

\chapter{Momentum - mass uncertainty relation}
\label{Ch4}
\section{Spectral decomposition of the mass operator}
\label{S14}

In standard theory the free mass operator commutes with the momentum operator. Therefore these
operators have a common set of eigenvectors. However, in dS theory the free
mass operator does not commute with the momentum operator. Therefore if a quasiclassical
wave function has a sharp maximum in the $n$-representations, this does not guarantee yet
that it has a sharp maximum in the mass representation. Since the operators
do not commute, there exists a momentum - mass uncertainty, i.e. the momentum and the
mass cannot be measured simultaneously. 

The decomposition of the eigenvector of the operator $W$ with the eigenvalue $\lambda$ over
the eigenvectors of the dS momentum operator is given by Eq. (\ref{67}), where in the
standard dS theory based on complex numbers, the limit $N\to \infty$ should be taken. 
However, if for some reasons, only momenta $n=0, 1, ... N$ are allowed, then, as follows from
Eqs. (\ref{65}) and (\ref{66}), $\chi(\lambda,N)$ will be the eigenvector of $W$ with the 
eigenvalue $\lambda$ if $\Delta^N(\lambda)=0$. Let $e_n=\Phi_n/||\Phi_n||$ be the 
normalized eigenvector of the momentum operator with the momentum $n$. Then, as follows from
Eqs. (\ref{61}), (\ref{66c}), (\ref{67b}) and (\ref{67c})
\begin{eqnarray}
&&\chi(\lambda,N)=\frac{(1+s)!}{(1+j)}
\{\prod_{l=1}^j[w_1+(2l+1)^2][w_1+(2l+1)^2]\}^{1/2}\nonumber\\
&&\frac{\Gamma((s+3-i\mu_1)/2)\Gamma((s+3-i\mu_2)/2)}
{\Gamma((s+3+i\sigma)/4)\Gamma((s+3-i(2\mu_1+2\mu_2+\sigma))/4)
\Gamma((s+3)/2)}\nonumber\\
&&\sum_{n=0}^N e_n\{\prod_{l=0}^{n-1}\frac{[\mu_1+i(2l+s+3)][\mu_2+i(2l+s+3)]}
{[w_1+(2l+s+3)^2]^{1/2}[w_2+(2l+s+3)^2]^{1/2}} \}\nonumber\\
&&(\frac{n+j+1}{C_{n+s+1}^{1+s}})^{1/2}\{\int_0^1(1-z)^{(s-1+i\sigma)/4}
z^{(s+1)/2}C_n^{1+j}(2z-1)\nonumber\\
&&F(\frac{s+3+i(2\mu_1+\sigma)}{4},\frac{s+3+i(2\mu_2+\sigma)}{4};\frac{s+3}{2};z)dz\}
\label{108}
\end{eqnarray}
As noted in Sect. \ref{S5} and in the preceding chapter, the de Sitter masses of particles and
the quantity $n$ are typically very large. Therefore, the gamma functions depending on the
de Sitter masses and kinetic energy $\sigma$ can be calculated by using the Stirling formula
and for the Gegenbauer polynomial we can take its asymptotic expression \cite{BE}
\begin{equation}
C_n^a(cos\theta)=2\frac{(a)_n}{n!}\frac{cos[(n+a)\theta-a\pi/2]}{(2sin\theta)^a}
\label{109}
\end{equation}

Our next task is to obtain the expression for the hypergeometric function in Eq. (\ref{108})
using the fact that the de Sitter masses are very large. 
This function can be written as
$F(a,b;c;z)$ where
\begin{equation}
a=\frac{s+3}{4}+i\alpha,\,\, b=\frac{s+3}{4}+i\beta,\,\, c = \frac{s+3}{2},
\,\, \alpha = \frac{2\mu_1+\sigma}{4},\,\, \beta = \frac{2\mu_2+\sigma}{4}
\label{110}
\end{equation}
We use the expression \cite{BE}
\begin{eqnarray}
&&F(a,b;c;z)=\Gamma(c)[\frac{\Gamma(c-a-b)}{\Gamma(c-a)\Gamma(c-b)}F(a,b;a+b+1-c;1-z)+\nonumber\\
&&\frac{\Gamma(a+b-c)}{\Gamma(a)\Gamma(b)}(1-z)^{c-a-b}F(c-a,c-b;c+1-a-b;1-z)]
\label{111}
\end{eqnarray}
Then by using the expression \cite{BE}
\begin{eqnarray}
&&F(a,b;a+b+1-c;1-z)=\frac{\Gamma(a+b+1-c)}{\Gamma(b)\Gamma(a+1-c)}\cdot\nonumber\\
&&\int_0^{\infty}t^{b-1}(1+t)^{c-b-1}(1+tz)^{-a}dt
\label{112}
\end{eqnarray}
Eqs. (\ref{110}), (\ref{111}) and the Stirling formula for the Gamma function, we have
\begin{eqnarray}
&&F(a,b;c;z)=\frac{\Gamma(c)}{\pi|\beta|^{(s+1)/2}}exp[-\frac{i}{2}(\alpha+\beta)ln(1-z)]\nonumber\\
&&Re\{exp[\frac{i}{2}(\alpha+\beta)ln(1-z)]\int_0^{\infty}\frac{[t(1+t)]^{(s-1)/4}}
{(1+tz)^{(s+3)/4}}\cdot\nonumber\\
&&exp[i\beta lnt-i\beta ln(1+t)-i\alpha ln(1+tz)]dt\}
\label{113}
\end{eqnarray}
Since the quantities $\alpha$ and $\beta$ are very large, we can calculate the last integral
by using the stationary phase method. The stationary point $t$ should satisfy the equation
\begin{equation}
\beta (1+tz)=\alpha zt(1+t)
\label{114}
\end{equation} 
and the only possible solution is
\begin{equation}
t=\frac{1}{2\alpha}\{\beta - \alpha +[(\alpha - \beta )^2+4\alpha\beta /z]^{1/2}\}
\label{115}
\end{equation} 

As a consequence, the result for $F(a,b;c;z)$ can be described as follows. Let 
$\phi(\alpha,\beta , z)$ be a function such that
\begin{equation}
\frac{d\phi(\alpha,\beta , z)}{dz}= -\frac{1}{2(1-z)}
[(\alpha - \beta )^2+4\alpha\beta /z]^{1/2},\quad \phi(\alpha,\beta , 0)=-\pi/4
\label{116}
\end{equation}
Then
\begin{eqnarray}
F(a,b;c;z)=\sqrt{\frac{2}{\pi}}\Gamma(c)\frac{exp[-\frac{i}{2}(\alpha+\beta)ln(1-z)]cos[\phi(\alpha,\beta , z)]}
{(\alpha\beta)^{(s+1)/4}z^{(s+2)/4}[4\alpha\beta +z(\alpha - \beta)^2]^{1/4}}
\label{117}
\end{eqnarray}

As follows from Eqs. (\ref{109}) and (\ref{117}), for calculating Eq. (\ref{108}), we
need to consider the integral
\begin{eqnarray}
&&I=\int_0^1cos[(n+j+1)\theta -\frac{\pi}{2}(j+1)]
\frac{cos[\phi(\alpha,\beta , z)]}{[\lambda -(\mu_1 - \mu_2)^2(1-z)]^{1/4}}\nonumber\\
&&exp[-\frac{i}{4}(\mu_1+\mu_2)ln(1-z)]\frac{dz}{z^{1/2}(1-z)^{3/4}}
\label{118}
\end{eqnarray}
where $cos\theta = 2z-1$. The product of two cosine functions in the integrand can be written
as a sum of four imaginary exponents and then for calculating the integral we can apply
the stationary phase method.
Let us note that in Sect. \ref{S11} we represented $\lambda$ as $(\mu_1+\mu_2+\sigma)^2$ only for
convenience of calculations. We assumed only that $\lambda > 0$ and therefore we can write
$\lambda$ as $\lambda=\xi^2$ where $\xi > 0$. As follows from Eqs. (\ref{61}), (\ref{65}) and 
(\ref{67}), the final
result for the decomposition of $\chi(\lambda , N)$ should not depend on $(\mu_1,\mu_2,\sigma)$ but
on $(w_1,w_2,\lambda)$ only.  Therefore for performing calculations in Eq. (\ref{108}) we can assume
for definiteness that $\mu_1>0,\,\mu_2>0,\,\alpha > 0$ and $\beta>0$. Then each contribution to
the integral can be considered in four cases: $(\mu_1+\mu_2>\xi ,\, n^2>\alpha\beta)$,
$(\mu_1+\mu_2<\xi ,\, n^2>\alpha\beta)$, $(\mu_1+\mu_2>\xi ,\, n^2<\alpha\beta)$ and 
$(\mu_1+\mu_2<\xi ,\, n^2<\alpha\beta)$.

The final result is as follows. Let 
\begin{eqnarray}
&&D=D(n,\lambda)=\{n^2\lambda-\frac{1}{16}[\lambda^2-2\lambda (w_1+w_2)+(w_1-w_2)^2]\}^{1/2}\nonumber\\
&&tg\gamma(n,\lambda)=\frac{4D}{\lambda-w_1-w_2-8n^2},\,\, 0<\gamma(n,\lambda)<\pi
\label{119}
\end{eqnarray}
and $\theta_0$ is a constant, which does not depend on $n$ and $\lambda$. As follows from the
expression for $D(n,\lambda)$, for any fixed value of $\lambda$, $n$ can be only in the range
$[n_{min},N]$ where 
\begin{equation}
n_{min} = n_{min}(\lambda)=\frac{1}{4\xi}[\lambda^2-2\lambda (w_1+w_2)+(w_1-w_2)^2]^{1/2}
\label{nmin}
\end{equation}
Then
\begin{equation}
\chi(\lambda,N) = C(\lambda)\sum_{n=n_{min}}^N \frac{1}{D^{1/2}}cos[arg(n,\lambda)]e_n
\label{120}
\end{equation}
where $C(\lambda)$ can be found from the normalization condition (see below) and
\begin{eqnarray}
&&arg(n,\lambda)=\frac{\xi}{4}ln|\frac{n\xi +D}{n\xi - D}| +
\frac{\mu_1}{4}ln|\frac{n(\lambda+w_1-w_2)-2\mu_1 D}{n(\lambda+w_1-w_2)+2\mu_1 D}|\nonumber\\ 
&&+\frac{\mu_2}{4}ln|\frac{n(\lambda+w_2-w_1)-2\mu_2 D}{n(\lambda+w_2-w_1)+2\mu_2 D}|-n\gamma+\theta_0
\label{121}
\end{eqnarray}
The function $arg(n,\lambda)$ indeed depends only on $(w_1,w_2,\lambda)$ since
it is invariant under the replacements $\mu_1\to -\mu_1,\,\mu_2\to -\mu_2,\, \xi\to -\xi$.

\begin{sloppypar}
As follows from these expressions, the asymptotics of $arg(n,\lambda)$ at large values of $n$ and $\xi > 0$ is 
\begin{eqnarray}
&&arg(n,\lambda) \approx \frac{\xi}{2}[ln(n\xi)+1+3ln2]-\frac{1}{4}(\xi+\mu_1+\mu_2)ln|\xi+\mu_1+\mu_2|-\nonumber\\
&& \frac{1}{4}(\xi-\mu_1-\mu_2)ln|\xi-\mu_1-\mu_2|-\frac{1}{4}(\xi+\mu_1-\mu_2)ln|\xi+\mu_1-\mu_2|-\nonumber\\
&&\frac{1}{4}(\xi-\mu_1+\mu_2)ln|\xi-\mu_1+\mu_2|+\theta_0 
\label{122}
\end{eqnarray}
This result is obviously invariant under the replacements $\mu_1\to -\mu_1,\,\mu_2\to -\mu_2$.
As noted above, if $N$ is finite then the condition $cos[arg(N+1,\lambda)]=0$ should be satisfied.
Therefore $arg(N+1,\lambda)=(k+1/2)\pi$ ($k=\pm 1, \pm 2,...$). If $N$ is finite then the
spectrum of the operator $W$ should be obviously discrete and, as follows from Eq. (\ref{122}), 
the distance $\Delta\xi$ between neighboring eigenvalues is such that
\begin{equation}
\frac{\Delta\xi}{2}ln\frac{8N\xi}{|(\xi+\mu_1+\mu_2)(\xi-\mu_1-\mu_2)
(\xi+\mu_1-\mu_2)(\xi-\mu_1+\mu_2)|^{1/2}}\approx\pi
\label{123}
\end{equation}
If $N\to\infty$ then $\Delta\xi\to 0$ as it should be since, as shown in Sect. \ref{S11}, in the
standard dS theory the spectrum of the operator $W$ is continuous. Note
that the spectrum is not equidistant and therefore $\Delta\xi$ is a function of $\xi$.
\end{sloppypar}

If $N$ is finite then the function $\chi(\lambda, N)$ can be normalized to one, i.e.
\begin{equation}
C(\lambda)^2 \sum_{n=n_{min}}^N \frac{1+cos[2arg(n,\lambda)]}{D(n,\lambda)}=1
\label{124}
\end{equation}
Suppose again that with a good accuracy we can replace summation by integration over $n$. 
As follows from Eqs. 
(\ref{119}), (\ref{nmin}) and (\ref{123})
\begin{equation}
\int_{n_{min}}^N \frac{dn}{D(n,\lambda})=\frac{2\pi}{\Delta\lambda}
\label{125}
\end{equation}
where $\Delta\lambda = 2\xi\Delta\xi$. Consider now the contribution of $cos[2arg(n,\lambda)]$ to
the integral in Eq. (\ref{124}). If $arg(n,\lambda)$ is formally treated as a function of 
continuous variables then a direct calculation gives
\begin{equation}
\frac{\partial arg(n,\lambda)}{\partial n}=-\gamma(n,\lambda),
\quad \frac{\partial arg(n,\lambda)}{\partial \xi}=\frac{1}{4}ln|\frac{n\xi +D}{n\xi - D}|
\label{deriv}
\end{equation}
As follows from the first expression in these formulas and from Eq. (\ref{119}), $cos[2arg(n,\lambda)]$
is a rapidly oscillating function in the integration range and therefore its contribution is
negligible.

We conclude that, as follows from Eq. (\ref{120}),
\begin{eqnarray}
&&\chi(\lambda,N) = \sqrt{\frac{\Delta\lambda}{2\pi}}\sum_{n=n_{min}}^N 
\frac{1}{D^{1/2}}cos[arg(n,\lambda)]e_n \nonumber\\
&&e_n = \sum_{\lambda=\lambda_{min}}^{\lambda_{max}} 
\sqrt{\frac{\Delta\lambda}{2\pi}}\frac{1}{D^{1/2}}cos[arg(n,\lambda)]\chi(\lambda,N)
\label{126}
\end{eqnarray}
where $\lambda_{min}=\lambda_{min}(n)$ and $\lambda_{max}=\lambda_{max}(n)$ are respectively the minimum and
maximum values of $\lambda$ at a fixed value of $n$. Note that since $\Delta\lambda$ is a function of
$\lambda$, we cannot take $\sqrt{\Delta\lambda /2\pi}$ out of the summation sign in the second 
expression.

To show that Eq. (\ref{126}) gives a one-to-one relation between the elements 
$\{e_n\}$ and $\{\chi(\lambda, N)\}$,
we have to prove that the basis $\{\chi(\lambda, N)\}$ is complete, i.e. the norm
of the second expression in Eq. (\ref{126}) equals one. We denote
\begin{equation}
\omega_1(n)=\sqrt{w_1+4n^2},\quad\omega_2(n)=\sqrt{w_2+4n^2}
\label{127}
\end{equation}
It is clear from the consideration in Sect. \ref{S5} that these quantities are the standard particle
energies in the Poincare invariant theory multiplied by $2R$. Then it follows from the expression for
$D$ in Eq. (\ref{119}) that
\begin{eqnarray}
&&\lambda_{min}(n)=|\omega_1(n)-\omega_2(n)|^2,\quad \lambda_{max}(n)=|\omega_1(n)+\omega_2(n)|^2\nonumber\\
&&D(n,\lambda)=\frac{1}{4}[(\lambda - \lambda_{min}(n))(\lambda_{max}(n)-\lambda)]^{1/2}
\label{128}
\end{eqnarray}
Note that when $n$ increases, $\lambda_{min}(n)$ decreases and $\lambda_{max}(n)$ increases. If
$N\gg \mu_1,\mu_2$ then the absolute minimum of $\lambda_{min}$ is approximately zero and
the absolute maximum of $\lambda_{min}$ approximately equals $4N^2$.  

In Poincare invariant theory the mass of the free two-body system equals a sum of the particle
energies in the c.m. frame. For this reason one might expect that in the de Sitter invariant
theory the operator $W$ has a spectrum in a small vicinity of $|\omega_1(n)+\omega_2(n)|^2$. As shown
in Chap. \ref{Ch2}, the de Sitter antigravity reduces the value of the mass but the reduction is small
since $r\ll R$. An important consequence of Eq. (\ref{128}) is that $|\omega_1(n)+\omega_2(n)|^2$ 
{\it is the maximum value of} $W$ while in general the values of $\lambda$ considerably less than
$|\omega_1(n)+\omega_2(n)|^2$ are possible. For example, if we define the mass operator as a square
root from the positive definite operator $W/4R^2$ then {\it there is no law prohibiting a free
nonrelativistic two-body system to be in a state where the mean value of the mass operator is
$\epsilon_1(p_0)+\epsilon_2(p_0)-Gm_1m_2/r_0$ where $p_0$ is the mean value of the
momentum in the c.m. frame and $r_0$ is the mean value of the distance between the particles.} 
The problem arises whether such a state can be quasiclassical and whether there are reasons making
such a value of the mass operator more preferable than the others.   

We are now in position to prove that the second expression in Eq. (\ref{126}) is indeed compatible
with the requirement $||e_n||=1$. As follows from this expression and Eq. (\ref{128}),
\begin{equation}
||e_n||^2=\frac{1}{\pi}\sum_{\lambda_{min}}^{\lambda_{max}}
\frac{\{1+cos[2arg(n,\lambda)]\}\Delta\lambda}{[(\lambda - \lambda_{min})(\lambda_{max}-\lambda)]^{1/2}} 
\label{129}
\end{equation} 
As follows from the second expression in Eq. (\ref{deriv}), $cos[2arg(n,\lambda)]$ is a rapidly
oscillating function in the summation region and therefore its contribution is negligible.
With a good accuracy we can replace summation by integration and, as a result
\begin{equation}
||e_n||^2=\frac{1}{\pi}\int_{\lambda_{min}}^{\lambda_{max}}
\frac{d\lambda}{[(\lambda - \lambda_{min})(\lambda_{max}-\lambda)]^{1/2}}=1 
\label{130}
\end{equation} 
This completes the proof that there exists a one-to-one relation between the elements 
$\{e_n\}$ and $\{\chi(\lambda, N)\}$ and this relation is described by Eq. (\ref{126}).

\begin{sloppypar}
\section{Quasiclassical wave functions in momentum and mass representations}
\label{S15}
\end{sloppypar}

As noted in Sect. \ref{S13}, one might expect that
the coefficients $c_n$ representing a quasiclassical wave function in the $n$-representation, 
have the form $c_n=a(n)cos[(n-n_0)r_0/R]$ or $c_n=a(n)exp[-i(n-n_0)r_0/R]$,
where the amplitude $a(n)$ has a sharp maximum at $n=n_0$. If the wave function can be written as
$\sum_{\lambda}b(\lambda)\chi(\lambda, N)$ then the coefficients $b(\lambda)$
represent the wave function in the $\lambda$-representation. From standard quasiclassical 
experience one might expect that in
quasiclassical approximation the function $b(\lambda)$ should have a sharp maximum at some
$\lambda_0$ close to $|\omega_1(n_0)+\omega_2(n_0)|^2$. Let us investigate whether this is
the case. 

For simplicity we assume that $c_n=a(n)exp[-i(n-n_0)r_0/R]$. Then, as follows from Eq. (\ref{126})
\begin{equation}
b(\lambda)=\sqrt{\frac{\Delta\lambda}{2\pi}}\sum_{n_{min}}^N \frac{a(n)}{D(n,\lambda)^{1/2}}
exp[-i(n-n_0)r_0/R]cos[arg(n,\lambda)]
\label{131}
\end{equation} 
We write $cos$ as a sum of two imaginary exponents and assume that with a good accuracy 
summation can be replaced by an integration (see the discussion in the end of this section). Then
\begin{eqnarray}
&&b(\lambda)=\sqrt{\frac{\Delta\lambda}{8\pi}}exp(in_0r_0/R)\int_{n_{min}}^N
\frac{a(n)}{D(n,\lambda)^{1/2}}\nonumber\\
&&exp\{i[arg(n,\lambda)-\frac{nr_0}{R}]\}+exp\{-i[arg(n,\lambda)+\frac{nr_0}{R}]\}dn
\label{132}
\end{eqnarray}
Since the exponents are rapidly oscillating, we can apply the stationary phase method. As
follows from Eq. (\ref{deriv}), the stationary point exists only in the second term and
is defined by the condition
\begin{equation}
\gamma(n,\lambda)=\frac{r_0}{R}
\label{133}
\end{equation}
Therefore the value of $\gamma(n,\lambda)$ in the stationary point is small but not zero.
As follows from Eqs. (\ref{119}) and (\ref{nmin}), the standard relativistic relation
$\lambda= |\omega_1(n)+\omega_2(n)|^2$ would imply $D=0$ and $n=n_{min}$ but this is
impossible since the stationary point should necessarily be inside the integration region. 

Since $\gamma(n,\lambda)$ is small, we can replace $tg\gamma$ by $\gamma$ in Eq. (\ref{119}).
Then, as follows from Eqs. (\ref{119}) and (\ref{128}), the condition (\ref{133}) can
be written as
\begin{equation}
4D(n,\lambda)=[(\lambda - \lambda_{min})(\lambda_{max}-\lambda)]^{1/2}=\frac{r_0}{2R}
(2\lambda - \lambda_{min}-\lambda_{max})
\label{134}
\end{equation}
Since $r_0\ll R$, it is clear from this expression that only a solution with $\lambda$ very
close to $\lambda_{max}$ is possible. The solution is given by
\begin{equation}
\lambda =\lambda_{max} -\frac{r_0^2}{R^2}\omega_1(n)\omega_2(n)=
|\omega_1(n)+\omega_2(n)|^2-\frac{r_0^2}{R^2}\omega_1(n)\omega_2(n)
\label{135}
\end{equation}
This expression defines the position of the stationary point $n=n(\lambda)$ as a function
of $\lambda$. It is easy to verify that in the nonrelativistic approximation Eq. 
(\ref{AG11}) follows from Eq. (\ref{135}).

For calculating integral in Eq. (\ref{132}) in the stationary phase method, we need to know
the second derivative of the phase in the stationary point. A direct calculation gives
\begin{equation}
-\frac{\partial^2\arg(n,\lambda)}{\partial n^2}=\frac{\partial\gamma(n,\lambda)}{\partial n}=
\frac{n[\lambda(\omega_1(n)^2+\omega_2(n)^2)-(w_1-w_2)^2]}{D(n,\lambda)\omega_1(n)^2\omega_2(n)^2}
\label{136}
\end{equation} 
As follows from Eq. (\ref{134}), the quantity $D$ in the stationary point is proportional
to $r_0/R$. Therefore when $\lambda$ satisfies Eq. (\ref{135}) and $r_0\ll R$, this condition
can be written as
\begin{equation}
\frac{\partial\gamma(n,\lambda)}{\partial n}=
\frac{8n\lambda_{max}}{D(n,\lambda)(\lambda_{max}-\lambda_{min})}=
\frac{64nR\lambda_{max}}{r_0(\lambda_{max}-\lambda_{min})^2}
\label{137}
\end{equation} 
As a consequence
\begin{eqnarray}
&&b(\lambda)=[\frac{\omega_1(n)\omega_2(n)\Delta\lambda}{8n\lambda_{max}(n)}]^{1/2}
a(n)\nonumber\\
&&exp[-i(arg(n,\lambda)+\frac{(n-n_0)r_0}{R}-\frac{\pi}{4})]
\label{138}
\end{eqnarray} 
where $n$ should be understood as $n(\lambda)$. A check for consistency of this result is to
verify that $b(\lambda)$ is normalized to one. We again replace a sum over $\lambda$ by integration
and note that, as follows from Eq. (\ref{135}), in the approximation $r_0\ll R$, 
$d\lambda=8n\lambda_{max}(n)dn/[\omega_1(n)\omega_2(n)]$. Then, as follows from Eq. (\ref{138})
\begin{equation}
\int_0^{4N^2}|b(\lambda)|^2d\lambda = \int_0^N |a(n)|^2dn = 1
\label{139}
\end{equation} 
as it should be.

Consider now the following question. We used the method of stationary phase assuming that 
summation over $n$ in Eq. (\ref{131}) can be approximated by integration over $n$. A justification 
of such an approximation is not obvious. For example one might expect that integration could
be a good approximation for summation if $arg(n,\lambda)$ does not change significantly when $n$
changes by one. In other words, we should have $|\partial arg(n,\lambda)/\partial n| \ll 1$.
At the same time, as follows from Eqs. (\ref{119}) and (\ref{deriv}), this is not the case for
all values of $n$. However, if the stationary phase method can be used then the main 
contribution is given by a small vicinity of the stationary point where $|\gamma(n,\lambda)|\ll 1$
and normalization is conserved. Therefore we need to justify the usage of the stationary point
method.  

As already noted, the stationary point cannot be at the value of
$n=n_{min}(\lambda)$ defined from the standard relativistic relation between the energy and
momentum. However, if $r_0\ll R$, the stationary point $n(\lambda)$ is close to that value.
Let us estimate the difference $\Delta n=n(\lambda)-n_{min}(\lambda)$. 
We denote $n_r=n_{min}(\lambda)$ where the subscript $r$ indicates that this quantity is
obtained from the standard relativistic expression
$\lambda =|\omega_1(n_r)+\omega_2(n_r)|^2$. 
Then it follows from Eq. (\ref{135}) that
\begin{equation}
\Delta n \approx \frac{[r_0\omega_1(n_r)\omega_2(n_r)]^2}
{8n_r[R\omega_1(n_r)\omega_2(n_r)]^2}
\label{140}
\end{equation}
In the stationary point method the main contribution to the integral is given by a small
vicinity of the stationary point defined by the second derivative of the exponent index.
Let $\delta n$ be the width of this vicinity. As follows from Eq. (\ref{136}), the value
of $\delta n$ can be estimated as
\begin{equation}
\delta n \approx \frac{\omega_1(n_r)\omega_2(n_r)}
{2[\omega_1(n_r)+\omega_2(n_r)]}\sqrt{\frac{r_0}{n_rR}}
\label{141}
\end{equation}
Then the stationary phase method can be used if the following conditions are satisfied: 
$\delta n\gg 1$, such that the vicinity of the stationary point giving the
main contribution contains many values of $n$; $\delta n \ll \Delta n$ such that this
vicinity is fully inside integration region. 

For simplicity we consider the nonrelativistic approximation. Then $\omega_1\approx 2Rm_1$,
$\omega_2\approx 2Rm_2$ and $n_r\approx Rm_{12}v$ where $m_{12}$ is the reduced mass and
$v$ is the relative velocity of particles 1 and 2. We now can rewrite Eqs. (\ref{140}) and
(\ref{141}) as follows
\begin{equation}
\Delta n \approx \frac{m_{12}r_0^2}{2Rv}, \quad \delta n = \frac{1}{2}\sqrt{\frac{m_{12}r_0}{v}}
\label{142}
\end{equation}
It is obvious that the conditions $\delta n\gg 1$ is always valid for macroscopic bodies
since $v\ll 1$ and $r_0$ is much greater than their Compton wave lengths. The condition
$\Delta n \gg \delta n$ is valid if 
\begin{equation}
\frac{m_{12}r_0}{v}(\frac{r_0}{R})^2 \gg 1
\label{143}
\end{equation}
One might conclude that this result is incorrect since in the Poincare limit $R\to \infty$ 
the stationary phase approximation should be undoubtedly valid for macroscopic particles.
However, as noted in Sect. \ref{S11}, we are working in the representation where there is no
direct transition to the Poincare invariant theory and we argued that this representation
is more physical than that described by Eqs. (\ref{9}) and (\ref{10}). 

It is clear that the greater the values of $m_{12}$ and $r_0$ are the higher is the
accuracy of Eq. (\ref{143}). Consider the Cavendish experiment where 
$m_1 = 348pounds$, $m_2 = 1.61pounds$ and $r_0= 9 inches$. In this case $m_{12}r_0=4.8\cdot 10^{41}$.
The maximum acceleration of the small ball is $Gm_1/r_0^2=2.0\cdot 10^{-7}m/s^{-2}$. Suppose
that equilibrium of the torsion balance apparatus is reached within a time of order $1s$. 
Then $v$ is of order $10^{-7}m/s$.
Therefore in units where $c=1$, $v=6.7\cdot 10^{-14}$ and $m_{12}r_0/v=7.2\cdot 10^{56}$.
If, for example, $R=10^{26}m$, then the l.h.s. of Eq. (\ref{143}) is $3.8\cdot 10^3$.
In the literature it is discussed at what conditions deviations from the Newton law of 
gravitation could be expected. The above example shows that at the conditions of the
Cavendish experiment the l.h.s. of Eq. (\ref{143}) is not anomalously large. Breaking of
Eq. (\ref{143}) would mean that the wave function cannot be simultaneously
quasiclassical in both momentum and mass representations, i.e. the state becomes essentially
quantum. It is clear that this happens much earlier than one could expect from the standard
quantum mechanical experience. In this case a measurement of the gravitational constant
might give a value considerably different from the expected one. 

Another consequence of Eq. (\ref{143}) is as follows. If the requirement of the existence
of quasiclassical states is such that the wave function should have a sharp maximum in
momentum and mass representations simultaneously then the cosmological constant $\Lambda$, which is
of order $1/R^2$, cannot be anomalously small. As noted in Chap. \ref{Ch1}, at present the 
conclusion that $\Lambda\neq 0$ is based on astronomical data. On the other hand, Eq. (\ref{143})
gives a restriction from the opposite side. If the wave function is simultaneously quasiclassical
in momentum and mass representations then the values of $R$ much greater than $10^{27}m$ are 
highly unlikely. As a consequence, the term "Poincare limit" should be understood not as 
a formal limit $R\to\infty$ but only as a situation when the distances in question are much smaller
than $R$ but $R$ cannot be infinite.

One might treat the results of this chapter as a proof that standard quantum mechanical intuition
works in the dS theory as well. Indeed, we have shown that not only the mean value of
the dS mass operator is in agreement with the Poincare limit and dS antigravity but it is
possible to construct wave functions which have a sharp maximum simultaneously in the momentum
and mass representation. Our construction was based on the assumption that the $r$ dependence
of the quasiclassical wave function in momentum representation contains $exp[-i(n-n_0)r/R]$
in analogy with the nonrelativistic quantum mechanics. On the other hand, we have noted that even
in relativistic theory there exists a serious problem with the position operator and,
to the best of our knowledge, the problem of the position operator in the dS theory
has not been discussed at all. Therefore the above assumption can be substantiated
only if an operator having the meaning of the spatial coordinate is constructed in the
framework of the dS theory.

\chapter{Relative distance operator in de Sitter invariant theory}
\label{Ch5} 
\section{Construction of relative distance operator}
\label{S16}

In Chaps. \ref{Ch1} and \ref{Ch2} we discussed problems in constructing a position operator
in dS theory. Strictly speaking, a measurable quantity is not a position by
itself but a relative position with respect to some other particle. In nonrelativistic quantum
mechanics the position operators for particles 1 and 2 in momentum representation are defined
as ${\bf r}_1=i\partial/\partial {\bf p}_1$ and ${\bf r}_2=i\partial/\partial {\bf p}_2$, respectively,
and the relative position operator is defined as ${\bf r}={\bf r}_1-{\bf r}_2$.

In the representation (\ref{9}) the form of
the Lorentz boost operator is the same as in relativistic theory. For spinless particles the
Lorentz boost operators in momentum representation can be written as
\begin{equation}
{\bf N}_r^{(1)}==-2i \epsilon_1({\bf p}_1)\frac{\partial}{\partial {\bf p}_1} \quad
{\bf N}_r^{(2)}==-2i \epsilon_2({\bf p}_2)\frac{\partial}{\partial {\bf p}_2}
\label{144}
\end{equation} 
where the subscript $r$ stands for "relativistic". Therefore, as already noted,
the Lorentz boost operator is proportional to the standard position operator only in nonrelativistic
approximation. Since the representation operators (\ref{9}) are the only operators in our
disposal, we conclude that in dS theory there are no physical operators 
proportional to $i\partial/\partial {\bf p}_1$ or $i\partial/\partial {\bf p}_2$. In  
relativistic theory this is a well known problem discussed in a wide literature. The most known
solution is the Newton - Wigner operator \cite{NW}, but it is recognized that even this
operator does not have all the required properties for the position operator. 

The idea of the 
Newton - Wigner and other constructions is to have an operator, which satisfies some minimum
requirements; in particular, this operator should be self-adjoint and proportional to the
nonrelativistic position operator in the nonrelativistic limit. For example, one could define
the relative position operator in relativistic theory as follows. Since the energy operator
$E_r^{(1)}$ for particle 1 is the operator of multiplication by $\epsilon_1({\bf p}_1)$ and analogously
for $E_r^{(2)}$, a possible definition is
\begin{equation}
{\bf D}_r=E_r^{(1)}{\bf N}_r^{(2)}-E_r^{(2)}{\bf N}_r^{(1)}
\label{145}
\end{equation}
This operator is indeed self-adjoint and proportional to the
relative position operator in the nonrelativistic limit. It is obvious that ${\bf D}_r$ does not
have the dimension of length. Since it should be treated as an operator related to the internal state 
of the two-body system, it
should commute with the total momentum operator ${\bf P}={\bf p}_1+{\bf p}_2$ and this
is indeed the case. 

Note that different components of ${\bf D}_r$ do not
commute with each other. A possibility of noncommutative space-time coordinates has been first
discussed by Snyder \cite{Snyder} and nowadays it is widely discussed in noncommutative
field theory and string theory. In our approach, noncommutativity of different
components of the operator ${\bf D}_r$ is not postulated from the beginning but follows 
from other requirements. As a consequence of noncommutativity, a part of the spectrum of the operator
${\bf D}_r$ might be not purely continuous and even the whole spectrum might be discrete.
To the best of our knowledge, in the literature on relativistic two-body systems 
(see eg. Refs. \cite{pack,rqm}) the operator ${\bf D}_r$ has not been investigated. 
Here a typical approach is that the relative momentum operator is constructed first and then, if
the coordinate description is required, the relative position operator is defined as canonically
conjugated with the relative momentum operator. Since in our approach there are no classical
manifolds at all, the notion of canonically conjugated operator is not natural. Our goal is to 
construct a dS analog of the operator ${\bf D}_r$ and investigate the spectrum of the dS
relative distance operator. 

In Sect. \ref{S11} we discussed the construction of the operator $W$, 
which is a reduction of the two-body operator $I_2$ onto the space $H_s$. 
The reduction is defined by Eq. (\ref{59}) and its meaning is as follows. The requirement
${\bf J}'x=0$ means that we treat ${\bf J}'$ as a dS analog of the total momentum operator
and consider a subspace of states with zero total momentum. The reduction of the operator
${\bf J}"$ onto that subspace is treated as an internal dS angular momentum operator, i.e. as a
spin operator for the two-body system. Then $H_s$ is defined as a subspace of eigenvectors
of the operator $J_z"$ with the eigenvalue $s$ and such that $J^{"+}x=0$. Therefore we
should define a dS relative position operator ${\bf D}$ in such a way that it commutes
with ${\bf J}'$ and is a vector operator with respect to ${\bf J}"$, i.e. 
\begin{equation}
[J^{'\alpha},D^{\beta}]=0,\quad [J^{"\alpha},D^{\beta}]=2ie_{\alpha\beta\gamma}D^{\gamma}
\label{146}
\end{equation}
where $(\alpha,\beta,\gamma)=1,2,3$, $e_{\alpha\beta\gamma}$ is the absolutely antisymmetric 
tensor and a sum over repeated indices is assumed. By using the relations (\ref{15}-\ref{17}),
one can directly verify that the operator
\begin{equation}
{\bf D}=E^{(1)}{\bf N}^{(2)}-E^{(2)}{\bf N}^{(1)}+{\bf N}^{(1)}\times {\bf N}^{(2)}
\label{147}
\end{equation}
satisfies Eq. (\ref{146}).
As follows from the
discussion in Sect. \ref{S5}, the operator ${\bf D}/2R$ becomes ${\bf D}_r$ in the formal
limit $R\to\infty$. The operator ${\bf D}$ is obviously dimensionless in agreement with our 
discussion in Sects. \ref{S3} and \ref{Comments}
that in dS theory there should be only dimensionless quantities. In the nonrelativistic
approximation ${\bf D}=\mu_1\mu_2{\bf r}/2R=2Rm_1m_2{\bf r}$ and therefore typically the
components of ${\bf D}$ are huge numbers. 

Different components of ${\bf D}$ do not commute with each other and the expressions for the
commutators are rather cumbersome. However, since $[{\bf D}^2, {\bf J}"]=0$, one can work in
the representation where the operators ${\bf J}^{"2}$, $J_z"$ and ${\bf D}^2$ are 
diagonal; in particular one can consider a reduction of ${\bf D}^2$ onto $H_s$. A direct
calculation using Eqs. (\ref{15}-\ref{17}) gives
\begin{equation}
{\bf D}^2=(E^{(1)2}+{\bf N}^{(1)2})(E^{(2)2}+{\bf N}^{(2)2})-
(E^{(1)}E^{(2)}+{\bf N}^{(1)}{\bf N}^{(2)})^2-16{\bf J}^{(1)'}{\bf J}^{(2)'}
\label{148}
\end{equation} 

We now consider a reduction of this operator onto the space $H_s$ defined by Eq. (\ref{59}).
The elements $\Phi_n$ defined by Eq. (\ref{60}) form a basis in this space. As follows from 
Eqs. (\ref{19}), (\ref{59}) and (\ref{60})
\begin{eqnarray}
&&[{\bf J}^{(1)'}]^2\Phi_n=[{\bf J}^{(1)"}]^2\Phi_n=[{\bf J}^{(2)'}]^2\Phi_n=
[{\bf J}^{(2)"}]^2\Phi_n=\nonumber\\
&&(n+j)(n+j+2)\Phi_n\quad {\bf J}^{'}\Phi_n=0\quad {\bf J}^{"2}\Phi_n=s(s+2)\Phi_n
\label{149}
\end{eqnarray}
Now by using Eqs. (\ref{17}), (\ref{22}), (\ref{23}), (\ref{62}) and (\ref{148}) one can 
show by a direct calculation that 
\begin{eqnarray}
&&{\bf D}^2\Phi_n=\{[w_1+9+4(n+j)(n+j+2)][w_2+9+4(n+j)\nonumber\\
&&(n+j+2)]-\frac{1}{4}{\tilde W}^2+16(n+j)(n+j+2)\}\Phi_n
\label{150}
\end{eqnarray}
where ${\tilde W}$ is a nondiagonal part of $W$, i.e. the operator with the matrix
elements ${\tilde W}_{nn}=0$, ${\tilde W}_{n,n+1}=W_{n,n+1}^s$ and 
${\tilde W}_{n+1,n}=W_{n+1,n}^s$.

Suppose there exist states which are quasiclassical in the momentum, mass and relative position
representations simultaneously. If a wave function has a sharp maximum in the momentum
representation at $n_0=Rq_0$, a sharp maximum in the $W$ representation at $w_0=4R^2M_0^2$
and a sharp maximum in the relative position representation at ${\bf D}_0$
then Eq. (\ref{150}) makes it possible to write down a relation between $q_0$, $M_0$ and 
${\bf D}_0$. We can neglect the commutators between $n$, $W$ and ${\bf D}^2$, take into 
account Eq. (\ref{65}) and the fact that in quasiclassical approximation $j\ll n$, $n\gg 1$
and $w_1,w_2\gg 1$. The result is
\begin{equation}
M_0^2=\epsilon_1(q)^2+\epsilon_2(q)^2+2\epsilon_1(q)\epsilon_1(q) 
[1-\frac{{\bf D}_0^2}{4R^2\epsilon_1(q)^2\epsilon_2(q)^2}]^{1/2}
\label{151}
\end{equation}
This is a general relation between $q_0$, $M_0$ and ${\bf D}_0$ without assuming
nonrelativistic approximation and that the distances are much less than $R$.
Suppose now that these assumptions are valid. Then, as noted above, ${\bf D}_0$
can be written as ${\bf D}_0=2Rm_1m_2{\bf r}_0$ and it is easy to verify that Eq. (\ref{151})
is compatible with Eq. (\ref{AG11}). 

We now use ${\cal D}$ to denote the operator, which in the normalized basis defined by Eq. (\ref{80})
has the matrix elements ${\cal D}_{nn}=0$, ${\cal D}_{n,n+1}=-iW(n,j)/2$ and 
${\cal D}_{n+1,n}=iW(n,j)/2$ where
$W(n,j)$ is defined in Eq. (\ref{81}). Then, as follows
from Eqs. (\ref{65}) and (\ref{127}), Eq. (\ref{150}) can be rewritten as
\begin{eqnarray}
&&{\bf D}^2e_n=\{{\cal D}^2+(\frac{[w_1+(2n+s+3)^2][w_2+(2n+s+3)^2]}{n+j+2}+\nonumber\\
&&\frac{[w_1+(2n+s+1)^2][w_2+(2n+s+1)^2]}{n+j})\frac{j(j+1)}{2(n+j+1)}+\nonumber\\
&&4[\omega_1(n+j+1)^2+\omega_2(n+j+1)^2+2]\}e_n
\label{152}
\end{eqnarray}
If $j\ll n$ and $n\gg 1$, this expression can be simplified as 
\begin{eqnarray}
{\bf D}^2e_n=\{{\cal D}^2+\frac{j(j+1)}{n^2}\omega_1(n)^2\omega_2(n)^2+
4[\omega_1(n)^2+\omega_2(n)^2]\}e_n
\label{153}
\end{eqnarray}
and the operator ${\cal D}$ acts as
\begin{equation}
{\cal D}\sum_n c(n)e_n = \frac{i}{2} \sum_n [c(n+1)W(n)-c(n-1)W(n-1)]e_n
\label{155}
\end{equation} 
where $W(n)=\omega_1(n)\omega_2(n)$.
 
Let us compare Eq. (\ref{152}) with the standard nonrelativistic expression 
(\ref{93}). As follows from Eqs. (\ref{127}) and (\ref{155}), in the nonrelativistic
approximation ${\cal D}$ is a finite difference version of the operator 
$i\mu_1\mu_2 \partial/\partial n$. Since $n=Rq$ then, in view of the relation between
${\bf D}$ and ${\bf r}$ in the nonrelativistic approximation, we conclude that the
term with ${\cal D}^2$ in Eq. (\ref{153}) is in agreement with the first term in Eq. (\ref{93}).
Analogously, the term with $j(j+1)$ in Eq. (\ref{153}) is in agreement with the 
centrifugal term in Eq. (\ref{93}). Finally, in the nonrelativistic limit the last term in 
Eq. (\ref{153}) is obviously negligible in comparison with the first and second terms. 

The above results give grounds to conclude that the operator ${\bf D}$ can be treated as
a relative position operator in dS theory and therefore ${\bf D}^2$ can be treated
as a relative distance operator squared. For quasiclassical states the values of the internal
angular momentum are typically very large and it is obvious that if $j\neq 0$, the centrifugal
term is much greater than the last term in Eq. (\ref{153}). Therefore in physically interesting
cases this term can be neglected. We consider first a model when $j=0$ and ${\bf D}^2={\cal D}^2$
and the case $j\neq 0$ will be discussed in Sect. \ref{S18}.

\section{Operators ${\cal D}$ and ${\cal D}^2$ in states with zero spin}
\label{S17}

As follows from the definition of the operator ${\cal D}$ and Eq. (\ref{81}), the action of 
the operator ${\cal D}$ is defined by 
\begin{equation}
{\cal D}_{n,n+1}=-{\cal D}_{n,n+1}=\frac{i}{2}\{[w_1+(2n+3)^2][w_2+(2n+3)^2]\}^{1/2}
\label{156}
\end{equation}
while all the other matrix elements are equal to zero. We consider a case when
the basis is formed by the elements $e_0,e_1,...e_N$, so the case of the
infinite dimensional space can be obtained in the limit $N\to\infty$. In view of 
the discussion in the preceding section and in Sect. \ref{S13}, one might think that
${\cal D}$ can be treated as a dS analog of the operator $id/dr$. As noted in Sect. \ref{S13},
the latter is not a well defined self-adjoint operator. At the same
time, ${\cal D}$ is not an operator defined on a space of functions of continuous
variable. It is obviously well defined when $N$ is finite but a problem arises whether 
${\cal D}$ is a good physical operator when $N\to\infty$. We will see below that in the
latter case only ${\cal D}^2$ is the operator with good physical properties while ${\cal D}$
is not. 

By analogy with Sect. \ref{S11}, in this chapter we use $\Delta_k^n(\eta)$
to denote the determinant of the matrix of the operator $({\cal D}-\eta)$ truncated
to a subspace with the basis $e_k,e_{k+1},...e_n$. We believe the usage of the same 
notation should not lead to misunderstanding. As follows from Eq. (\ref{156}), an analog of 
Eq. (\ref{66}) is
\begin{eqnarray}
&&\Delta^{n+1}(\eta)=-\eta\Delta^n(\eta)-a_n\Delta^{n-1}(\eta),\quad a_n= 
{\cal D}_{n,n+1}{\cal D}_{n+1,n}= \nonumber\\
&&\frac{1}{4}[w_1+(2n+3)^2][w_2+(2n+3)^2]
\label{157}
\end{eqnarray}
where $\Delta^n(\eta)\equiv \Delta_0^n(\eta)$. 
As follows from Eqs. (\ref{156}) and (\ref{157})
\begin{equation}
E(\eta)=\sum_{n=0}^N (-1)^n\Delta^{n-1}(\eta)e_n/\prod_{k=0}^{n-1}{\cal D}_{k,k+1}
\label{158}
\end{equation}
is the eigenvector of the operator ${\cal D}$ with the eigenvalue $\eta$ if
\begin{equation}
\Delta^N(\eta)=0
\label{159}
\end{equation}
Our next goal is to find solutions of this equation. 

It is obvious that $\Delta^0(\eta)=-\eta$ and $\Delta^1(\eta)=\eta^2-a_0$. Therefore Eq. (\ref{157})
makes it possible to calculate $\Delta^n(\eta)$ for any $n$. In particular, it is clear that the
expression for $\Delta^n(\eta)$ contains only odd powers of $\eta$ if $n$ is even and only even powers
of $\eta$ if $n$ is odd. Therefore
\begin{equation}
\Delta^n(\eta) =\sum_{l=0}^{n+1}b(l,n)\eta^l
\label{160}
\end{equation}
where the coefficients $b(l,n)$ are such that $b(l,n)=0$ if $n+l$ is even. If $n$ is odd, it is easy to
show by induction that
\begin{equation}
b(0,n)=(-1)^{(n+1)/2}a_0a_2\cdots a_{n-1}
\label{161}
\end{equation}
and, as follows from Eq. (\ref{157})
\begin{eqnarray}
&&b(0,n)=(-1)^{(n+1)/2}4^{3(n+1)/2}\cdot\nonumber\\
&&|\frac{\Gamma((-i\mu_1+2n+5)/4)\Gamma((-i\mu_2+2n+5)/4)}
{\Gamma((-i\mu_1+3)/4)\Gamma((-i\mu_2+3)/4)}|^2
\label{162}
\end{eqnarray}

As follows from the Jacobi formula for the derivative of the determinant
\begin{equation}
b(l,n)=\frac{1}{l!}\frac{d^l}{d\eta^l}[\Delta^n(0)]
\label{163}
\end{equation}
Therefore by using the fact that the matrix of the operator ${\cal D}$ is threediagonal and all the
diagonal matrix elements of the the operator ${\cal D}-\eta$ are equal to $-\eta$, one obtains
\begin{equation}
b(l,n)=\sum_{i_1i_2...i_l}\Delta_0^{i_1-1}(0)\Delta_{i_1+1}^{i2-1}(0)\cdots \Delta_{i_l+1}^n(0)
\label{164}
\end{equation}
where $0\leq i_1<i_2...<i_l\leq n$. 

Consider, for example, a case when $n$ is odd. Since $\Delta_l^k(0)=0$ if $k-l$ is even, 
the nonzero coefficients can be written as
\begin{equation}
b(2l,n)=\sum_{j_1j_2...j_{2l}}\Delta_0^{2j_1-1}(0)\Delta_{2j_1+1}^{2j_2}(0)
\Delta_{2j_2+2}^{2j_3-1}(0)\cdots \Delta_{2j_{2l}+2}^n(0)
\label{165}
\end{equation}
where $0\leq j_1\leq j_2<j_3\leq j_4<j_5...\leq j_{2l}\leq (n-1)/2$. By analogy with Eq. (\ref{162}),
one can show that
\begin{eqnarray}
&& \Delta_{2j_1+1}^{2j_2}(0)=(-1)^{j_2-j_1}4^{3(j_2-j_1)}\cdot\nonumber\\
&&|\frac{\Gamma((-i\mu_1+4j_2+5)/4)\Gamma((-i\mu_2+4j_2+5)/4)}
{\Gamma((-i\mu_1+4j_1+5)/4)\Gamma((-i\mu_2+4j_1+5)/4)}|^2\nonumber\\
&& \Delta_{2j_2+2}^{2j_3-1}(0)=(-1)^{j_2-j_1+1}4^{3(j_2-j_1-1)}\cdot\nonumber\\
&&|\frac{\Gamma((-i\mu_1+4j_3+3)/4)\Gamma((-i\mu_2+4j_3+3)/4)}
{\Gamma((-i\mu_1+4j_2+7)/4)\Gamma((-i\mu_2+4j_2+7)/4)}|^2
\label{166}
\end{eqnarray}
and, as a consequence
\begin{eqnarray}
&&b(2l,n)=(-1)^{\frac{n+1}{2}-l}4^{3(\frac{n+1}{2}-l)}\cdot\nonumber\\
&&|\frac{\Gamma((-i\mu_1+2n+5)/4)\Gamma((-i\mu_2+2n+5)/4)}
{\Gamma((-i\mu_1+3)/4)\Gamma((-i\mu_2+3)/4)}|^2\nonumber\\
&&\sum_{j_1j_2...j_{2l}}A(j_1)B(j_2)\cdots A(j_{2l-1})B(j_{2l})
\label{167}
\end{eqnarray}
where 
\begin{eqnarray}
&&A(j)=|\frac{\Gamma((-i\mu_1+4j+3)/4)\Gamma((-i\mu_2+4j+3)/4)}
{\Gamma((-i\mu_1+4j+5)/4)\Gamma((-i\mu_2+4j+5)/4)}|^2\nonumber\\
&&B(j)=|\frac{\Gamma((-i\mu_1+4j+5)/4)\Gamma((-i\mu_2+4j+5)/4)}
{\Gamma((-i\mu_1+4j+7)/4)\Gamma((-i\mu_2+4j+7)/4)}|^2
\label{168}
\end{eqnarray}

So far, no approximation has been made and the general expression (\ref{167}) is rather complicated.
It can be simplified as follows. We can use the fact that if $z$ is large then 
$\Gamma(z+1/2)\approx \Gamma(z)z^{1/2}$ and take into account that $A(j)$ and $B(j)$ typically enter 
Eq. (\ref{167}) with large values of $j$. Therefore
\begin{equation}
A(j)\approx B(j)\approx 4[(w_1+16j^2)(w_2+16j^2)]^{-1/2} 
\label{169}
\end{equation} 
The next approximation is as follows. If $l\ll n$, the contribution to the sum in Eq. (\ref{167}) of 
the terms where some of the $j's$ are equal to each other, is much smaller than the contribution of
the terms with $j_1<j_2<...j_{2l}$. Therefore
\begin{equation}
\sum_{j_1j_2...j_{2l}}A(j_1)B(j_2)\cdots A(j_{2l-1})B(j_{2l})\approx 
\frac{1}{(2l)!}u(\frac{n-1}{2})^{2l}
\label{170}
\end{equation}
where the function $u(k)$ is defined as
\begin{equation}
u(k)=\sum_{j=0}^k [(w_1+16j^2)(w_2+16j^2)]^{-1/2}
\label{171}
\end{equation}
Since $w_1,w_2\gg 1$, with a good accuracy $u(k)$ can be treated as a function of a continuous variable
$k$ defined as
\begin{equation}
u(k)=\int_0^k [(w_1+16x^2)(w_2+16x^2)]^{-1/2}dx
\label{171a}
\end{equation}

With these approximations, the expression for $\Delta^n(\eta)$ when $n$ is odd, is proportional to the sum
$$1-\frac{1}{2!}(2\eta u(K))^2 + \frac{1}{4!}(2\eta u(K))^4\pm \frac{1}{(2K)!}(2\eta u(K))^{2K}$$
where $K=(n-1)/2$. A question arises whether this sum can be replaced by $cos(2\eta u(K))$. The matter
is that the argument of the cosine is typically very large. For example, in the nonrelativistic
approximation, $u(K)=n/2\mu_1\mu_2$, $\eta=2\mu_1\mu_2 r/$ and therefore $2\eta u(K)=rn/R=rq$ as
expected. For macroscopic bodies this quantity is indeed very large. On the other hand, if $x$ is a
large positive value, the Taylor series for $cos(x)$ gives a good approximation if the number of
terms in the series is of order $x$. Therefore the above sum can be replaced by $cos(2\eta u(K))$ if
$r\ll R$.

The case when $n$ is even can be considered analogously and the final result is
\begin{eqnarray}
&&\Delta^n(\eta)=(-1)^{(n+1)/2}4^{3(n+1)/2}cos(2\eta u(\frac{n-1}{2}))\cdot\nonumber\\
&&|\frac{\Gamma((-i\mu_1+2n+5)/4)\Gamma((-i\mu_2+2n+5)/4)}
{\Gamma((-i\mu_1+3)/4)\Gamma((-i\mu_2+3)/4)}|^2\nonumber\\
&&\Delta^n(\eta)=(-1)^{n/2+1}4^{3(n+1)/2}sin(2\eta u(\frac{n-1}{2}))\cdot\nonumber\\
&&|\frac{\Gamma((-i\mu_1+2n+5)/4)\Gamma((-i\mu_2+2n+5)/4)}
{\Gamma((-i\mu_1+3)/4)\Gamma((-i\mu_2+3)/4)}|^2
\label{172}
\end{eqnarray}  
if $n$ is odd and even, respectively. Therefore solutions of Eq. (\ref{159}) can be written as
\begin{eqnarray}
\eta_1(N)=\pm (l+\frac{1}{2})\pi/2u(\frac{N-1}{2}),\quad
\eta_2(N)= \pm l\pi/2u(\frac{N}{2})
\label{173}
\end{eqnarray}
where $l=0,1,2...$ and $N$ is odd and even, respectively. As follows from Eqs. (\ref{156}), (\ref{158})
and (\ref{172}), the eigenvector of the operator ${\cal D}$ with the eigenvalue $\eta$ corresponding to
$r\ll R$ can be written as
\begin{eqnarray}
&&E(\eta)=E_1(\eta)+E_2(\eta)\nonumber\\
&&E_1(\eta )=const\sum_{k=0}^{[N/2]}\frac{cos[2\eta u(k-1)]e_{2k}}
{[(w_1+16k^2)(w_2+16k^2)]^{1/4}}\nonumber\\
&&E_2(\eta )=const\sum_{k=0}^{[(N-1)/2]}\frac{sin[2\eta u(k)]e_{2k+1}}
{[(w_1+16k^2)(w_2+16k^2)]^{1/4}}
\label{174}
\end{eqnarray}
where $[N/2]$ is the integer part of $N/2$ and the values of $const$ should be determined
from the normalization condition. Since we have calculated the coefficients assuming that $n\gg 1$,
then strictly speaking, the lower limit in these sums should be not zero but some quantity $k_0\gg 1$.

We see that the spectrum of the operator ${\cal D}$ is pure discrete, at least in the region
corresponding to $r\ll R$. Let us estimate the quantum of the distance. If $N\gg \mu_1$ 
and $N\gg \mu_2$, the
upper limit in Eq. (\ref{171a}) can be replaced by $\infty$ and the integral becomes the
complete elliptic integral of the first kind \cite{BE} 
\begin{equation}
u(\infty)\equiv u_{\infty}=\frac{\pi}{8\mu_1}F(\frac{1}{2},\frac{1}{2};1;\frac{\mu_1^2-\mu_2^2}{\mu_1^2})
\label{ellip}
\end{equation}
Therefore, if $\mu_1$ and $\mu_2$ are of the same order, $u_{\infty}$ is of order $\pi/8\mu$, where $\mu$
is of order $\mu_1$ and $\mu_2$. If $\mu_1\gg \mu_2$, the hypergeometric function can be estimated by
using its integral representation and the result is
\begin{equation}
u_{\infty}=\frac{1}{4\mu_1}ln\frac{2\mu_1^2}{\mu_2^2}
\label{ellip2}
\end{equation}
In the nonrelativistic approximation $\eta = \mu_1\mu_2r/R$. Therefore if the masses are of the same order,
the quantum of $r$ is $R/\mu=1/m$ while if $\mu_1\gg\mu_2$, the quantum of $r$ is of order 
$1/[m_2ln(m_1/m_2)]$. This is in agreement with what was noted about the position operator
in Chaps. \ref{Ch1} and \ref{Ch2}. On the other hand, as already noted, in dS theory
any operator with the dimension of length is artificial and we treat $\eta$ as the dS analog of the
length.  

As follows from Eq. (\ref{173}), if $\eta \neq 0$ belongs to the spectrum of the operator ${\cal D}$
then $-\eta$ also belongs to the spectrum. This does not mean yet that the operator ${\cal D}$ is
unphysical since it is ${\cal D}^2$, which is treated as the operator of the distance squared.
However, if one is working with the operator ${\cal D}$, the following problems arise. First of all,
if one requires that the basis is formed by the eigenvectors of the operator ${\cal D}^2$ which are
simultaneously the eigenvectors of ${\cal D}$, then each eigenvalue of the operator ${\cal D}^2$
is doubly degenerated and therefore such eigenvalues do not represent a full set of quantum
numbers. Another problem is related to the behavior of the coefficients in Eq. (\ref{174}) when
$k\to\infty$. This limit is not defined uniquely since, depending on whether $N$ is odd or even,
in the limit $N\to\infty$ one obtains two different sets of eigenvalues. Suppose, for example, that
$N\to\infty$ and $N$ is odd. Then, if $E(\eta)=\sum (c_{2k}e_{2k}+c_{2k+1}e_{2k+1})$, it follows from 
Eqs. (\ref{173}) and (\ref{174}) that at large $k$,
$c_{2k}=O(1/k)$ while $c_{2k+1}=O(1/k^2)$. Analogously, if $N$ is even then $c_{2k}=O(1/k^2)$ 
and $c_{2k+1}=O(1/k)$. Such a behavior of the coefficients is unphysical since it would mean that at
large momenta a half of the coefficients can be neglected. In addition, we will see below that the falloff
of the coefficients as $O(1/k)$ leads to infinite mean value of the mass operator.

The vectors $E_1(\eta)\pm E_2(\eta)$ are the eigenvectors of the operator ${\cal D}$ with the eigenvalues
$\pm \eta$, respectively but $E_1(\eta)$ and $E_2(\eta)$ are not the eigenvectors of ${\cal D}$.
One can represent the Hilbert space $H$ as a direct sum $H=H_1+H_2$ where $H_1$ is the space with
the basis $\{e_{2k}\}$ and $H_2$ is the space with the basis $\{e_{2k+1}\}$. As follows from Eq. (\ref{156}),
${\cal D}$ has nonzero matrix elements only for transitions between $H_1$ and $H_2$ and therefore each
subspace, $H_1$ and $H_2$, is invariant under the action of ${\cal D}^2$. In particular, 
$E_1(\eta)$ and $E_2(\eta)$ are the eigenvectors of ${\cal D}^2$ with the eigenvalue $\eta^2$. 
If $x_1\in H_1$ and $x_2\in H_2$ are nonzero vectors then $x_1\pm x_2$ can be the eigenvector of ${\cal D}^2$
only if $x_1$ and $x_2$ are the eigenvectors of ${\cal D}^2$ with the same eigenvalue. However, if we
do not require that each eigenvector of ${\cal D}^2$ should be the eigenvector of ${\cal D}$, we can
seek eigenvectors of ${\cal D}^2$ in the subspaces $H_1$ and $H_2$ independently. Suppose, for example,
that $N$ is even. Then $E_1(\eta_1(N))$ is the eigenvector of ${\cal D}^2$ with the eigenvalue
$\eta_1(N)^2$ and $E_2(\eta_2(N-1))$ is the eigenvector of ${\cal D}^2$ with the eigenvalue
$\eta_2(N-1)^2$. These eigenvalues are different and therefore the set of eigenvalues of the operator
${\cal D}^2$ is a full set of quantum numbers. When $n\to\infty$, the decomposition coefficients of
the both vectors, $E_1$ and $E_2$ over the basis $\{e_n\}$ have the behavior $O(1/n^2)$. Therefore
indeed, ${\cal D}^2$ has good physical properties while ${\cal D}$ has not. 

If $E_1(\eta)$ and $E_2(\eta)$ are treated as independent eigenvectors of the operator ${\cal D}^2$
then $||E_1(\eta)||=||E_2(\eta)||=1$ and the value of $const$ in Eq. (\ref{174}) can be 
calculated as follows. Since only large values of $k$ are important, one can use Eq. (\ref{171a})
and $u(k-1)\approx u(k)$. Then
\begin{eqnarray}
||E_{1,2}||^2=\frac{1}{2}const^2\int_0^{u(N/2)}[1\pm cos(4\eta u)]du\approx \frac{1}{2}const^2 u(N/2)
\label{const}
\end{eqnarray}
since the contributin of $cos(4\eta u)$ is negligible. Therefore
\begin{eqnarray}
&&E_1(\eta )=[\frac{2}{u(N/2)}]^{1/2}\sum_{k=0}^{[N/2]}\frac{cos[2\eta u(k)]e_{2k}}
{[(w_1+16k^2)(w_2+16k^2)]^{1/4}}\nonumber\\
&&E_2(\eta )=[\frac{2}{u(N/2)}]^{1/2}\sum_{k=0}^{[(N-1)/2]}\frac{sin[2\eta u(k)]e_{2k+1}}
{[(w_1+16k^2)(w_2+16k^2)]^{1/4}}
\label{const1}
\end{eqnarray}
The vectors $E_1(\eta)$ and $E_2(\eta)$ are orthogonal since they belong to orthogonal subspaces,
$H_1$ and $H_2$, respectively. Since ${\cal D}^2E_{1,2}(\eta)=\eta^2E_{1,2}(\eta)$, 
$$(E_1(\eta_1),E_1(\eta_2))=(E_1(\eta_1),E_1(\eta_2))=0$$
if $\eta_1^2\neq \eta_2^2$. The orthogonality can also be proved explicitly since, as follows
from Eqs. (\ref{171a}), (\ref{const1}) and (\ref{173})
\begin{eqnarray}
&&(E_{1,2}(\eta_1),E_{1,2}(\eta_2))=\frac{2}{u(N/2)}\int_0^{u(N/2)}
[cos(2(\eta_1-\eta_2)u]\pm\nonumber\\
&&cos(2(\eta_1+\eta_2)u)du =0
\label{orth}
\end{eqnarray}

Let us note an important difference between discrete and continuous problems. In the
eigenvalue problem for a differential operator defined on a space of functions $f(x)$, $x\in [a,b]$,
one should require $f(a)=f(b)=0$. In particular, as noted in Chap. \ref{Ch3}, the radial wave function
should be equal to zero both, at $r=0$ and $r\to\infty$. In the discrete case the coefficients
$c(n)$ defining the vectors $E_1(\eta)$ and $E_2(\eta)$, are such that $c(n)\to 0$ when $n\to\infty$.
However, in this case there is no requirement $c(n)\to 0$ when $n\to 0$. As seen from Eq. (\ref{const1}),
this is the case only for coefficients defining $E_2(\eta)$. One might think that physical states
should be described only by vectors belonging to $H_2$. However, as noted in Sect. \ref{S12}, the
standard expression for the mean value of the mass operator can be obtained only when $c(n)\approx c(n-1)$.
This observation will be important in the next sections where we consider a case of nonzero spin.

\section{Operator ${\bf D}^2$ with centrifugal contribution}
\label{S18}

As follows from Eqs. (\ref{65}), (\ref{81}), (\ref{127}) and (\ref{150}), without any approximations
the action of the operator ${\bf D}^2$ can be written as
\begin{eqnarray}
&&{\bf D}^2e_n=-\frac{1}{4}\{[w_1+(2{\tilde n}+5)^2][w_2+(2{\tilde n}+5)^2]
[w_1+(2{\tilde n}+3)^2]\nonumber\\
&&[w_2+(2{\tilde n}+3)^2][1-\frac{j(j+1)}{({\tilde n}+2)({\tilde n}+3)}]
[1-\frac{j(j+1)}{({\tilde n}+1)({\tilde n}+2)}]\}^{1/2}e_{n+2}\nonumber\\
&&-\frac{1}{4}\{[w_1+(2{\tilde n}+1)^2][w_2+(2{\tilde n}+1)^2][w_1+(2{\tilde n}-1)^2]\nonumber\\
&&[w_2+(2{\tilde n}-1)^2][1-\frac{j(j+1)}{{\tilde n}({\tilde n}+1)}][1-\frac{j(j+1)}
{{\tilde n}({\tilde n}-1)}]\}^{1/2}e_{n-2}\nonumber\\
&&+\frac{1}{4}\{[w_1+(2{\tilde n}+3)^2][w_2+(2{\tilde n}+3)^2][1+
\frac{j(j+1)}{({\tilde n}+1)({\tilde n}+2)}]+\nonumber\\
&&[w_1+(2{\tilde n}+1)^2][w_2+(2{\tilde n}+1)^2][1+\frac{j(j+1)}{{\tilde n}({\tilde n}+1)}]+\nonumber\\
&&16[w_1+w_2+8({\tilde n}+1)^2]\}e_n
\label{176}
\end{eqnarray}   
where ${\tilde n}=n+j$. In macroscopic systems the value of $j$ is very large (for example, in the 
Earth - Moon system $j$ is of order $10^{68}$) and $n\gg 1$. Therefore, if $E(\eta)=\sum_n c(n)e_n$
is an eigenvector of the operator ${\bf D}^2$ with an eigenvalue $\eta^2$ then, with a high accuracy
\begin{eqnarray}
&&-\frac{1}{4}(w_1+4{\tilde n}^2)(w_2+4{\tilde n}^2)[1-\frac{j(j+1)}{{\tilde n}^2}]\{[c(n+2)+c(n-2)-\nonumber\\
&&2c(n)]+8{\tilde n}[(\frac{1}{w_1+4{\tilde n}^2}+\frac{1}{w_2+4{\tilde n}^2}][c(n+2)-c(n-2)]\}+\nonumber\\
&&(w_1+4{\tilde n}^2)(w_2+4{\tilde n}^2)\frac{j(j+1)}{{\tilde n}^2}c(n)=\eta^2 c(n)
\label{177}
\end{eqnarray}
Here we take into account that if $j\gg 1$ then 
$j(j+1)(w_1+4{\tilde n}^2)(w_1+4{\tilde n}^2)/{\tilde n}^2 \gg w_1,w_2,{\tilde n}^2$.
As already noted, in macroscopic systems $n\gg j$. Therefore if ${\tilde n}$ is formally treated
as a continuos variable and $c({\tilde n})$ does not change significantly when ${\tilde n}$ changes 
by one, the above finite difference equation can be replaced by
\begin{eqnarray} 
&&(w_1+4{\tilde n}^2)(w_1+4{\tilde n}^2)[-\frac{d^2c({\tilde n})}{d{\tilde n}^2}-8{\tilde n}
(\frac{1}{w_1+4{\tilde n}^2}+\frac{1}{w_2+4{\tilde n}^2})\frac{dc({\tilde n})}{d{\tilde n}}+\nonumber\\
&&\frac{j(j+1)}{{\tilde n}^2}c(n)]=\eta^2 c({\tilde n})
\label{178}
\end{eqnarray} 

Here comes an important difference between the standard and dS theories.
Compare Eqs. (\ref{lneq0}) and (\ref{178}). In the former the centrifugal contribution at large
momenta is negligible while in the latter it is much greater than the r.h.s. As a consequence,
as follows from Eq. (\ref{178}), if at large ${\tilde n}$, 
$c({\tilde n})=O({\tilde n}^k)$ and 
$c({\tilde n})\to 0$ then $k=-(j+2)$, i.e. $c({\tilde n})$ is a function rapidly 
decreasing at infinity. Below this
question is discussed in greater details. We will also see that at small momenta the behavior of
wave functions in standard and dS theories is essentially different as well.

If $c({\tilde n})=b({\tilde n})[(w_1+4{\tilde n}^2)(w_2+4{\tilde n}^2)]^{-1/4}$ then Eq. (\ref{178}) can
be written as
\begin{equation}
-\frac{1}{4}\frac{d^2b(u)}{du^2}+(w_1+4{\tilde n}^2)(w_2+4{\tilde n}^2)
\frac{j(j+1)}{{\tilde n}^2}b(u)=\eta^2 b(u)
\label{179}
\end{equation} 
where $u$ is now the independent variable and ${\tilde n}$ is treated as a function of $u$ defined from 
$u = u({\tilde n}/2)$ (see Eq. (\ref{171a})). Here we again assume that $j\gg 1$. 

Eq. (\ref{179}) can be considerably simplified if $\mu_1=\mu_2=\mu$ and we first
consider this case. In terms of the variable $\theta = 8\mu u$ we have 
${\tilde n}=(\mu/2) tan^{-1} (\theta/2)$ and Eq. (\ref{179}) can be written as
\begin{equation}
-\frac{d^2 b(\theta)}{d\theta^2} +\frac{j(j+1)}{sin^2\theta}b(\theta) = \frac{\eta^2}{16\mu^2}b(\theta)
\label{180}
\end{equation} 
This is a special case of the associated Mathieu equation, which can be written as an equation
for the spheroidal function $f(\theta)=b(\theta)/(sin\theta)^{1/2}$: 
\begin{equation}
\frac{d^2 f(\theta)}{d\theta^2} +cotan\theta \frac{df(\theta)}{d\theta}+
[(\frac{\eta^2}{16\mu^2}-\frac{1}{4})-\frac{(j+1/2)^2}{sin^2\theta}]f(\theta) = 0
\label{181}
\end{equation} 
  
In summary, one can consider an approximation when in the case of equal masses the variable ${\tilde n}$
is replaced by a continuous variable $\theta$ such that the normalization condition for the wave function
is given by
\begin{equation}
\sum_n |c(n)|^2 =\frac{1}{4\mu} \int_{\theta_{min}}^{\pi}|f(\theta)|^2sin\theta d\theta
\label{182}
\end{equation}
where $f(\theta)$ is a solution of Eq. (\ref{181}) and the value of $\theta_{min}$ will be discussed later.

In the literature Eq. (\ref{181}) is written in the form 
\begin{equation}
\frac{d^2 f(\theta)}{d\theta^2} +cotan\theta \frac{df(\theta)}{d\theta}+
[\nu(\nu+1)-\frac{\mu^2}{sin^2\theta}]f(\theta) = 0
\label{183}
\end{equation} 
and its two linearly independent solutions are the associated Legendre functions $P_{\nu}^{\mu}(x)$
and $Q_{\nu}^{\mu}(x)$ where $x=cos\theta$. If $\nu$ or $\mu$ are integers, they are written as
$n$ and $m$, respectively. When both $\nu$ and $\mu$ are integers, $P_n^m(x)$ is the
associated Legendre polynomial, which describes the $\theta$ dependence of the spherical wave function
$Y_{nm}(\theta,\varphi)$. Since it is an eigenfunction of a differential operator
defined on a space of functions $f(\theta)$, $\theta \in [0,\pi]$, it should satisfy the condition
$P_n^m(0)=P_n^m(\pi)=0$ if $n\neq 0$ (see the remark at the end of the preceding section). Such a
polynomial can be constructed only if the magnetic quantum number $m$ is an integer.
 
Since we reserve $\mu$ for other purposes, we will write solutions of Eq. (\ref{181}) as
$P_l^k(x)$ and $Q_l^k(x)$ where $k=j+1/2$ and $\eta^2 = 16\mu^2(l+1/2)^2$. One can immediately notice
that, in contrast with the standard case, the magnetic quantum number $k$ is now not an integer but a
half-integer. Therefore one might wonder whether it is possible to find solutions which becomes zero
at $\theta=0$ and $\theta=\pi$. However, as already noted, since we consider Eq. (\ref{181}) as an
approximation for a discrete problem, it is not necessary to require that the solution should be
zero at $\theta=0$. At the same time, $\theta=\pi$ corresponds to $n\to \infty$ and therefore the requirements
$P_l^k(x)\to 0$ and $Q_l^k(x)\to 0$ when $x\to -1$ are necessary.

Since (see e.g. Ref. \cite{BE}) 
\begin{equation}
P_l^k(x)=\frac{1}{\Gamma(1-k)}(\frac{1+x}{1-x})^{k/2}F(-l,l+1;1-k;\frac{1-x}{2})
\label{184}
\end{equation}
and $k$ is a half-integer, the above expression can be regular at $x\to -1$ only if $l$ or $-(l+1)$ is a
positive integer and these cases are essentially the same. So we will assume that $l$ is a
positive integer. Then the series for the hypergeometric function becomes a finite polynomial
and therefore $P_l^k(x)=O((1-x)^{k/2})$ when $x\to -1$. It is easy to see that this is in agreement
with the above remark that $c(n)=O(1/n^{j+2})$ when $n\to\infty$. At the same time, in contrast with
the standard case, the function $P_l^k(x)$ is very singular when $\theta\to 0$ and it is not possible
to avoid the singularity when $k$ is a half-integer. However, as noted above, the continuous approximation
can be valid only if $\theta$ is not anomalously small.  

Consider now a case when $Q_l^k(x)$ is chosen as a solution. When $k$ is a half-integer (see e.g. Ref.
\cite{BE}), 
\begin{equation}
Q_l^k(x)=\frac{\Gamma(1+l+k)\Gamma(-k)}{2\Gamma(1+l-k)}(\frac{1-x}{1+x})^{k/2}F(-l,l+1;1+k;\frac{1-x}{2})
\label{185}
\end{equation}
We can use the relation (see e.g. Ref. \cite{BE})
\begin{equation}
F(a,b;c;z)= (1-z)^{c-a-b}F(c-a,c-b;c;z)
\label{186}
\end{equation} 
and rewrite Eq. (\ref{185}) as
\begin{eqnarray}
&&Q_l^k(x)=\frac{\Gamma(1+l+k)\Gamma(-k)}{2^{1+j}\Gamma(1+l-k)}[(1-x)(1+x)]^{k/2}\nonumber\\
&&F(1+k+l,k-l;1+k;\frac{1-x}{2})
\label{187}
\end{eqnarray}
This expression can be regular at $x\to -1$ only when the hypergeometric function becomes a finite
polynomial. This happens when $l-k$ or $-(1+l+k)$ are positive integers. We will assume that $l$ is 
a positive half-integer and such that
$l-k \geq 0$. In this case the gamma functions are well defined and $Q_l^k(x)$ is regular at both,
$x\to 1$ and $x\to -1$. So $Q_l^k(cos\theta)$ is a true eigenfunction of the differential operator
defined by Eq. (\ref{181}) if $l$ and $k$ are positive half-integers.

Since $\eta^2=16\mu^2 (l+1/2)^2$ and $\eta^2$ is the eigenvalue of the operator ${\bf D}^2$,
we cannot choose a solution, which is a linear combination of $P_l^k(x)$ and $Q_l^k(x)$ since
the values of $\eta^2$ are different when the quantities $l$ are integers or half-integers.
So the only possible solutions are either $P_l^k(x)$ or $Q_l^k(x)$. What is a correct choice? 

As noted in the preceding section, the Hilbert space in question is a direct sum of $H_1$ and
$H_2$ which are invariant subspaces of the operator ${\bf D}^2$. We can seek eigenfunctions of
${\bf D}^2$ in $H_1$ and $H_2$ independently and the above discussion can be applied in the
both cases. Also it has been noted that if the eigenvalues of ${\bf D}^2$ in $H_1$ and $H_2$
are different, they represent a full set of quantum numbers. Therefore in view of the above
discussion we have only two possibilities: either the solution $P_l^k(x)$ is chosen in $H_1$
and $Q_l^k(x)$ in $H_2$ or vice versa. To choose a correct possibility one might consider the
eigenvalue problem beyond the continuous approximation. However, a simpler way is as follows.
In the region where the centrifugal contribution is much less than the term with $\eta^2$ in
Eq. (\ref{179}), we can use the results of the preceding section that the solution for $b(\theta)$ in 
Eq. (\ref{179}) should be proportional to $cos(2\eta u)$ in $H_1$ and to $sin(2\eta u)$ in 
$H_2$. On the other hand, we can use the asymptotics of Legendre 
functions when $l$ is very large \cite{BE}:
\begin{eqnarray}
&&P_l^k(cos\theta)=const[\frac{1}{sin\theta}]^{1/2}
sin[(l+\frac{1}{2})\theta+\frac{(2k+1)\pi}{4}]\nonumber\\
&&Q_l^k(cos\theta)=const[\frac{1}{sin\theta}]^{1/2}
cos[(l+\frac{1}{2})\theta+\frac{(2k+1)\pi}{4}]
\label{188}
\end{eqnarray}
Therefore if $j$ is even, one should choose the solution $P_l^k$
in $H_1$ and $Q_l^k$ in $H_2$ and vice versa if $j$ is odd.  

Consider now a case when the masses of the particles are not necessarily equal to each 
other. Suppose for simplicity that the system is nonrelativistic. This implies that if
${\bf v}_0$ is the quasiclassical value of the relative velocity and $v_0=|{\bf v}_0|$
then $v_0\ll 1$. Let ${\bf r}_0$ be the quasiclassical value of the relative distance and 
$r_0=|{\bf r}_0|$. Then $j$ is of order $m_{12}|{\bf r}_0\times {\bf v}_0|$ where 
$m_{12}$ is the reduced mass. Let us compare the contributions of the second and last terms
in Eq. (\ref{179}). Suppose that $j$ is of order $m_{12}r_0v_0$ and $\eta$ is of order 
$\mu_1\mu_2 r_0/R$. Then one can make the following conclusions. When ${\tilde n}\ll m_{12}Rv_0$,
the centrifugal term is much greater than the one with $\eta^2$. When ${\tilde n}$ increases and
becomes of order $m_{12}Rv_0$, the terms become of the same order and then the one with $\eta^2$
becomes much greater. They again become of the same order in the ultrarelativistic region
where ${\tilde n}$ is of order $(m_1+m_2)R/v_0$ and finally the centrifugal term becomes much
greater if ${\tilde n}\gg (m_1+m_2)R/v_0$. As already noted, there is no analog of the last case 
in the standard theory.

If the centrifugal contribution is small, we can use the results of the preceding section. 
Then, as noted above, $b(u)$ should be proportional to $cos(2\eta u)$ in $H_1$ 
and to $sin(2\eta u)$ in $H_2$. Consider first
the case when ${\tilde n}$ is of order $(m_1+m_2)R/v_0$ or greater. Then, as follows from 
Eq. (\ref{171a}), $u_{\infty}-u=y=1/8{\tilde n}$ and Eq. (\ref{179}) can be rewritten as 
\begin{equation}
y^2\frac{d^2b(y)}{dy^2}+[4\eta^2y^2-j(j+1)]b(y)=0
\label{189}
\end{equation} 
This is the Riccati - Bessel equation, which has two solutions. Since a necessary requirement is
$b(u)\to 0$ when ${\tilde n}\to \infty$, the only possible choice is
\begin{equation}
b(y) = const\, (2\eta  y)^{1/2}J_k(2\eta y)
\label{190}
\end{equation} 
where $J_k$ is the Bessel function of the order $k=j+1/2$. This solution should be compatible with 
the one in the region where the centrifugal contribution is small. Suppose that in this region we can use
the asymptotic expression for the Bessel function and thus
\begin{equation}
b(u) = const\, cos[2\eta u_{\infty}-\frac{(j+1)\pi}{2}-2\eta u]
\label{191}
\end{equation}
If $j$ is even then this solution is proportional to $cos(2\eta u)$ if 
$2\eta  u_{\infty}=(l+\frac{1}{2})\pi$ and to $sin(2\eta u)$
if $2\eta  u_{\infty}= l\pi$, where $l=0,1,2,...$ (compare with Eq. (\ref{173})). 
Analogously, if $j$ is odd, the solution is proportional to $sin(2\eta u)$
if $2\eta u_{\infty}= l\pi$ and to $cos(2\eta u)$ if $2\eta u_{\infty}= (l+\frac{1}{2})\pi$. 
Therefore if $j$ is even, one should choose the solution $2\eta  u_{\infty}=(l+\frac{1}{2})\pi$ in
$H_1$, $2\eta  u_{\infty}= l\pi$ in $H_2$ and vice versa if $j$ is odd. Hence the spectrum of
the operator ${\bf D}^2$ is the same as the spectrum of the operator ${\cal D}^2$ discussed
in the preceding section. 

Consider now a case when ${\tilde n}$ is of order $m_{12}Rv_0$ or less but the continuos 
approximation is still possible. This is a nonrelativistic
region and, as follows from Eq. (\ref{171a}), $u\approx {\tilde n}/(2\mu_1\mu_2)$. Therefore
Eq. (\ref{179}) can be represented as
\begin{equation}
u^2\frac{d^2b(u)}{du^2}+[4\eta^2u^2-j(j+1)]b(u)=0
\label{192}
\end{equation}
It is interesting to note that in dS theory the both opposite cases, when ${\tilde n}$ is very small
and very large, are described by the Riccati - Bessel equation. Eq. (\ref{192}) has two solutions
\begin{equation}
b(u) = const\, (2\eta  u)^{1/2}J_{\pm k}(2\eta u)
\label{193}
\end{equation} 
(compare with Eq. (\ref{190})) and, by analogy with the above discussion of the case with equal 
masses, we cannot conclude that only the first case is possible. Eq. (\ref{193}) 
should be compatible with the result in the region where the centrifugal contribution is small.
By using the asymptotic expression of the Bessel function, we obtain that
\begin{eqnarray}
&&(2\eta  u)^{1/2}J_k(2\eta u)\approx const\, cos[2\eta u - \frac{(j+1)\pi}{2}] \nonumber\\
&&(2\eta  u)^{1/2}J_{-k}(2\eta u)\approx const\, cos[2\eta u + \frac{j\pi}{2}]
\label{194}
\end{eqnarray}
Therefore if $j$ is even, one has to choose the solution with $J_{-k}$ in $H_1$ and with $J_k$ in $H_2$, 
while if $j$ is odd, the opposite choice should be made. It is easy to
see that this conclusion is compatible with the above result for equal masses. 

The above results can be summarized as follows. For analogy with the case of equal masses, we denote 
$\theta=\pi u/u_{\infty}$. Let $\theta_1$ be a value of $\theta$ when the continuous approximation is still 
valid. As noted above, this value should 
correspond to ${\tilde n}$, which is still greater than $j$. Therefore $\theta_1\geq \pi j/(\mu_1\mu_2 u_{\infty})$.
This is a very small value. For example, when $\mu_1$ and $\mu_2$ are of the same order, $\theta_1$ is
of order $v_0r_0/$R. Let $\theta_2$ be a value of $\theta$ corresponding to ${\tilde n}$ of order $(\mu_1+\mu_2)/v_0$.
Then the centrifugal contribution is dominant when $\theta_2\leq \theta\leq \pi$.
The value of $\pi-\theta_2$ is of order $v_0/[(\mu_1+\mu_2)u_{\infty}]$. When $\mu_1$ and $\mu_2$ are of the same order,
$\pi-\theta_2$ is of order $v_0$. Let $\theta_3$ b such that $\theta_1<\theta_3<\theta_2$ and at $\theta=\theta_3$
the centrifugal contribution can be neglected. Then the solutions at $\theta_1\leq \theta \leq \theta_3$ and
$\theta_3\leq \theta \leq \theta_2$ can be joined as follows. 
Consider first the case when $j$ is even. Then the eigenvalues $\eta^2$ of the operator ${\bf D}^2$ in $H_1$ are such that 
$2\eta  u_{\infty}=(l+\frac{1}{2})\pi$ ($l=0,1...$) and the corresponding 
eigenfunctions are given by
\begin{eqnarray}
&&b_l(\theta)=(-1)^{j/2} [\frac{\pi}{2} (l+\frac{1}{2})\theta]^{1/2}
J_{-(j+1/2)}((l+\frac{1}{2})\theta)\quad (\theta_1\leq \theta \leq \theta_3)\nonumber\\
&&b_l(\theta)=(-1)^{l+j/2} [\frac{\pi}{2} (l+\frac{1}{2})(\pi-\theta)]^{1/2}
J_{(j+1/2)}((l+\frac{1}{2})(\pi-\theta))\nonumber\\
&& (\theta_3\leq \theta \leq \pi)
\label{195}
\end{eqnarray}
In this case $b_l(\theta)=cos((l+1/2)\theta)$ in the region where the centrifugal contribution can be 
neglected. The eigenvalues $\eta^2$ in $H_2$ are such that 
$2\eta  u_{\infty}=l\pi$ ($l=0,1...$) and the corresponding 
eigenfunctions are given by 
\begin{eqnarray}
&&b_l(\theta)=(-1)^{j/2} [\frac{\pi}{2} l\theta]^{1/2}J_{(j+1/2)}(l\theta)
\quad (\theta_1\leq \theta \leq \theta_3)\nonumber\\
&&b_l(\theta)=(-1)^{l+j/2} [\frac{\pi}{2} l(\pi-\theta)]^{1/2}
J_{(j+1/2)}(l(\pi-\theta))\nonumber\\
&& (\theta_3\leq \theta \leq \pi)
\label{196}
\end{eqnarray}
In this case $b_l(\theta)=sin(l\theta)$ in the region where the centrifugal contribution can be 
neglected. Finally, when $j$ is odd, the solutions in $H_1$ and $H_2$ should be interchanged. 

The following important remarks are in order. While in the case of equal masses the solutions
(\ref{188}) describe the eigenfunctions in the region $\theta_1\leq \theta\leq \pi$, in the general case
we did not succeed in finding solutions describing the eigenfunctions at all those values of $\theta$.
We tried to join the solutions in different regions assuming that if $\eta u \gg j$ or 
$\eta (u_{\infty}-u)\gg j$ (such that the terms with $\eta$ in Eqs. (\ref{189}) and (\ref{192})
are much greater than the centrifugal contribution) then one can take asymptotic expressions for the
corresponding Bessel functions. This is the case when $j$ is not very large. However, when both,
$z$ and $\nu$ are large, the condition $z\gg \nu$ is not sufficient for guaranteeing that for
the Bessel functions $J_{\nu}(z)$ and $J_{-\nu}(z)$ one can take their asymptotic expressions.
This can be guaranteed only if $z\gg \nu^2$. Since as already noted, in real situations the value
of $j$ is very large, the problem of finding solutions in the case of unequal masses requires
a further study. 

Another very important feature of the above results is that it is not possible to avoid solutions, 
which are anomalously large at small values of $\theta$.
When $j$ is very large, the functions $P_l^{(j+1/2)}(\theta)$ and $J_{-(j+1/2)}((l+1/2)\theta)$
are very singular at $\theta\to 0$. The existence of these solutions is a consequence of the fact
that the cosmological constant is finite and therefore there are no analogs of such solutions
in the standard theory. As already noted, such solutions are valid only in the continuous approximation
when $\theta\geq \theta_1$, and the very problem of finding eigenvalues and eigenvectors of the
operator ${\bf D}^2$ is such that singularities are not possible. Nevertheless a question arises
whether the contribution of $\theta \leq \theta_1$ can be neglected. If $\theta\ll 1$ then $n\ll \mu_1,\mu_2$.
We also assume that $j\ll \mu_1,\mu_2$. Then, if $E(\eta)=\sum_n a(n)e_n$ is the eigenvector of the
operator ${\bf D}^2$ with the eigenvalue $\eta^2$, it follows from Eqs. (\ref{81}) and (\ref{150}) that
\begin{eqnarray}
&&-\frac{a(n+2)}{4(n+j+2)}[\frac{(n+1)(n+2)(n+2j+2)(n+2j+3)}{(n+j+1)(n+j+3)}]^{1/2}-\nonumber\\
&&-\frac{a(n-2)}{4(n+j)}[\frac{n(n-1)(n+2j)(n+2j+1)}{(n+j-1)(n+j+1)}]^{1/2}+
a(n)[1- \frac{\eta^2}{w_1w_2}-\nonumber\\
&&\frac{(n+1)(n+2j+2)}{4(n+j+1)(n+j+2)}-\frac{n(n+2j+1)}{4(n+j)(n+j+1)}]=0
\label{197}
\end{eqnarray} 
When $n$ is of order $j$ or less, one can neglect the term with $\eta$ since $\eta^2/w_1w_2$ is
of order $r_0^2/R^2$.
If $j\gg 1$ and $n\ll j$ then it follows from this expression that
\begin{equation}
a(n+2)\approx \frac{2j}{[(n+1)(n+2)]^{1/2}}a(n)
\label{198}
\end{equation} 
Therefore at these conditions the coefficients $a(n)$ sharply increase when $n$ increases. The speed
of the increase becomes slower when $n$ increases but nevertheless, when $n$ becomes of order $j$,
the values of $a(n)/a(0)$ in $H_1$ and $a(n)/a(1)$ in $H_2$ are huge numbers of order 
$[2j]^{j/2}/\sqrt{j!}\approx (2e)^{j/2}/(2\pi j)^{1/4}$. This might be an indication
that the region $n\leq j$ is important. It is clear that in the standard theory the relation 
$n\leq j$ is impossible since $j$ does not depend on $R$, $n$ is proportional to $R$ and the standard 
theory corresponds to the case $R\to\infty$. 

\section{A possible mechanism of gravity}
\label{S19}

Let us consider the internal Hilbert space at fixed $j$ as a direct sum $H=H_1+H_2$, where the basis in 
$H_1$ is $(e_0,e_2,...e_{2k}...)$ and the basis in $H_2$ is $(e_1,e_3,...e_{2k+1}...)$.
Consider the mean value of the mass operator over the state
$$\Psi = \sum_{k=0}^{\infty} [a(k)e_{2k}+b(k)e_{2k+1}]$$
We assume that the main contribution is given by $k\gg 1$ and the coefficients $a(k)$ and $b(k)$
do not change significantly when $k$ changes by one. Then Eq. (\ref{84}) can be written in the form  
\begin{eqnarray}
&&(\Psi,M^2\Psi)=\sum_{k=0}^{\infty} 
\{[m_1^2+m_2^2+8(k/R)^2][|a(k)|^2+|b(k)|^2] +\nonumber\\
&&4[m_1^2+4(k/R)^2]^{1/2}[m_2^2+4(k/R)^2]^{1/2}Re[a(k)b(k)^*]\}
\label{199}
\end{eqnarray}
As explained in Chap. \ref{Ch3}, the value of $2k/R$ has the meaning of the standard relative
momentum $q$ and, if the coefficients $c_n$ in Eq. (\ref{84}) do not change significantly
when $n$ changes by one, this equation results in the standard expression for the
mean value of the mass operator $(\Psi,M\Psi)=M_0(q_0)=\epsilon_1(q_0)+\epsilon_2(q_0)$ where 
$\epsilon_1(q)=\sqrt{m_1^2+q^2},\,\,\epsilon_2(q)=\sqrt{m_2^2+q^2}$. The effect
that the phase of $c_n$ slightly changes when $n$ changes by one results in the correction of
order $r_0^2/R^2$, which is the dS antigravity. 

The last term in Eq. (\ref{199}) shows that the result depends on interference between the wave
functions in $H_1$ and $H_2$. It is not possible to obtain the standard expression for the mean
value if $\Psi$ belongs only to $H_1$ or $H_2$. Indeed, Eq. (\ref{199}) can be rewritten as 
\begin{eqnarray}
&&(\Psi,M^2\Psi)=\sum_{k=0}^{\infty}\{M_0(2k/R)^2[|a(k)|^2+|b(k)|^2] -\nonumber\\
&&2\epsilon_1(2k/R)\epsilon_2(2k/R)|a(k)-b(k)|^2\}
\label{standard}
\end{eqnarray}
This expression clearly demonstrates that the standard result can be obtained only in
approximation $a(k)\approx b(k)$. Meanwhile, the results of this chapter give strong
indications that the  functions $a(k)$ and $b(k)$ might be considerably different.  

Consider first the case $j=0$. In view of the results of Sects. \ref{S17} and \ref{S18}, we will 
use $E_1(l)$ to denote the eigenvector of the operator 
${\cal D}^2$ in $H_1$ with the eigenvalue $\eta^2$ such that $2\eta u_{\infty}=(l+1/2)\pi$ and  
$E_2(l)$ to denote the eigenvector of the operator ${\cal D}^2$ in $H_2$ with the eigenvalue 
$\eta^2$ such that $2\eta u_{\infty}=l\pi$. Then if $\theta(k)=\pi u(k)/u_{\infty}$, Eq. 
(\ref{const1}) at $N\to\infty$ can be written as 
\begin{eqnarray}
&&E_1(l)=[\frac{2}{u_{\infty}}]^{1/2}\sum_{k=0}^{\infty}\frac{cos[(l+1/2)\theta (k)]e_{2k}}
{[(w_1+16k^2)(w_2+16k^2)]^{1/4}}\nonumber\\
&&E_2(l)=[\frac{2}{u_{\infty}}]^{1/2}\sum_{k=0}^{\infty}\frac{sin[l\theta(k)]e_{2k+1}}
{[(w_1+16k^2)(w_2+16k^2)]^{1/4}}
\label{200}
\end{eqnarray}
If $\Psi=\sum_l[A(l)E_1(l)+B(l)E_2(l)]$ then the functions $A(l)$ and $B(l)$ can be called the
coordinate wave functions in $H_1$ and $H_2$, respectively, while $a(k)$ and $b(k)$ can be called
the momentum wave functions in $H_1$ and $H_2$, respectively. It is clear that there exists a one to
one correspondence between the functions $A(l)$ and $a(k)$ and, analogously, a 
one to one correspondence between the functions $B(l)$ and $b(k)$.

As noted in Sect. \ref{S13}, in view of the wave function reduction principle, it is natural to
believe that the coordinate wave function of a quasiclassical state has a finite supporter.
For this reason we will assume that the functions $A(l)$ and $B(l)$ are not equal to zero only
if $l\in [L_1,L_2-1]$. Then as follows from Eq. (\ref{200})
\begin{eqnarray}
&&a(k)=\frac{[2/u_{\infty}]^{1/2}}{[(w_1+16k^2)(w_2+16k^2)]^{1/4}}
\sum_{l=L_1}^{L_2-1}A(l)cos[(l+1/2)\theta (k)]\nonumber\\
&&b(k)=\frac{[2/u_{\infty}]^{1/2}}{[(w_1+16k^2)(w_2+16k^2)]^{1/4}}
\sum_{l=L_1}^{L_2-1}B(l)sin[l\theta(k)]
\label{201}
\end{eqnarray}
These expressions are analogs of the relations between the coordinate and momentum wave functions
in the standard theory. It is clear from Eq. (\ref{201}) that, in contrast with the standard theory,
it is not possible to obtain an arbitrary falloff of the momentum wave function at $k\to\infty$
since $\theta(k)\to\pi$ when $k\to\infty$. As follows from Eqs. (\ref{171}) and (\ref{201}), when 
$k\to\infty$, $a(k)=O(1/k^2)$ and $b(k)=O(1/k^2)$.

In view of the above remarks, it is important to understand for which coordinate wave functions the
momentum wave functions $a(k)$ and $b(k)$ are as close to each other as possible. In the standard
theory the coordinate wave function should be differentiable and therefore it is natural to assume
that the functions $A(l)$ and $B(l)$ should not significantly differ from each other. As an analog
of the wave function (\ref{103}), we consider 
\begin{equation}
A(l)=\frac{exp[i(l+1/2)\theta(k_0)]}{[2(L_2-L_1)]^{1/2}}\quad 
B(l)=-i\frac{exp[il\theta(k_0)]}{[2(L_2-L_1)]^{1/2}}
\label{202}
\end{equation}
if $l\in [L_1,L_2-1]$ and $A(l)=B(l)=0$ otherwise. Then the wave function is normalized to one and we 
will see below that such a choice of the phase factors is needed to obtain as close dependence of 
$a(k)$ and $b(k)$ on $k$ as possible. The meaning of $k_0$ is that $2k_0/R=q_0$ where $q_0$ is the
classical relative momentum of the two-body system. 

As follows from Eqs. (\ref{201}) and (\ref{202})
\begin{eqnarray}
&&a(k)=\frac{1}{2}[u_{\infty}(L_2-L_1)]^{-1/2}\frac{exp[\frac{i}{2}(L_1+L_2)\theta(k_0)]}
{[(w_1+16k^2)(w_2+16k^2)]^{1/4}}\nonumber\\
&&\{exp[\frac{-i}{2}(L_1+L_2)\theta(k)]\frac{sin[\frac{L_2-L_1}{2}(\theta(k)-\theta(k_0))]}
{sin[\frac{1}{2}(\theta(k)-\theta(k_0))]}+\nonumber\\
&&\{exp[\frac{i}{2}(L_1+L_2)\theta(k)]\frac{sin[\frac{L_2-L_1}{2}(\theta(k)+\theta(k_0))]}
{sin[\frac{1}{2}(\theta(k)+\theta(k_0))]}\}\nonumber\\
&&b(k)=\frac{1}{2}[u_{\infty}(L_2-L_1)]^{-1/2}\frac{exp[\frac{i}{2}(L_1+L_2-1)\theta(k_0)]}
{[(w_1+16k^2)(w_2+16k^2)]^{-1/4}}\nonumber\\
&&\{exp[\frac{-i}{2}(L_1+L_2-1)\theta(k)]\frac{sin[\frac{L_2-L_1}{2}(\theta(k)-\theta(k_0))]}
{sin[\frac{1}{2}(\theta(k)-\theta(k_0))]}-\nonumber\\
&&\{exp[\frac{i}{2}(L_1+L_2-1)\theta(k)]\frac{sin[\frac{L_2-L_1}{2}(\theta(k)+\theta(k_0))]}
{sin[\frac{1}{2}(\theta(k)+\theta(k_0))]}\}
\label{203}
\end{eqnarray} 
Since our coordinate wave function is normalized to 1/2 in $H_1$ and $H_2$ (and the total
probability equals 1) it is clear that 
$$\sum_{k=0}^{\infty}|a(k)|^2=\sum_{k=0}^{\infty}|b(k)|^2=1/2$$
by construction. 

Consider the normalization of $a(k)$ if only the first term in Eq. (\ref{203})
is taken into account. By using Eq. (\ref{171a}) and the definition of the variable $\theta$, we can 
replace summation over $k$ by integration over $\theta$:
\begin{equation}
\sum_{k=0}^{\infty} |a(k)|^2 =\frac{1}{4\pi (L_2-L_1)}\int_0^{\pi}
|\frac{sin[\frac{L_2-L_1}{2}(\theta-\theta_0)]}
{sin[\frac{1}{2}(\theta-\theta_0)]}|^2d\theta
\label{204}
\end{equation} 
where $\theta_0=\theta(k_0)$. Let us assume that only a small vicinity of $\theta$ near $\theta_0$
gives an important contribution. Then we can replace $sin[(\theta-\theta_0)/2]$ by
$(\theta-\theta_0)/2$ and formally integrate over $\theta$ from $-\infty$ to $\infty$. Then by using
Eq. (\ref{105}) we obtain that indeed $a(k)$ is normalized to 1/2. This result justifies our assumption
and shows that the main contribution is given by the values of $\theta$ such that 
$|\theta-\theta_0|\leq 1/(L_2-L_1)$. 
 If $\eta$ is of order $\mu_1\mu_2 r/R$ and the masses are of order $m$ then $l$ is
of order $mr$ and $L_2-L_1$ is of order $m\Delta r$ where $\Delta r$ is the uncertainty of $r$.  
For macroscopic bodies this value is very large. 

A simple explanation that the main contribution is
given by $|\theta-\theta_0|\leq 1/(L_2-L_1)$ is as follows. At such values of $\theta$, 
$|sin[(L_2-L_1)(\theta-\theta_0)/2]/sin[(\theta-\theta_0)/2]|^2$ is of order $(L_2-L_1)^2$,
the factor $1/(L_2-L_1)$ arises from the fact that $|\theta-\theta_0|\leq 1/(L_2-L_1)$ and therefore
the overall contribution is of order unity. The contribution of such values of $\theta$ where
$sin[(L_2-L_1)(\theta-\theta_0)/2]$ cannot be replaced by $(L_2-L_1)(\theta-\theta_0)/2$ is order
$1/(L_2-L_1)$ since the integral is of order unity. This explanation also makes it clear that for
macroscopic bodies the main contributions to $a(k)$ and $b(k)$ are given by the first terms in
the corresponding expressions in Eq. (\ref{203}) and therefore $a(k)\approx b(k)$ as expected.
In addition, it follows from Eq. (\ref{171}) that at least in the nonrelativistic approximation
the condition $|\theta-\theta_0|\leq 1/(L_2-L_1)$ is equivalent to the standard uncertainty
relation $\Delta q\Delta r \geq 1$.

Consider now whether the mean value of the mass operator is close to $M(q_0)$ where $q_0=2k_0/R$.
It is clear from the above discussion that the contribution of $k's$ not close to $k_0$
is suppressed by a factor $1/(L_2-L_1)$. The question arises whether the extent of suppression
is sufficient to exclude a contribution of large values of $k$. As already noted, when $k$ is
large, $a(k)=O(1/k^2)$ and $b(k)=O(1/k^2)$. Therefore the sum for the mean value of the mass
operator converges, in contrast with the situation discussed in Sect. \ref{S13}. However, one
can notice that each of two terms for $a(k)$ and $b(k)$ in Eq. (\ref{203}) behaves as $O(1/k)$.
The terms with $O(1/k)$ cancel out only at $k\geq (L_2-L_1)/u_{\infty}$. This is an 
extremely large value. For example, if $m_1$ and $m_2$ are of the same order $m$ then
$(L_2-L_1)/u_{\infty}$ is of order $mR(L_2-L_1)$. Therefore the mean value of the mass operator
is anomalously large.

As noted in the preceding section, for quasiclassical particles the value $j=0$ is not realistic
and in practice the value of $j$ is very large. It has been also noted that at large values of
$k$ the eigenstates of the operator ${\bf D}^2$ drop as $O(k^{-(j+2)})$ beginning from $k$'s
of order $(m_1+m_2)/v_0$. Therefore in this case the factor $1/(L_2-L_1)$ is sufficent to
suppress the contribution of large $k$'s. At the same time, a problem with an anomalously large 
contribution from small values of $k$ arises. Let us estimate this contribution. Suppose that $j$ is
even. Then, as follows from Eqs. (\ref{195}) and (\ref{196}), the contribution of small $k$'s to
the eigenvectors of the operator ${\bf D}^2$ can be written as  
\begin{eqnarray}
&&E_1(l)=const\sum_k [(l+1/2)\theta (k)]^{1/2}N_{(j+1/2)}((l+1/2)\theta (k))e_{2k}\nonumber\\
&&E_2(l)=const\sum_k [l\theta(k)]^{1/2}J_{(j+1/2)}(l\theta(k))e_{2k+1}
\label{205}
\end{eqnarray}
where $N_{(j+1/2)}=(-1)^{j+1}J_{-(j+1/2)}$ is the Bessel function of the second kind (the Neumann
function). As an analog of Eq. (\ref{202}) for the coordinate wave functions one can take
\begin{eqnarray}
&&A(l)= const [(l+1/2)\theta (k_0)]^{1/2}H_{(j+1/2)}^{(1)}((l+1/2)\theta (k_0))\nonumber\\
&&B(l)= const [l\theta(k_0)]^{1/2}H_{(j+1/2)}^{(1)}(l\theta(k_0))
\label{206}
\end{eqnarray}
if $l\in [L_1,L_2-1]$, where $H_{(j+1/2)}^{(1)}$ is the Bessel function of the third kind (the Hankel function).
Indeed, in situations when for the Hankel functions one can take their asymptotic expressions \cite{BE},
Eq. (\ref{206}) is compatible with Eq. (\ref{202}). 
As follows from Eqs. (\ref{205}) and (\ref{206}),
\begin{eqnarray}
&&a(k)=const[\theta (k)\theta (k_0)]^{1/2}\sum_{l=L_1}^{L_2-1} (l+1/2) 
N_{(j+1/2)}((l+1/2)\theta (k))\nonumber\\
&&H_{(j+1/2)}^{(1)}((l+1/2)\theta (k_0))\nonumber\\
&&b(k)=const[\theta(k)\theta(k_0)]^{1/2}\sum_{l=L_1}^{L_2-1} l J_{(j+1/2)}(l\theta(k))\nonumber\\
&&H_{(j+1/2)}^{(1)}(l\theta(k_0))
\label{207}
\end{eqnarray}
Suppose that the sums in these expressions can be replaced by integrals. Then by using the
formulas for the integrals involving Bessel functions of different kinds \cite{BE} we obtain
\begin{eqnarray}
&&a(k)=const\frac{[\theta (k)\theta (k_0)]^{1/2}}{\theta(k)^2-\theta (k_0)^2}\nonumber\\
&&[(l+1/2)\theta(k)N_{(j+3/2)}((l+1/2)\theta (k))H_{(j+1/2)}^{(1)}((l+1/2)\theta (k_0))-\nonumber\\
&&(l+1/2)\theta(k_0)N_{(j+1/2)}((l+1/2)\theta (k))H_{(j+3/2)}^{(1)}((l+1/2)\theta (k_0))]_{l=L_1}^{l=L_2-1}\nonumber\\
&&b(k)=const\frac{[\theta (k)\theta (k_0)]^{1/2}}{\theta(k)^2-\theta (k_0)^2}
[l\theta(k)J_{(j+3/2)}(l\theta (k))H_{(j+1/2)}^{(1)}(l\theta (k_0))-\nonumber\\
&&l\theta(k_0)J_{(j+1/2)}(l\theta (k))H_{(j+3/2)}^{(1)}(l\theta (k_0))]_{l=L_1}^{l=L_2-1}
\label{208}
\end{eqnarray}
In situations when the Bessel functions can be replaced by their asymptotic expressions, these
expressions are compatible with Eq. (\ref{203}). However, the expression for $a(k)$ is formally
divergent when $k\to 0$ and it is not possible to avoid the divergency if the coordinate wave function
has a finite supporter. As noted in the preceding section, in fact there is no infinite contribution to
$a(k)$ since the continuous approximation is not valid when $k\leq j$. 

Let us summarize the discussion in this chapter. Since the internal two-body space $H$ can be represented
as a direct sum $H=H_1+H_2$ such that $H_1$ and $H_2$ are invariant subspaces of the operator ${\bf D}^2$,
the internal two-body state can be represented as a direct sum of the states in $H_1$ and $H_2$.
If $a(k)$ and $b(k)$ are the momentum wave functions in $H_1$ and $H_2$, respectively, then the standard
expression for the mean value of the mass operator can be obtained only in the approximation 
$a(k)\approx b(k)$. At the same time, the above discussion shows that the behavior of $a(k)$ and $b(k)$
at small values of $k$ is essentially different. As argued in the preceding section, one might expect
that the continuous approximation is not sufficient at small $k$'s and the contribution of the
region $k\leq j$ might be very important. As noted in the preceding section, there is no analog of such
a contribution in the standard theory. The investigation of the contribution of small $k$'s requires 
further study but in any case it seems extremely unlikely that $a(k)\approx b(k)$ and thus the
contribution of the last term in Eq. (\ref{standard}) is negligible. Any nonnegligible contribution
to the mean value of the mass operator implies that on classical level there exists an effective
interaction. If one accepts that important effective interaction does arise then the only reasonable
assumption is that this effective interaction is just gravity.

\chapter{Discussion}
\label{Ch6}

We postulate that on quantum level the dS invariance implies that
the expressions in Eq. (\ref{2}) are valid. They do not contain any free parameters and, in particular, 
they do not contain the cosmological constant. Therefore in such a formulation the cosmological 
constant problem does not arise. We argue in Chap. \ref{Ch2} that the cosmological constant arises only
if one wishes to describe the results in terms of Poincare invariant theories or classical dS space. 
We also mention the discussion about fundamental constants in Ref. \cite{Okun} and give additional
arguments in favor of the statement of the first author that a fundamental physical 
theory should not contain dimensionful constants at all. In particular, we argue that the gravitational 
and cosmological constants are not fundamental and that there are no physical arguments in favor of 
an approach where Poincare invariance is treated as fundamental and de Sitter invariance - as a result 
of nonzero vacuum energy in the theory with the Poincare background. 

In Chap. \ref{Ch2} we also discuss in detail the well known example of de Sitter antigravity. Although
the de Sitter antigravity has been discussed by numerous authors, it seems rather strange that important
features of this phenomenon have not been discussed. Although on classical level the dS antigravity
means a repulsion between particles in the dS space, a problem arises whether the dS antigravity can
be treated as an interaction at all. The phenomenon of the dS antigravity does not obey the rule of
QFT that any interaction is a result of exchange of some virtual particles. The classical dS antigravity
is even more universal than gravity since the relative acceleration (or rather retardation) of particles 
in the dS space 
does not depend on their masses and depends only on the radius of the de Sitter space $R$. However,
the fact that the acceleration is proportional to $1/R^2$ does not mean that $1/R^2$ can be treated
as an interaction constant. The dS antigravity is simply a consequence of dS invariance and it can 
be called an interaction only if we accept by definition that relative acceleration implies interaction.
However, such a notion about interactions is based only on our experience in Poincare invariant
theories. 

The phenomenon of the dS antigravity also poses the problem whether our understanding of other
interactions should be revisited. Indeed, at first glance the dS antigravity might be important only at
cosmological distances and therefore it is not important for quantum physics. However, consider, for
example, any interacting two-body system in dS theory. Since at large distances the dS antigravity is
much greater than all the other known interactions, the dS Hamiltonian cannot contain bound states,
its spectrum is the same as the spectrum of the free Hamiltonian and the free and interacting
Hamiltonians are unitarily equivalent \cite{jpa1}. In Poincare invariant theory it is also a
possible situation when the bound states are absent and therefore the free and interacting
Hamiltonians are unitarily equivalent. However, in this case the criterion of interaction is whether 
the S matrix is an identity operator or not. In dS theory there is no S matrix (or the notion of
the S matrix can be only approximate) and therefore the problem arises whether unitarily equivalent
representations are equivalent physically.

Also a question arises whether, by analogy with the dS antigravity, all interactions in nature manifest
themselves only as interactions in Poincare terms while in fact they are simply a consequence of
a higher symmetry. We believe that gravity might be the first candidate on the role of such an "interaction".
The phenomenon of gravity is known only on macroscopic level at nonrelativistic momenta and large
distances. So is very natural to consider two-body quasiclassical wave functions in dS theory, 
trying to understand whether gravity might be simply a consequence of de Sitter invariance.
Although such a problem statement seems to be extremely simple and natural, we are not aware of the 
literature where such a possibility has been discussed.  

In Chap. \ref{Ch3} we note that although quasiclassical approximation in standard quantum mechanics
has been discussed in a wide literature, it is not quite clear yet what are typical quasiclassical
wave functions for macroscopic bodies. A possible definition of quasiclassical states might be
as follows. Let $A$ be an operator of a physical quantity and a system is in a state where the
mean value of $A$ is $A_0$ and the uncertainty of $A$ is $\Delta A$. Then the system is quasiclassical
in $A$ if $|\Delta A|\ll |A_0|$. The system is typically meant to be quasiclassical if it is 
simultaneously quasiclassical in all the coordinates, momenta and energy. A typical
example when this is the case is when the system wave function is Gaussian. Since the Fourier 
transform of the Gaussian wave function is again a Gaussian wave function, such a wave function 
has a good behavior at infinity in both, the coordinate and momentum representations. The question
arises whether Gaussian wave functions are realistic quasiclassical wave functions. Indeed, if
we accept the reduction wave function postulate in the framework of the Copenhagen formulation then
after each measurement of the coordinate the wave function is not equal to zero only in a small
range of coordinates. 

Consider, for example, the Sun - Earth system. We know that the relative wave function in
the coordinate space has a very sharp maximum at $r\approx 150\cdot 10^6km$. There is no
doubt that the probability to find the Earth on the Venus or Mars orbits is extremely small. But is 
this probability exactly zero? One might think that this question is of academic interest only. 
However, if we accept the Copenhagen formulation, then after
each measurement of the Sun - Earth distance, the wave function of the Sun - Earth
system is not equal to zero only in a very small vicinity of $r\approx 150\cdot 10^6km$. 
In the Copenhagen formulation, the measurement is treated as an interaction with a
classical object. So it is not quite clear how often the relative distance in the Sun - Earth system 
is measured. 

This example indicates that realistic wave functions of macroscopic bodies might have only
a small finite supporter. But then, as noted in Sect. \ref{S13}, it is not obvious what types of
wave functions ensure that the system is quasiclassical in all the coordinates, momenta and energy. 
In particular, it is not difficult to construct wave functions where uncertainties of coordinates
and momenta are such that $\Delta r \Delta p$ is of order $\hbar$, but we should also guarantee
that such wave functions are quasiclassical in energy.

In dS theory the operators of de Sitter energy and de Sitter momenta do not commute with each other.
Therefore there is the energy-momentum uncertainty relation and the problem arises when a system
is simultaneously quasiclassical in energy and momentum. This problem is discussed in Chap. \ref{Ch4}.
We derive a relation (\ref{143}), which shows that a system can be simultaneously quasiclassical 
in dS energy and momentum only if the cosmological constant is not anomalously small. We also
estimate the quantities entering this expression for the case of the famous Cavendish experiment.
At the same time, our conclusions are based on the assumption that the coordinate dependence of
quasiclassical wave functions in dS theory is similar to the standard one. To understand whether
this is the case, one needs to explicitly construct the relative distance operator in dS theory.

Such a construction is carried out in Chap. \ref{Ch5}, which contains the main results of the
paper. It is shown that when the relative momentum is not asymptotically small or asymptotically large,
our standard intuition works. However, the behavior of eigenstates at very large and very small
momenta considerably differs from the standard one. At large momenta the falloff of the eigenstates
of the relative distance operator is much faster than in standard theory. The modern approaches to
gravity assume that on quantum level gravity manifests itself at Planck distances, which are
associated with very large momenta. On the other hand, as already noted, the phenomenon of gravity 
has been observed so far only at nonrelativistic momenta and large distances. So the question
arises whether indeed quantum gravity is sought where it really is. Since the approaches based on 
QFT, string theory, loop quantum gravity etc. are now dominant, this question usually is not posed. 
On the other hand, our result poses the question whether the region of large 
momenta is indeed relevant for explaining quantum gravity. Let us stress that we consider only 
two-body de Sitter kinematics and no dynamics is assumed. The only uncertainty in our consideration
is whether our relative distance operator is indeed the physical relative distance operator.
Although we argue that it is, a problem arises what is the
uncertainty in choosing the form of the relative distance operator in dS theory. In any case
our result seems to be more natural than the standard assumption.

Our calculations also show that an anomalously large (but finite) contribution arises from the region
of extremely small momenta. There is no analog of such a contribution in the standard theory.
The region in question is such that the de Sitter momentum $n$, which is dimensionless, is less
or smaller than the spin of the two-body system $j$. Indeed, the quantity $n$ is related to the
standard relative momentum $q$ as $n=qR$, where $R$ is the radius of the de Sitter world.
Since the standard theory corresponds to the formal limit $R\to\infty$, the relation $n\leq j$
in the standard theory is impossible. In orther words, the contribution of the region $n\leq j$
is possible only in the case when the cosmological constant is finite. It is clear that in this
region the standard intuition does not work. As noted in Sect. \ref{S18}, the contribution of
this region might be large since the decomposition coefficients describing the eigenstates of
the relative distance operator rise exponentially when $n$ is rising but is still in the region
$n\leq j$. Since typically small momenta are related to large distances, these results is an
additional argument (see the above discussion) that the phenomenon of the dS antigravity is 
important not only at cosmological distances. 

In Sect. \ref{S19} we argue that the existence of anomalously large contribution from the
region of small momenta implies that some effective interaction arises. The nature of this interaction
has to be investigated but a natural assumption is that it is just gravity. In other words,
the dS antigravity and gravity might be only different aspects of dS invariance. 
The problem of investigating the region of small momenta is that here the continuous approximation
does not work and discreteness of the de Sitter momentum becomes important. The fact that the
decomposition coefficients rise exponentially and become huge numbers might be an indication
that gravity might be explained as a manifestation of the fact that the ultimate quantum theory
is in fact based not on the field of complex numbers but on a Galois field. This possibility has
been proposed in our Refs. \cite{jmp0,lev3,FF} and a further study is required. 

We would like to conclude the paper with the following remark. Although quantum theory and de
Sitter invariance are already known for many years, such a seemingly simple problem as 
de Sitter invariant quantum mechanics of the free two-body system is far from being understood.
The above paper gives a strong indication that the investigation of this problem might
considerably change our understanding of quantum theory.

{\it Acknowledgements: } The author is grateful to Sergey Dolgobrodov, Alik Makarov,
Mikhail Borisovich Mensky, Ulrich Mutze, Volodya Netchitailo, Michel Planat, Metod Saniga, 
Skiff Nikolaevich Sokolov and Teodor Shtilkind for stimulating discussions.

\end{document}